\documentclass[aps,preprint,preprintnumbers,nofootinbib,showpacs,prd]{revtex4}
\usepackage{graphicx,color}
\usepackage{amsmath}
\usepackage{amssymb}
\usepackage{multirow}
\usepackage{ulem}
\newcommand{\lsim}{\mathrel{\mathop{\kern 0pt \rlap
  {\raise.2ex\hbox{$<$}}}
  \lower.9ex\hbox{\kern-.190em $\sim$}}}
\newcommand{\gsim}{\mathrel{\mathop{\kern 0pt \rlap
  {\raise.2ex\hbox{$>$}}}
  \lower.9ex\hbox{\kern-.190em $\sim$}}}
\newcommand{\ie}{\textit{i.e.}, }

\newcommand{\gev}{{\,{\rm GeV}}}

\newcommand{\ifb}{{\,{\rm fb}^{-1}}}
\newcommand{\neu}{\tilde{\chi}^0}
\newcommand{\neuo}{{\tilde{\chi}^0_1}}

\newcommand{\chao}{{\tilde{\chi}^\pm_1}}
\newcommand{\chaop}{{\tilde{\chi}^+_1}}
\newcommand{\chaom}{{\tilde{\chi}^-_1}}

\newcommand{\ww}{W^+W^-}
\newcommand{\chat}{{\tilde{\chi}^\pm_2}}
\newcommand{\smu}{{\tilde{\mu}}}
\newcommand{\smur}{\tilde{\mu}_R}
\newcommand{\smul}{\tilde{\mu}_L}

\newcommand{\casea}{\texttt{Case-A}}
\newcommand{\caseb}{\texttt{Case-B}}
\newcommand{\casec}{\texttt{Case-C}}

\newcommand{\gm}{{\gamma}}

\newcommand{\Gm}{{\Gamma}}

\newcommand{\beq}{\begin{equation}}
\newcommand{\eeq}{\end{equation}}
\newcommand{\bea}     {\begin{eqnarray}}
\newcommand{\eea}     {\end{eqnarray}}
\newcommand{\bit}{\begin{itemize}}
\newcommand{\eit}{\end{itemize}}
\newcommand{\ben}{\begin{enumerate}}
\newcommand{\een}{\end{enumerate}}

\newcommand{\no}{{\nonumber}}

\newcommand{\me}{{\rlap/{\!E}}}

\newcommand{\cosmax}{|\cos\Theta|_{\rm max} }
\newcommand{\maa}{m_{aa}}
\newcommand{\maac}{m^{\rm cusp}_{aa}}
\newcommand{\maamax}{m^{\rm max}_{aa}}
\newcommand{\maamin}{m^{\rm min}_{aa}}
\newcommand{\eamax} {E_a^{\rm max}}
\newcommand{\eamin} {E_a^{\rm min}}
\newcommand{\eaamax} {E_{aa}^{\rm max}}
\newcommand{\eaacusp} {E_{aa}^{\rm cusp}}
\newcommand{\eaamin} {E_{aa}^{\rm min}}
\newcommand{\exxmax} {E_{XX}^{\rm max}}
\newcommand{\exxcusp} {E_{XX}^{\rm cusp}}
\newcommand{\exxmin} {E_{XX}^{\rm min}}
\newcommand{\mllc}{m^{\rm cusp}_{\mu\mu}}

\newcommand{\mllmax}{m^{\rm max}_{\mu\mu}}
\newcommand{\mllmin}{m^{\rm min}_{\mu\mu}}
\newcommand{\elmax} {E_\mu^{\rm max}}
\newcommand{\elmin} {E_\mu^{\rm min}}
\newcommand{\ellmax} {E_{\mu\mu}^{\rm max}}
\newcommand{\ellcusp} {E_{\mu\mu}^{\rm cusp}}
\newcommand{\ellmin} {E_{\mu\mu}^{\rm min}}
\newcommand{\mwwc}{m^{\rm cusp}_{WW}}
\newcommand{\mwwmax}{m^{\rm max}_{WW}}
\newcommand{\mwwmin}{m^{\rm min}_{WW}}
\newcommand{\ewmax} {E_W^{\rm max}}
\newcommand{\ewmin} {E_W^{\rm min}}
\newcommand{\ewwmax} {E_{WW}^{\rm max}}
\newcommand{\ewwcusp} {E_{WW}^{\rm cusp}}
\newcommand{\ewwmin} {E_{WW}^{\rm min}}

\newcommand{\mrec}{m_{\rm rec}}
\newcommand{\erec}{E_{\rm rec}}
\newcommand{\mrecmin}{m_{\rm rec}^{\rm min}}
\newcommand{\mrecc}{m_{\rm rec}^{\rm cusp}}
\newcommand{\mrecmax}{m_{\rm rec}^{\rm max}}

\newcommand{\mpt}{\rlap/p_T}

\newcommand{\key}[1]{}

\definecolor{pink}{rgb}{1,0.2,1}

\begin{document}
\preprint{PITT-PACC 1402}

\title{Determining the Dark Matter Particle Mass\\
through Antler Topology Processes\\
at Lepton Colliders}
\author{Neil D. Christensen$^{1,2}$, Tao Han$^{2,3}$,  Zhuoni Qian$^{2}$,  Josh Sayre$^{2}$,  Jeonghyeon Song$^4$,
and Stefanus$^{2}$}
\affiliation{\footnotesize{$^{1}$Department of Physics, Illinois State University, Normal, IL 61790 USA\\
$^{2}$Pittsburgh Particle physics
Astrophysics and Cosmology Center, Department of Physics and Astronomy, University of Pittsburgh, Pittsburgh, PA 15260 USA\\
$^3$ Korea Institute for Advanced Study, Seoul 130-722, Korea\\
$^4$ Division of Quantum Phases \& Devices, School of Physics,
KonKuk University, Seoul 143-701, Korea}}
\begin{abstract}
We study the kinematic cusps and endpoints of processes with the ``antler topology''
as a way to measure the masses of the parity-odd missing particle and the intermediate parent at a high energy lepton collider. The fixed center of mass energy at a lepton collider
makes many new physics processes suitable for the study of the antler decay topology.  It also
provides new kinematic observables with cusp structures, optimal for the missing mass determination.
We also study realistic effects on these observables, including initial state radiation, beamstrahlung, acceptance cuts, and detector resolution.
We find that the new observables, such as the reconstructed invariant mass of invisible particles and the summed energy of the observable final state particles, appear to be more stable than the commonly considered energy endpoints against realistic factors and are very efficient at measuring the missing particle mass.
For the sake of illustration,
we study smuon pair production and chargino pair production
within the framework of the minimal supersymmetric standard model.
We adopt the log-likelihood method  to optimize the analysis. We find that at the 500 GeV ILC,
 a precision of approximately $0.5\gev$ can be achieved in the case of smuon production  with a leptonic final state, and approximately $2\gev$ in the case of chargino production  with a hadronic final state.
\end{abstract}
\maketitle


\section{Introduction}
\label{sec:introduction}

With the monumental discovery of the Higgs boson at the LHC~\cite{Higgs},
all of the fundamental particles in the standard model (SM)  have been discovered. The SM as an effective field theory can be valid up to a very high scale.
Nevertheless, there are strong indications that the SM is incomplete.
Certain observed particle physics phenomena
cannot be accounted for within the SM.
Among them, the discovery and characterization of the dark matter (DM) particle may be one of the most pressing issues.

The existence of dark matter has been well established
through a combination of galactic velocity rotation
curves~\cite{dm:rotation curve},
the cosmic microwave background~\cite{Komatsu:2010fb},
Big Bang nucleosynthesis~\cite{bbn},
gravitational lensing~\cite{grav-lense},
and the bullet cluster \cite{Clowe:2006eq}.
As a result of these observations, we know that dark matter is non-baryonic, electrically neutral
and composes roughly 23\% of the energy
and 83\% of the matter of the universe.

Among the many possibilities for dark matter \cite{Feng:2010gw},
weakly interacting massive particles (WIMPs) are arguably the most attractive because of the so-called WIMP miracle:
to get the relic abundance right, a WIMP mass is roughly
\bea
M_{\rm WIMP} \lsim {g^2\over 0.3}\ 1.8~{\rm TeV},
\eea
which miraculously coincides with the new physics scale expected from the ``naturalness'' argument for electroweak physics. Therefore, there is a high hope that the search for a dark matter particle may be intimately related to the discovery of TeV scale new physics.

Direct searches of weak scattering of dark matter off nuclear targets in underground labs have been making
great progress in improving the sensitivity to the DM mass and couplings, most recently by the XENON \cite{xenon}, LUX \cite{null} and SuperCDMS \cite{SuperCDMS} collaborations.
WIMPs can also be produced at colliders either directly in pairs or from cascade decays of other heavier particles. Since a WIMP is non-baryonic and electrically neutral, it does not leave any trace in the detectors and thus only appears as missing energy.
In order to establish a DM candidate convincingly,
it is ultimately important to reach consistency between direct searches and collider signals for the common parameters of mass, spin and coupling strength.

\begin{figure}[!tb]
\centering
  \includegraphics[height=4.7cm]{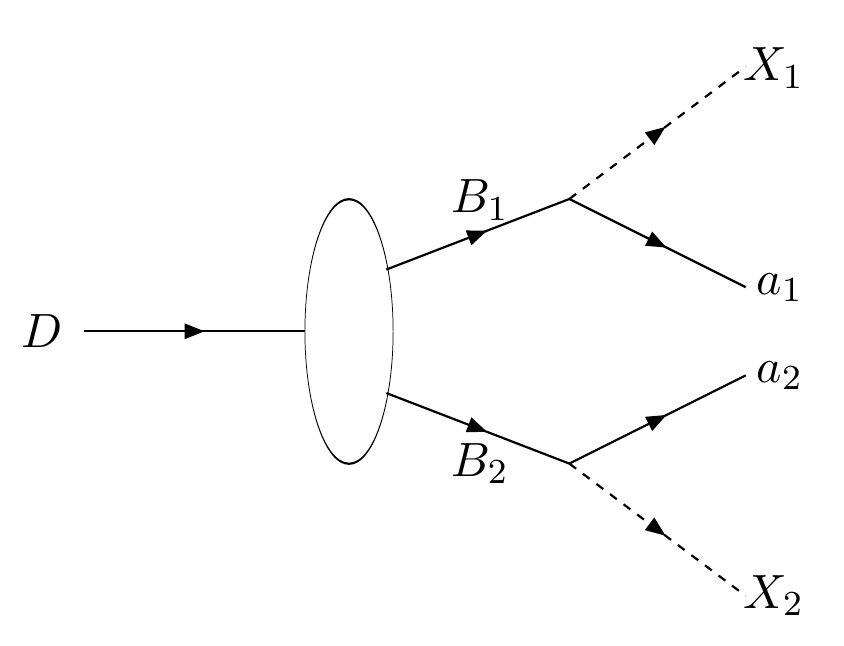}
  \caption{\label{fig:feyn:antler}
The antler decay diagram of a heavy particle $D$ into two visible particles
$a_1$ and $a_2$ and two invisible particles $X_1$ and $X_2$
through on-shell intermediate particles $B_1$ and $B_2$.}
\end{figure}

It is very challenging to determine the missing particle mass at colliders due to the under-constrained kinematical system with two missing particles in an event. It is particularly difficult at hadron colliders
because of the unknown partonic c.m.~energy and frame. There exist many attempts  to determine the missing particle mass at the LHC, such as endpoint methods~\cite{endpoint}, polynomial methods~\cite{polynomial},
$M_{T2}$ methods~\cite{MT2}, and the matrix element method \cite{MEM}.
Recently, we studied the ``antler decay'' diagram \cite{Han:short:antler},
as illustrated in Fig.~\ref{fig:feyn:antler} with
a resonant decay of a heavy particle $D$
into two parity-odd particles ($B_1$ and $B_2$) at the first step,
followed by each $B_i$'s decay into a missing
particle $X_i$ and a visible particle $a_i$.
We found that a resonant decay through the antler diagram
develops cusps in some kinematic distributions
and the cusp positions along with the endpoint positions
determine the missing particle mass as well as the intermediate particle mass~\cite{Han:short:antler,Han:long:antler,Agashe1}.



In this article, we focus on lepton colliders \cite{ILC,TLEP,CLIC,muon collider}, in which the antler topology applies.
The initial state is well-defined with fixed c.m.~energy and c.m.~frame.
This allows various antler processes without going through a resonant decay of a heavy particle $D$.
We consider kinematic variables such as the angle and the energy of a visible particle
for the mass determination.
We also show that the invariant mass of two invisible particles,
which can be indirectly reconstructed using the recoil mass technique,
is crucial for the mass measurement and the SM background suppression.
The energy sum of the two visible particles or of the two invisible particles will also be shown to be equally powerful.
At a linear $e^+ e^-$ collider, the available beam polarization
can additionally be used to suppress the SM background and enhance the
sensitivity of the mass measurement.

Two common methods of the missing mass measurement have been studied in the literature for $e^+e^-$ collisions:
\begin{enumerate}
\item The lepton energy endpoints in cascade decays \cite{Martyn:El};
\item The photon energy endpoint in the direct WIMP pair production associated with a photon \cite{eeXXgamma}.
\end{enumerate}
In comparison, we find that our results from the antler topology can be at least comparable to the energy endpoint method and do much better than the single photon approach.
For the sake of illustration,
we will concentrate on the minimal supersymmetric standard model (MSSM) and consider the scenario where the lightest neutralino $\neuo$ is the lightest supersymmetric particle (LSP) and, therefore, stable in the framework of a R-parity conserving scenario.
We consider two MSSM processes that satisfy the antler topology: pair production of scalar muons (smuons) and that of charginos. In order to be as realistic as possible with the kinematical construction, we analyze the effects of
the initial state radiation (ISR), beamstrahlung, acceptance cuts, and detector resolutions on the observables.
We adopt the log-likelihood method based on Poisson statistics to quantify the
precision of the mass measurements. We find that this method optimizes the sensitivity to the mass parameters in the presence of these realistic effects.

We note that the scanning through the pair production threshold could give a much more accurate determination for the intermediate parent mass \cite{threshold}. With this as an input, one could improve the measurement of the missing particle mass by the energy endpoint method or by the Antler technique. However, the threshold scan would require {\it a priori} knowledge of the intermediate particle mass, and would need more integrated luminosity to reach such a high sensitivity \cite{threshold}.
Our proposed method does not assume to know  any masses, and our outputs would benefit the design of the threshold scan.

The rest of the paper is organized as follows. In section \ref{sec:review},
we review the kinematic cusps and endpoints of antler processes.
We present the analytic expressions for six kinematic variables in terms of the masses.
For a benchmark scenario, we first show smuon pair production as an example of massless visible particles
in section \ref{sec:massless}.
We reproduce the expected kinematical features numerically
and illustrate the effects of the acceptance cuts on the final state observable particles.
Other realistic effects including full spin correlation, SM backgrounds,  ISR, beamstrahlung, and detector resolutions are considered.
Adopting the log-likelihood method based on the Poisson probability density, we quantify the accuracy with which the missing particle mass measurement may be determined in section \ref{sec:LL}.
In section \ref{sec:massive}, chargino pair production is studied,
as an example of massive visible particles with a hadronic final state.
In section \ref{sec:conclusions}, we give a summary and draw our conclusions.


\section{Cusps and endpoints of the antler process}
\label{sec:review}


We start from a state with a fixed c.m.~energy $\sqrt s$, which produces two massive particles $B_1$ and $B_2$,
followed by each $B$'s decay into a visible particle $a$ and an invisible heavy particle $X$,
 as depicted in Fig.~\ref{fig:feyn:antler}.
In $e^{+}e^-$ collisions, it is realized as
\bea
\label{eq:antler}
e^+ e^- & \to& B_1 +B_2,
\\ \no  &&
B_1 \to a_1+ X_1, \quad
B_2 \to a_2 +X_2.
\eea
For simplicity, we further assume that 
$B_1$ and $B_2$ ($X_1$ and $X_2$)
are identical particles to each other:
\bea
m_{B_1} =m_{B_2} \equiv m_B,\quad
m_{X_1} = m_{X_2}=m_X.
\eea
The kinematics is conveniently expressed by the rapiditiies $\eta_j$ (equivalent to the speed $\beta=|\vec{p}\,|/E$),
which specifies the four-momentum of a massive particle $j$ from a two-body decay of $i \rightarrow j + k$ in the rest frame of the parent particle $i$ as $p_{j}^{(i)} = m_j \left( \cosh\eta_{j},\ \hat{p}^{(i)}_j \sinh\eta_{j}  \right)$.
In general, the kinematics of Eq.(\ref{eq:antler}) is determined by three rapidities of the intermediate particle $B$,
the visible particle $a$, and
the missing particle $X$, given by
\bea
\label{eq:rapidity:W}
\cosh\eta_B = \frac{ \sqrt{s} }{2 m_B}, \ \
\cosh \eta_a = \frac{m_B^2 - m_X^2 + m_a^2}{2 m_a m_B}, \ \
\cosh \eta_X = \frac{m_B^2 - m_W^2 + m_X^2}{2 m_X m_B}.
\eea
Note that in the massless visible particle case $(m_a=0)$  
the rapidity $\eta_a$ goes to infinity.

We find the distributions of the following six kinematic variables informative:
\bea
m_{aa}, \quad \mrec, \quad \cos\Theta, \quad E_a, \quad E_{aa}, \quad
E_{XX}.
\eea

\begin{table}[t!]
\centering
{
\renewcommand{\arraystretch}{1.1}
\begin{tabular}{|c|c|c|c|}
\hline
        & ${\cal R}_1:~\eta_{B} < \frac{\eta_{a}}{2}$ &
        ~~${\cal R}_2:~\frac{\eta_{a}}{2} < \eta_{B} < \eta_{a}$~~ &
	${\cal R}_3:~\eta_{a} < \eta_{B}$ \\ \hline
~~$\maamin$~~ & \multicolumn{2}{c|}{$2 m_{a}$} &
                ~~$2 m_{a} \cosh (\eta_{B} -\eta_{a})$~~~ \\ \hline
$\maac$ & ~~$2 m_{a} \cosh(\eta_{B} -\eta_{a})$~~ &
     \multicolumn{2}{|c|}{$2 m_{a} \cosh \eta_{B}$}  \\ \hline
$\maamax$ & \multicolumn{3}{c|}{$2 m_{a} \cosh(\eta_{B} + \eta_{a})$} \\ \hline
\end{tabular}
\caption{\label{table:cusp:massive}
The cusp and endpoints of the invariant mass distribution
$m_{aa}$ in the three regions of c.m.~energy and parameter space.
}
}
\end{table}

\noindent
(\textit{i}) \textit{$m_{aa}$ distribution}:
$m_{aa}$ is the invariant mass of the two visible particles.
This distribution
accommodates three singular points: a minimum, a cusp, and a maximum.
Their positions are not uniquely determined by the involved masses.
They differ according to the relative scales of masses.
There are three regions~\cite{Han:long:antler}
\bea
{\cal R}_1: \eta_{B} < \frac{\eta_{a}}{2},
\quad
{\cal R}_2: \frac{\eta_{a}}{2} < \eta_{B} < \eta_{a},
\quad
{\cal R}_3: \eta_{a} < \eta_{B}.
\eea
The cusps and endpoints in the three regions are given in Table \ref{table:cusp:massive}.
The minimum endpoint is the same for ${\cal R}_1$ and ${\cal R}_2$ but
different for ${\cal R}_3$.
The cusp is the same  for ${\cal R}_2$ and ${\cal R}_3$,
which is different for ${\cal R}_1$.
The maximum endpoints are the same for all three regions.
The absence of {\it a priori} knowledge of the masses
gives us ambiguity among ${\cal R}_1$, ${\cal R}_2$,
and ${\cal R}_3$.
For example
we do not know
whether the measured
$\maamin$ is $2 m_a$ or
$2 m_a \cosh (\eta_{B} - \eta_a)$.

In the massless visible particle case, however,
three singular positions are uniquely determined as
\begin{eqnarray}
\label{eq:maa:min:c:max}
\maamin &=& 0 \,,\\ \no
\maac &=&
m_B \left( 1 - \frac{m_X^2}{m_B^2}\right) e^{-\eta_B}\,,\\ \no
\maamax &=&
m_B \left( 1 - \frac{m_X^2}{m_B^2}\right) e^{\eta_B} \,.
\end{eqnarray}
According to the analytic function for the $m_{aa}$ distribution~\cite{Han:short:antler},
the $m_{aa}$ cusp is sharp only when the $B$ pair production is near threshold,
\ie when $0.443 \sqrt{s} < m_B  < 0.5\sqrt{s}$.

\noindent
(\textit{ii}) \textit{$\mrec$  distribution}:
The invariant mass of two invisible particles, denoted by $\mrec$,
can be measured through the relation
\bea
\mrec^2 \equiv m_{XX}^2= s - 2 \sqrt{s} \left( E_{a_1} + E_{a_2}\right) + m_{aa}^2.
\eea
The $\mrec$ distribution is related to
the invariant mass distribution of massive visible particles
because of the symmetry of the antler decay topology.
It also has
three singular points,
$\mrecmin$, $\mrecc$, and $\mrecmax$.
Their positions are as in Table \ref{table:cusp:massive},
with replacement of $m_a \to m_X$ and $\eta_a \to \eta_X$.

\noindent
(\textit{iii}) \textit{$E_a$ distribution}:
The energy distribution of one visible particle 
in the lab frame also provides important information about the masses.
If the intermediate particle $B$ is a scalar particle like a slepton, its decay
is isotropic
and thus produces
a flat rectangular distribution.
Two end points,
$\eamin$ and $\eamax$,
are determined by the masses:
\bea
E_a^{\max,\min}
=
\frac{\sqrt{s}}{4}
\left(1-
\frac{m_X^2-m_a^2}{m_{B}^2}
\right)
\left(
1\pm\beta_B
\sqrt{1- \frac{4 m_a^2 m_{B}^2}{(m_B^2 + m_a^2 - m_X^2)^2}}
\right),
\eea
where $\beta_B$ is defined by
\bea
\label{eq:beta}
\beta_B = \sqrt{1-\frac{4 m_B^2}{s}}.
\eea
Note that if $m_B \ll \sqrt{s}/2$ or $m_X \approx m_B$,
then $\eamin$ can be very small, even below the experimental acceptance for observation.

\noindent
(\textit{iv}) \textit{$E_{a a}$ distribution}:
The distribution of the combined energy of the $a_1 a_2$ system,
$E_{aa} \equiv E_{a_1}+E_{a_2}$,
is triangular, leading to three singular positions, $\eaamin$, $\eaacusp$, and $\eaamax$,
which are in terms of masses
\bea
\label{eq:EWW:min}
E_{aa}^{\rm max,mix }&=& 2 m_a \cosh(\eta_a \pm \eta_B),  
\\
\eaacusp &=& 2 m_a \cosh\eta_a \cosh \eta_B.
\no
\eea
For $m_a=0$, we have simpler expressions as
\bea
\label{eq:Eaa:min:cusp:max}
\left. E_{aa}^{\rm max,mix }\right|_{m_a=0} &=& \frac{\sqrt{s}}{2} \left( 1-\frac{m_X^2}{m_B^2}\right) (1\pm\beta_B),
\\ \no
\left. \eaacusp\right|_{m_a=0}  &=&\frac{\sqrt{s}}{2} \left( 1-\frac{m_X^2}{m_B^2}\right).
\eea

\noindent
(\textit{v}) \textit{$E_{X X}$ distribution}:
Although the energy of one invisible particle is not possible to measure,
the sum of two invisible particle energies can be measured
through
\bea
E_{XX} \equiv E_{X_1}+ E_{X_2} = \sqrt{s} - E_{aa}.
\eea
The distribution of $E_{X X}$
is a mirror image of the $E_{aa}$ distribution,
which is triangular with a sharp cusp.

\noindent
(\textit{vi}) \textit{$\cos\Theta$ distribution}:
Here $\Theta$ is the angle between the momentum direction of one visible particle (say $a_1$)
in the c.m.~frame of $a_1$ and $a_2$
and the c.m.~moving direction of the pair in the lab frame.
For $m_a \neq 0$, the $\cos\Theta$ distribution 
does not present a sharp cusp or endpoint~\cite{Han:long:antler}.
If $m_a=0$, however,
the distribution has a simple functional form as
\bea
\label{eq:cosTheta}
\left. \frac{d\Gm}{d \cos\Theta} \right|_{m_a=0}
\propto
\left\{
\begin{array}{ll}
\dfrac{1}{\sin^3\Theta}, & \hbox{~~~for } \left|\cos\Theta\right| < \beta_B,\\[7pt]
0, &\hbox{~~~otherwise},
\end{array}
\right.
\eea
which accommodates two pronounced peaks
where the cusp and the maximum endpoint meet at $\cos\Theta=\pm\beta_B$.


\section{Massless visible particle cases: smuon pair production}
\label{sec:massless}

For the massless observable particles $a_{1}$ and $a_{2}$, we now present the general feature based on the previous discussions and demonstrate the observable aspects for the missing mass measurements at the ILC. 
Throughout this paper, we choose to show the results for the c.~m.~energy $$\sqrt{s}=500\gev.$$


\subsection{The kinematics of cusps and endpoints}
\label{sec:slepton}

A lepton collider is an ideal place to probe the charged slepton sector of the MSSM.
To illustrate the basic features of cusps and endpoints at the ILC,
we consider smuon pair production. In principle, the scalar nature of the smuon
can be determined by the shape
of the total
cross section
near threshold
and the angular distributions of the final muons \cite{Christensen:2013sea}.
There are two kinds of smuons, $\smul$ and $\smur$,
scalar partners of the left-handed and right-handed muons respectively.
A negligibly small mass of the muon
suppresses the left-right mixing and thus makes $\smul$ and $\smur$
the mass-eigenstates.
The smuon pair production in $e^+ e^-$ collisions
is via $s$-channel diagrams mediated by a photon or a $Z$ boson.
Since the exchanged particles are vector bosons,
the helicities of $e^+$ and $e^-$ are opposite to each other,
and only two kinds of pairs, $\smur^+\smur^-$
and $\smul^+\smul^-$,
are produced.
If the lightest neutralino $\neuo$ has a dominant Bino component,
$\smur$ predominantly decays into $\mu \neuo$.
The decay of $\smul \to\mu \neuo$ is also sizable.
At the ILC, the process $e^+ e^- \to \smur\smur/\smul\smul \to \mu\neuo+\mu\neuo$
has a substantial rate.  The final state we observe is
\bea
\label{eq:final}
e^+ e^- \to \mu^+ \mu^- + \me.
\eea
This is one good example of the antler process. However, we note that the leading SM process,
$W^{+}W^{-}$ production followed by $W \to \mu\nu_\mu$, is also of the antler structure.

\begin{table}[t]
\centering
{
\renewcommand{\arraystretch}{1.1}
\begin{tabular}{|c||c|c||c|c|c|c|c|c|}
\hline
Label &  $\smur$ & $\smul$& $\neuo$ & $\neu_2$ &
	$\neu_3$ & $\neu_4$ & $\chao$ & $\chat$ \\ \hline
~~\casea~(\caseb)~~&  ~~~158~~~ & ~636~(170)& ~141~ &
~529~ & ~654~ & ~679~ &
~529~ & ~679~ \\ \hline
\casec &  $-$ & $-$ & ~139~ &
~235~ & ~504~ & ~529~ &
~235~ & ~515~ \\ \hline
\end{tabular}
\caption{\label{table:sps1b:mass}
Illustrative SUSY mass spectrum for $\casea$, $ \caseb$ (as introduced in Sec.~\ref{sec:slepton})
and $\casec$ (as introduced in Sec.~\ref{sec:massive}).
All of the masses are in units of GeV.}
}
\end{table}

For illustrative purposes of the signals, we consider two benchmark points for the MSSM parameters,
called \casea~and \caseb, as listed in Table~\ref{table:sps1b:mass}.
These two cases have the same mass spectra,
except for the $\smul$ mass.
In \casea, $\smul$ is too heavy for the pair production
at $\sqrt{s}=500\gev$.
We have a simple situation
where the new physics signal for the final state in Eq.~(\ref{eq:final})
involves only $\smur\smur$ production.
In \caseb, the $\smul$ mass comes down close to
the $\smur$ mass,
with a mass gap of about 10 GeV.
In this case with $m_{\smur} \simeq m_{\smul}$,
the cross section of $\smur\smur$ production
is compatible with that of $\smul\smul$ production.
This is because the left-chiral and right-chiral couplings of the smuon to the $Z$ boson,
say $g^L_{\smu\smu Z}$ and $g^R_{\smu\smu Z}$ respectively,
are accidentally similar in size:
\bea
g^L_{\smu\smu Z} = \frac{-1 + 2 \sin^2\theta_W}{2 \sin\theta_W\cos\theta_W} \approx -0.64,
\quad
g^R_{\smu\smu Z} = \frac{\sin\theta_W}{\cos\theta_W} \approx 0.55.
\eea
In \caseb,
three signals from  $\smur\smur$, $\smul\smul$, and $\ww$
all  have the same antler decay topology. The goal is to disentangle the information and achieve the mass measurements of $\smur$, $\smul$, and $\neuo$.

It is noted that the LHC searches for slepton direct production
does not reach enough sensitivity with the current data yet \cite{LHCslept}
and would be very challenging in Run-II as well for the parameter choices under consideration, due to the small signal cross section, large SM backgrounds, and the disfavored kinematics of the small mass difference.
On the other hand, once crossing the kinematical threshold at a lepton collider,
the slepton signal could be readily established.

In Table \ref{table:cusp:values}, we list the values of various kinematic cusps and
endpoints for the five variables discussed above.
The mass spectra of the $\smur\smur$ antler and the $\ww$ antler
apply to both  \casea~and \caseb,
while that of $\smul\smul$ applies only to \caseb.
With the given masses, all of the minimum, cusp, and maximum positions
are determined.
They are considerably different from each other, indicating important complementarity
of these kinematic variables.

{\renewcommand{\arraystretch}{1.1}
\begin{table}[t!]
\centering
\begin{tabular}{|c||c|c|c|}
\hline
$\sqrt{s}$ & \multicolumn{3}{|c|}{$500\gev$} \\ \hline
~~~Production channel~~~& $\smur\smur$ & $\smul\smul$  & $\ww$ \\ \hline
input $(m_B,m_X)$ & $(158, 141)$ & $(170, 141)$ & $(m_W,0)$ \\ \hline\hline
$\cosmax $ & 0.77 & 0.73 & 0.95\\
\hline
$(\mllmin,\mllc,\mllmax)$ & $(0, 12,91)$ & $(0,21,137)$ & $(0,13,487)$\\
\hline
$(\mrecmin,\mrecc,\mrecmax)$ & ~~$(408,445,488)$~~ & ~~$(363,413,479)$~~ & $(0,13,487)$\\
\hline
$(\elmin,\elmax)$ & $(6,46)$ & $(11,69)$ & $(7,243)$ \\
\hline
$(\ellmin,\ellcusp,\ellmax)$ & $(12,52,92)$ & $(21, 79,137)$ & ~~$(13,250,487)$~~  \\ \hline
\end{tabular}
\caption{\label{table:cusp:values}
The values of various kinematic cusps and endpoints as seen
in Fig.~\ref{fig:cusp:generic},
for the mass parameters in Table \ref{table:sps1b:mass}.
All of the masses and energies are in units of GeV.
}
\end{table}
}


\begin{figure}[t!]
\centering
\includegraphics[width=.47\textwidth]{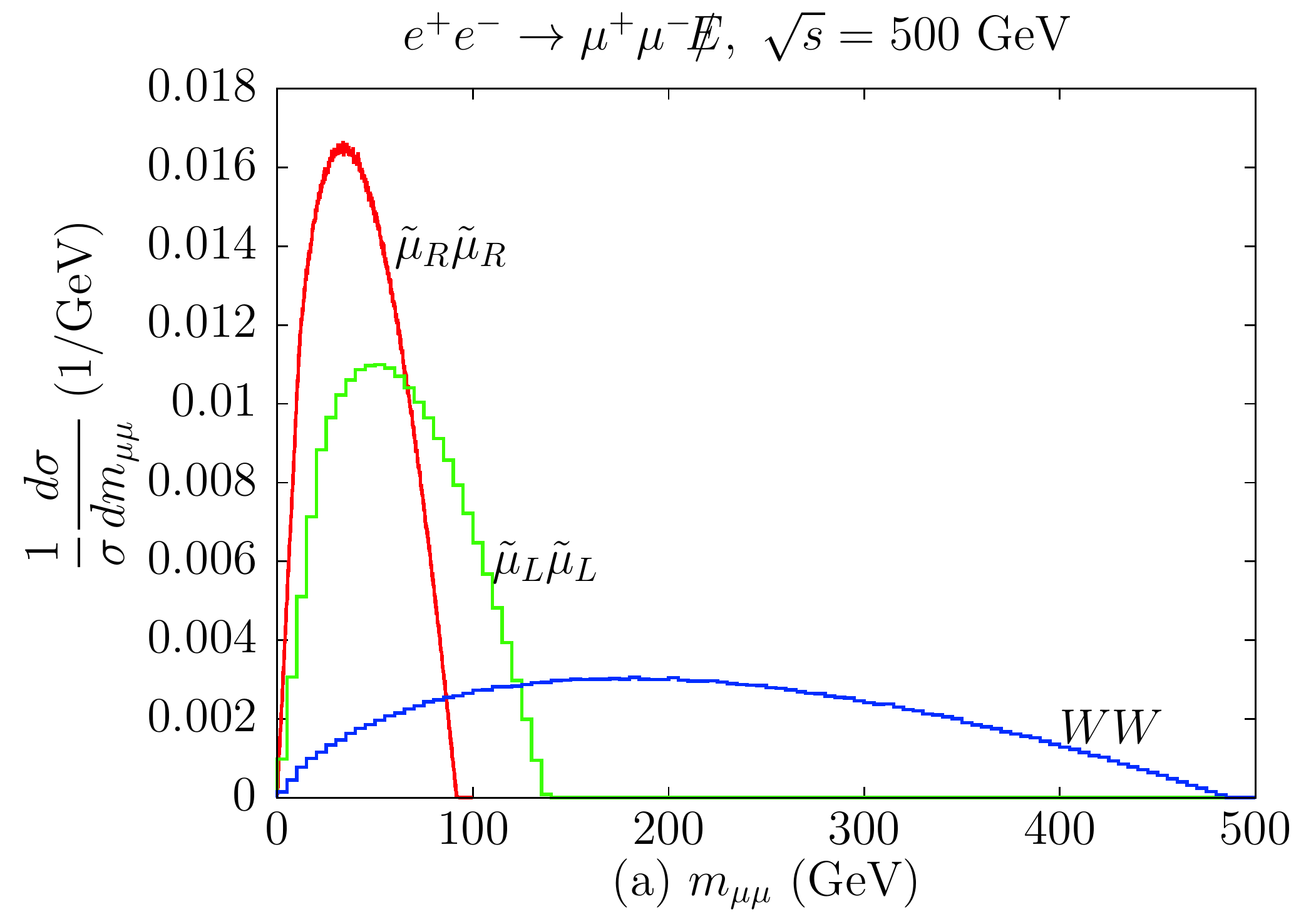}
~~
\includegraphics[width=.47\textwidth]{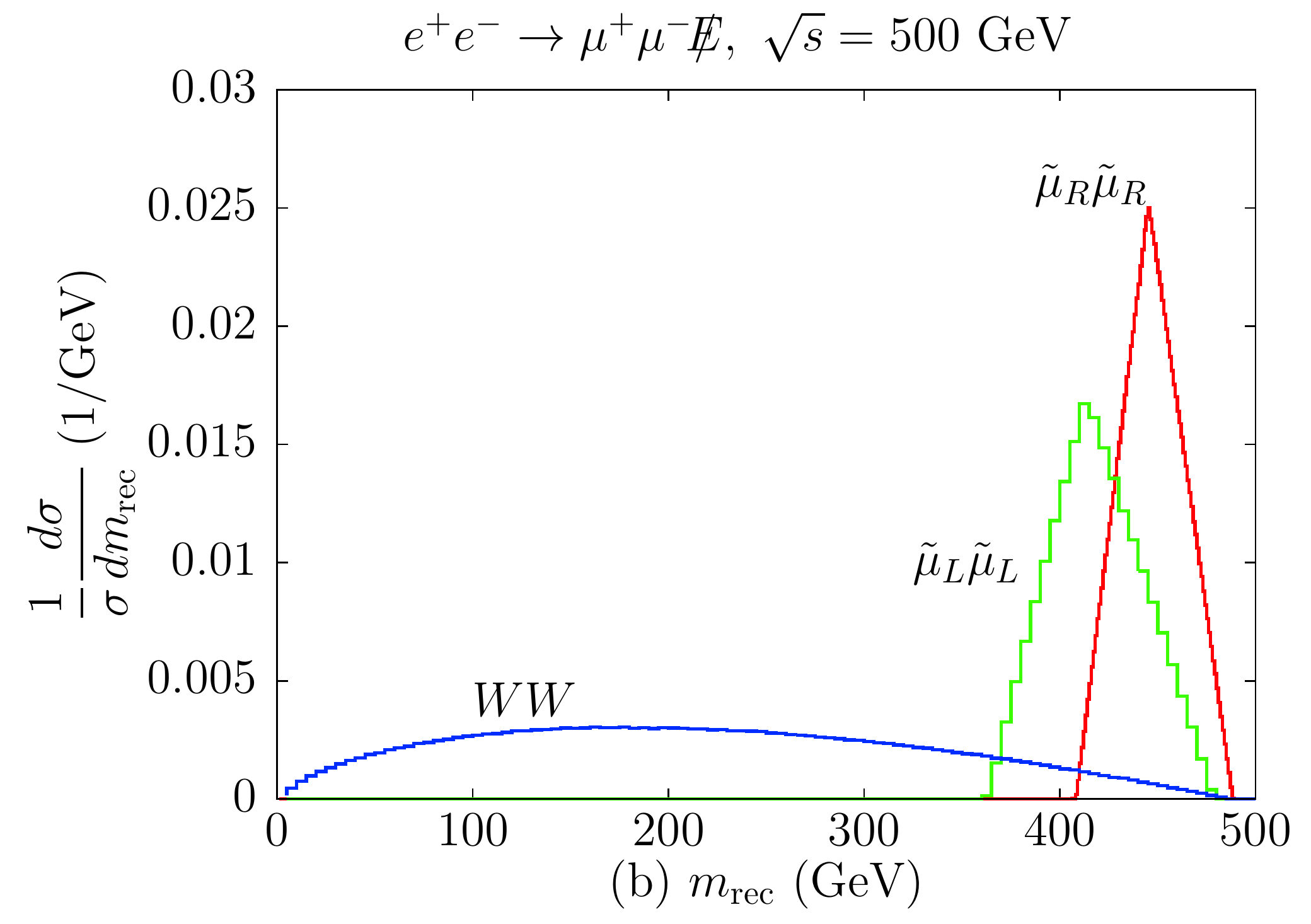}
\\
\includegraphics[width=.47\textwidth]{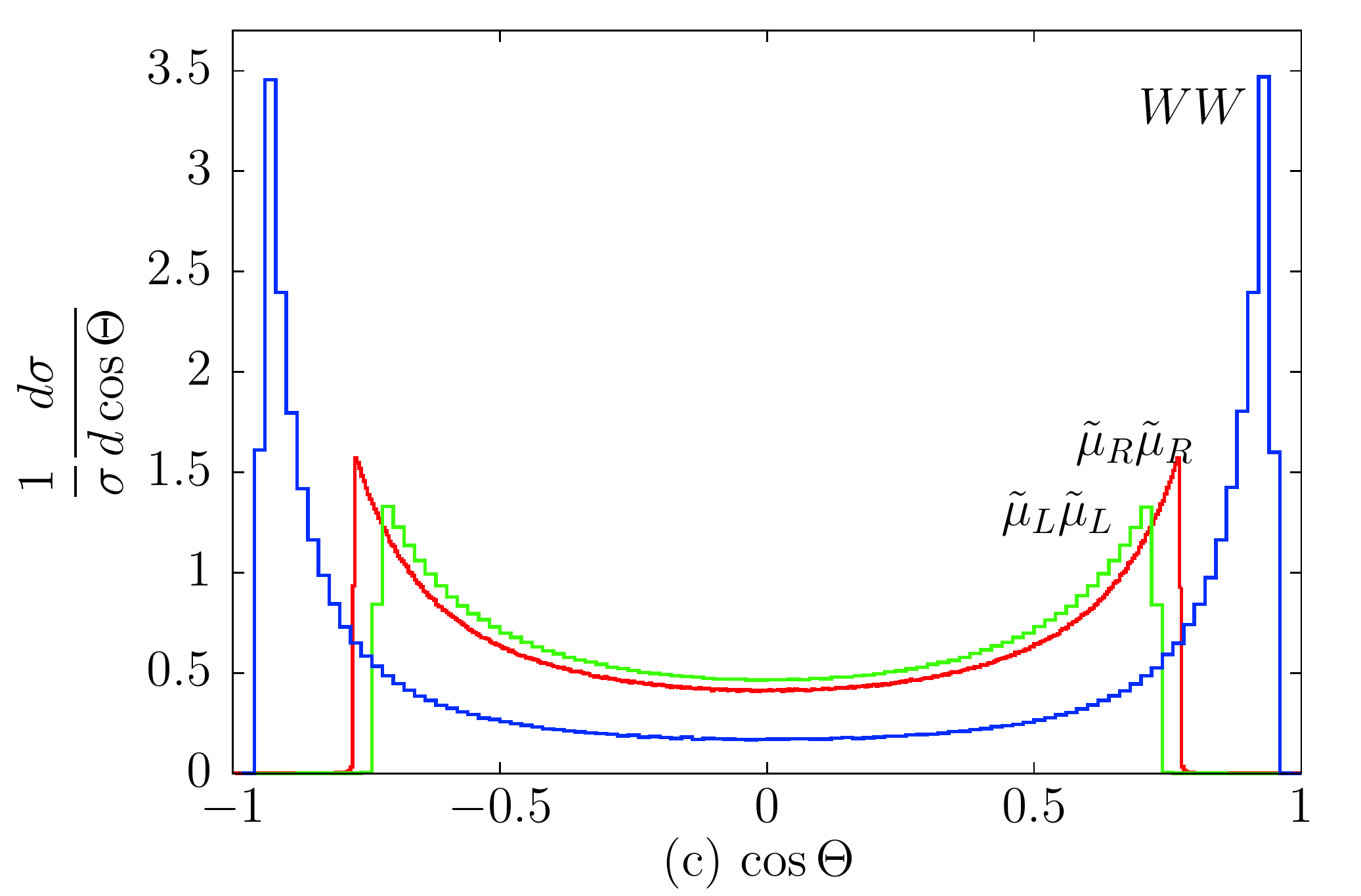}
~~
\includegraphics[width=.47\textwidth]{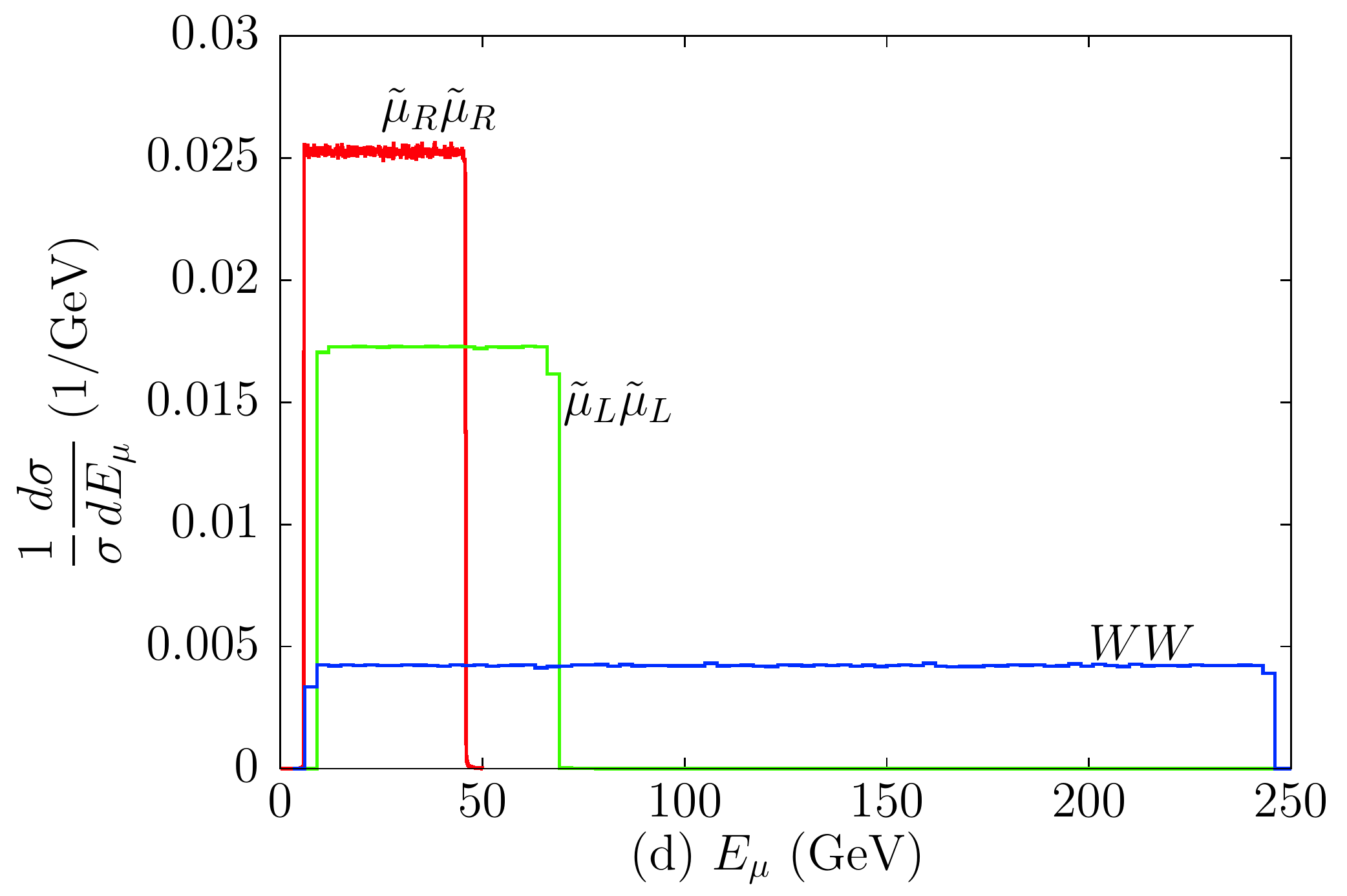}
\\
\includegraphics[width=.47\textwidth]{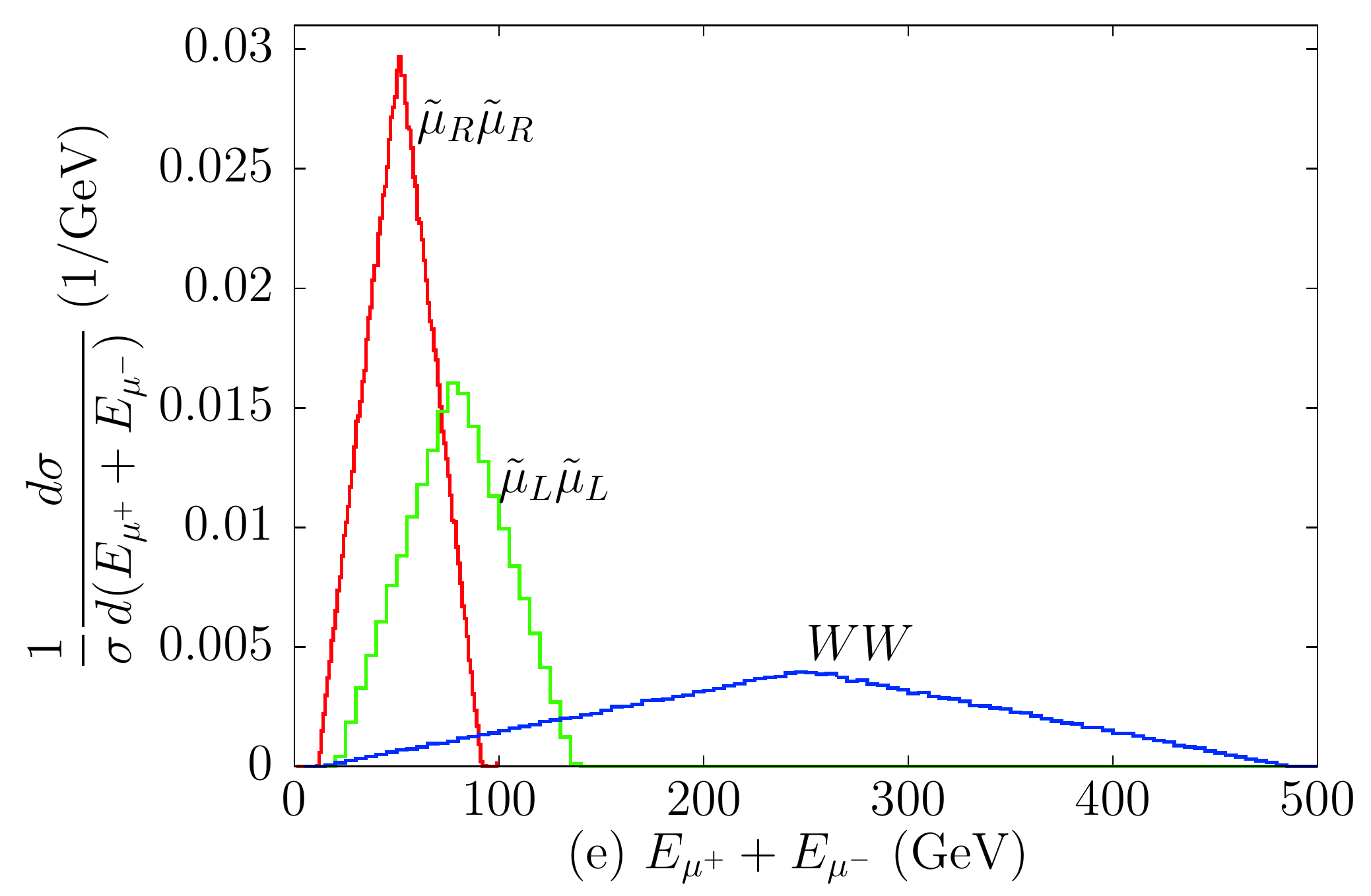}
\caption{\label{fig:cusp:generic}
The normalized distributions of (a) $m_{\mu\mu}$, (b) $\mrec$,
(c) $\cos\Theta$, (d)  $E_\mu$  and (e) $E_{\mu^+}+E_{\mu^-}$
for the three cases in Table \ref{table:cusp:values},
\ie for $\smur\smur$, $\smul\smul$ and $\ww$ production at $\sqrt{s}=500\gev$.
Here we consider only the kinematics without spin correlations.
}
\end{figure}

In Fig.~\ref{fig:cusp:generic}, we show the normalized distributions of (a) $m_{\mu\mu}$, (b) $\mrec$,
(c) $\cos\Theta$, (d) $E_\mu$,
and (e) $E_{\mu^+}+E_{\mu^-}$ for $\smur\smur$, $\smul\smul$, and $\ww$ production
at the ILC with a c.m.~energy of $500\gev$.
To appreciate the striking features of the distributions, we have only considered the kinematics here.
The full results including spin correlations, initial state radiation (ISR),
beamstrahlung, and detector smearing effects will be shown, beginning in
section  3.3.
First, the $m_{aa}$ distributions for $\smur\smur$, $\smul\smul$, and $\ww$ production do not
show a clear cusp. This is because the c.m.~energy is too high compared with the intermediate mass to reveal the $m_{aa}$ cusp, which would become pronounced when $m_B > 0.44 \sqrt{s}$~\cite{Han:short:antler}.
For $B=\smur$,  a sharp $m_{aa}$ cusp requires $\sqrt{s}\lsim 360\gev$.
On the contrary, the $\mrec$ distributions for $\smur\smur$ and $\smul\smul$
in Fig.~\ref{fig:cusp:generic}(b) are of the shape of a sharp triangle.
This is attributed to the massive $X$.
For $\ww$ production, the missing particles are massless neutrinos, therefore,
the $m_{aa}$ distribution is the same as the $\mrec$ distribution.

The $\cos\Theta$ distributions of $\smur\smur$, $\smul\smul$, and $\ww$
in Fig.~\ref{fig:cusp:generic}(c) present
the same functional behavior,
proportional to $1/\sin^3\Theta$.
There are two sharp points where the cusp and the maximum merge,
which correspond to $\pm|\cos\Theta|_{\rm max} $.
The $\smur\smur$ and $\smul\smul$ processes
have similar values of $|\cos\Theta|_{\rm max} $,
while the $\ww$ process peaks at a considerably larger value.
Figure \ref{fig:cusp:generic}(d) shows the energy distribution of one visible particle $\mu$.
The distributions for the smuon signals are flat due to their scalar nature, while the flat distribution for the $\ww$ channel is artificial due to the neglect of spin correlation. We will include the full spin effects from  section  \ref{sec:realistic considerations} and on.

In principle, the two measurements of $\elmin$ and $\elmax$
can determine the two unknown masses $m_B$ and $m_X$.
However the minimum of $E_a$ can be below the detection threshold as in the $\smur$ case of $\elmin \simeq 5.8\gev$.
One may thus need another independent observable to determine all the masses.
In addition, over-constraints
on the involved masses are very useful in establishing the new physics model.

The distribution of $E_{\mu\mu} (\equiv E_{\mu^+} +E_{\mu^-})$ in Fig.~\ref{fig:cusp:generic}(e)
is different from the individual energy distribution:
the former is triangular while the latter is rectangular.
For $\smur\smur$ and $\smul\smul$,
the $E_{aa}$ distributions are localized
so that the pronounced cusp is easy to identify.
For $\ww$, however,
the $E_{aa}$ distribution is widespread.

In order to further understand the singular structure,
we examine four representative configurations in terms of $(\cos\theta_1,\cos\theta_2)$,
where $\theta_1$ and $\theta_2$ are the polar angle of $a_1$ and $a_2$
in the rest frame of their parent particles $B_1$ and $B_2$, respectively.
The correspondence of each corner to a singular point is as follows:
\bea
\label{eq:1d:configuration}
{\renewcommand{\arraystretch}{1.2}
\begin{array}{c|c c c c }
\hbox{1D configuration } & m_{aa} & \mrec & E_{aa} & E_{XX}\\ \hline
~~(i)~~\stackrel{a_2}{\Longleftarrow}
~~\stackrel{B_2}{\longleftarrow}
~~\stackrel{{e^+}{e^-}}{\bullet}~~
\stackrel{B_1}{\longrightarrow}
~~
\stackrel{a_1}{\Longrightarrow}~
& ~~~\hbox{ max }~~~ & ~~~\hbox{ min }~~~& ~~~\hbox{ max }~~~& ~~~\hbox{ min }~~~ \\
(ii)~\stackrel{a_2}{\Longrightarrow}
~~\stackrel{B_2}{\longleftarrow}
~~\stackrel{{e^+}{e^-}}{\bullet}~~
\stackrel{B_1}{\longrightarrow}
~~
\stackrel{a_1}{\Longleftarrow}
& \hbox{ cusp } & \hbox{ max } & ~~~\hbox{ min }~~~ & \hbox{ max } \\
(iii)~\stackrel{a_2}{\Longrightarrow}
~~\stackrel{B_2}{\longleftarrow}
~~\stackrel{{e^+}{e^-}}{\bullet}~~
\stackrel{B_1}{\longrightarrow}
~~
\stackrel{a_1}{\Longrightarrow}
& \hbox{ min } & \hbox{ cusp } & \hbox{ cusp }& \hbox{ cusp }\\
(iv)~\stackrel{a_2}{\Longleftarrow}
~~\stackrel{B_2}{\longleftarrow}
~~\stackrel{{e^+}{e^-}}{\bullet}~~
\stackrel{B_1}{\longrightarrow}
~~
\stackrel{a_1}{\Longleftarrow}
& \hbox{ min } & \hbox{ cusp } & \hbox{ cusp }& \hbox{ cusp }\\[3pt]
\end{array}
}
\eea

%


\subsection{The effects of acceptance cuts}
\label{sec:cut:effect}

\begin{figure}[t!]
\centering
\includegraphics[width=.47\textwidth]{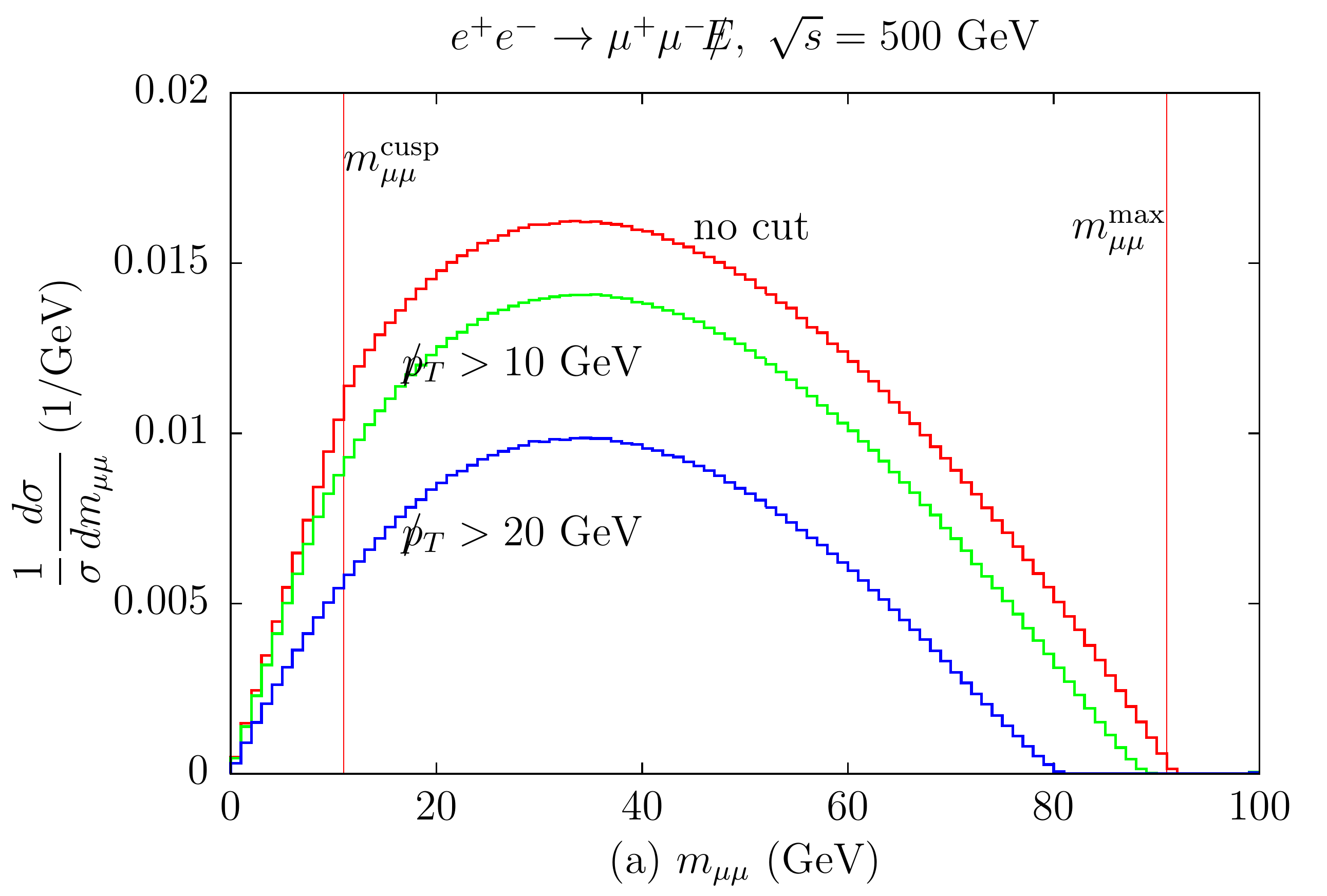}
~~
\includegraphics[width=.47\textwidth]{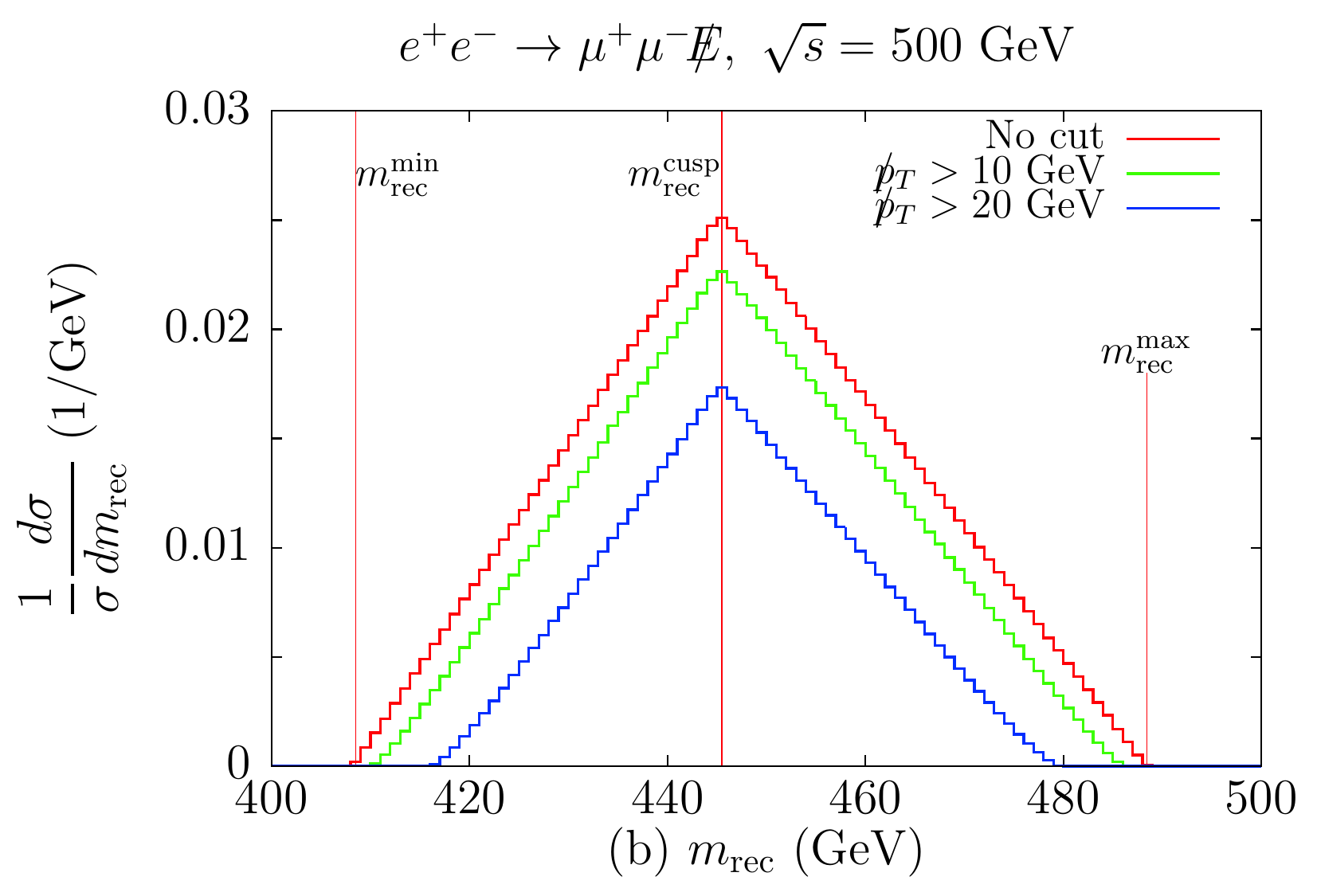}
\\[10pt]
\includegraphics[width=.45\textwidth]{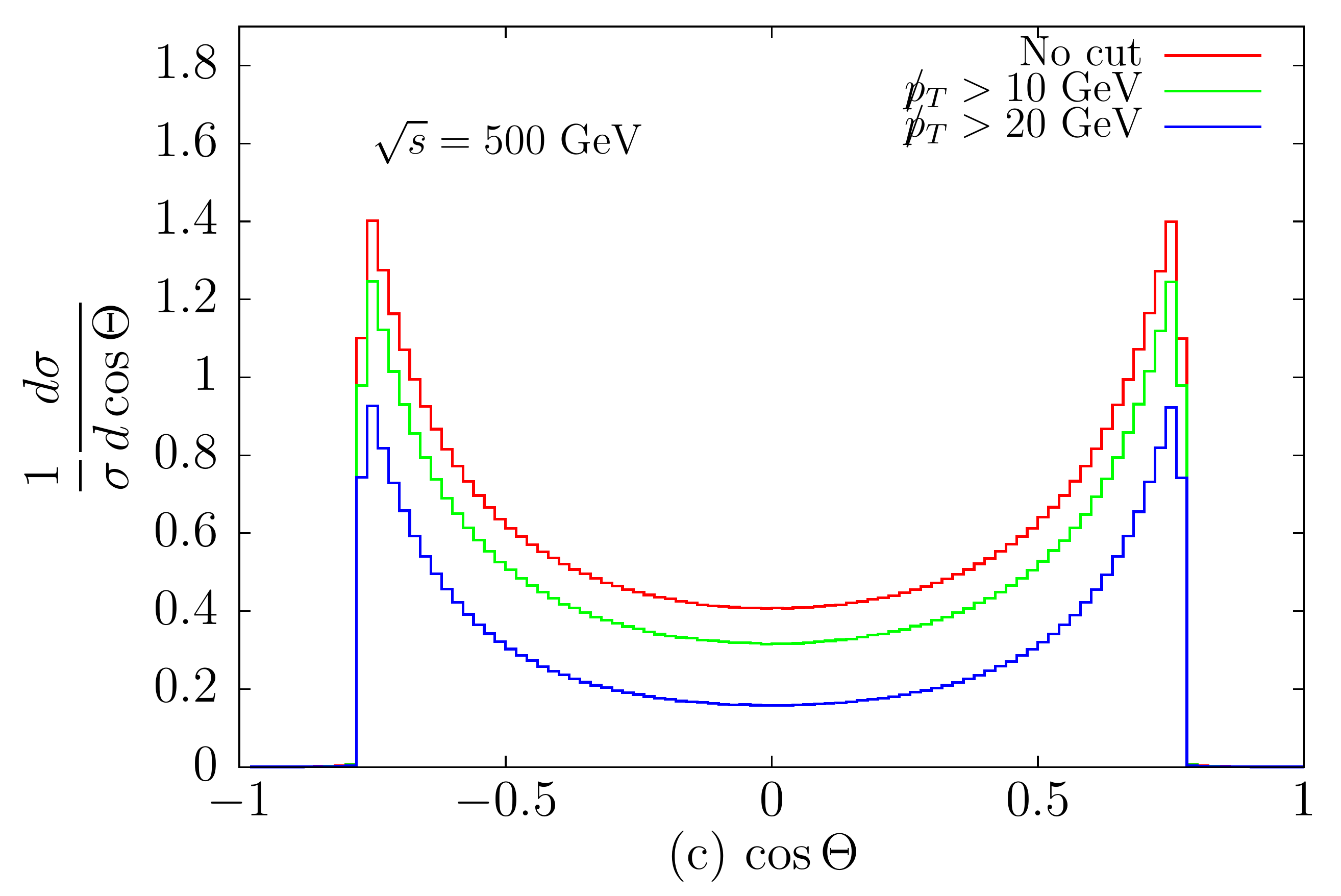}
~~
\includegraphics[width=.47\textwidth]{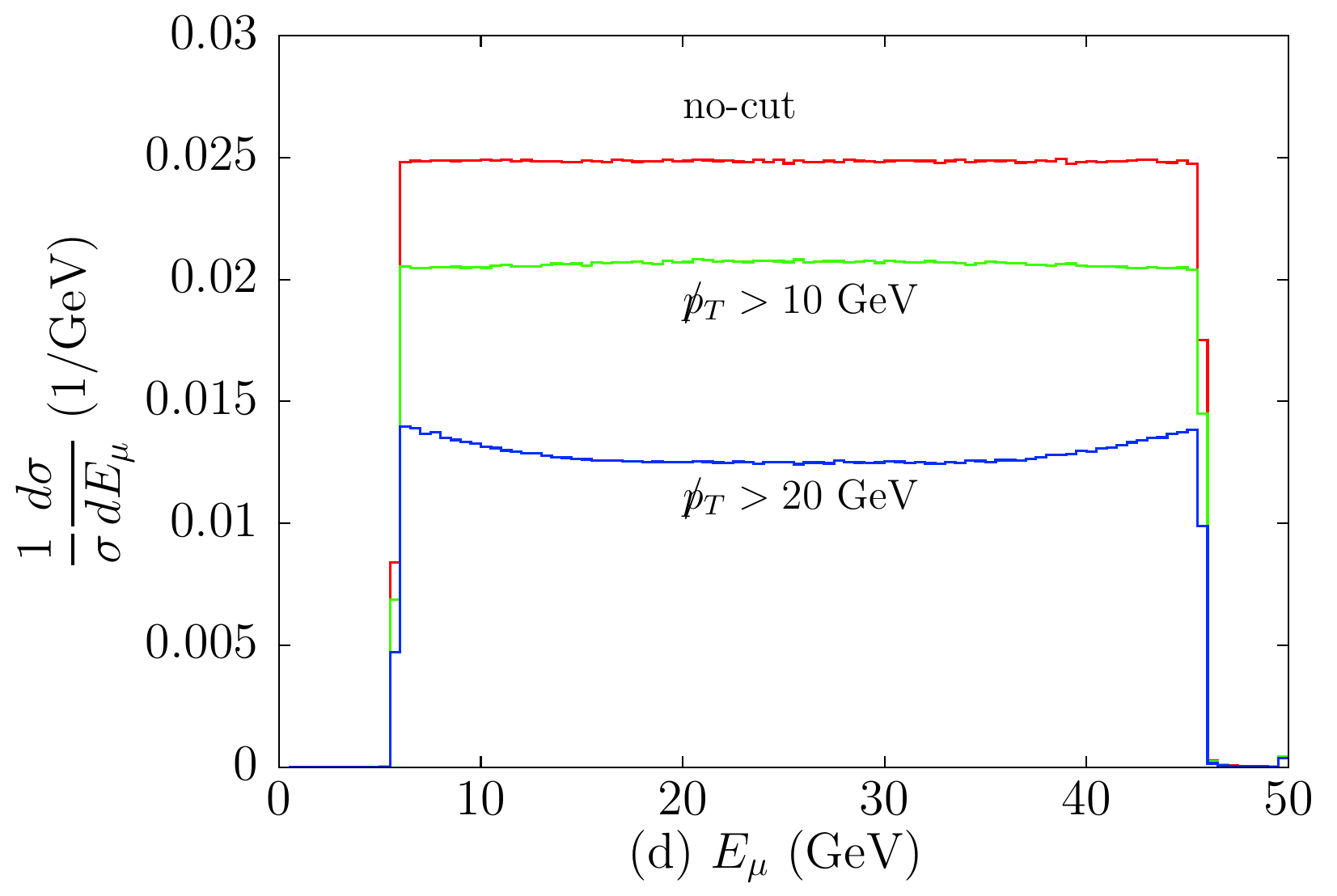}
\\[10pt]
\includegraphics[width=.47\textwidth]{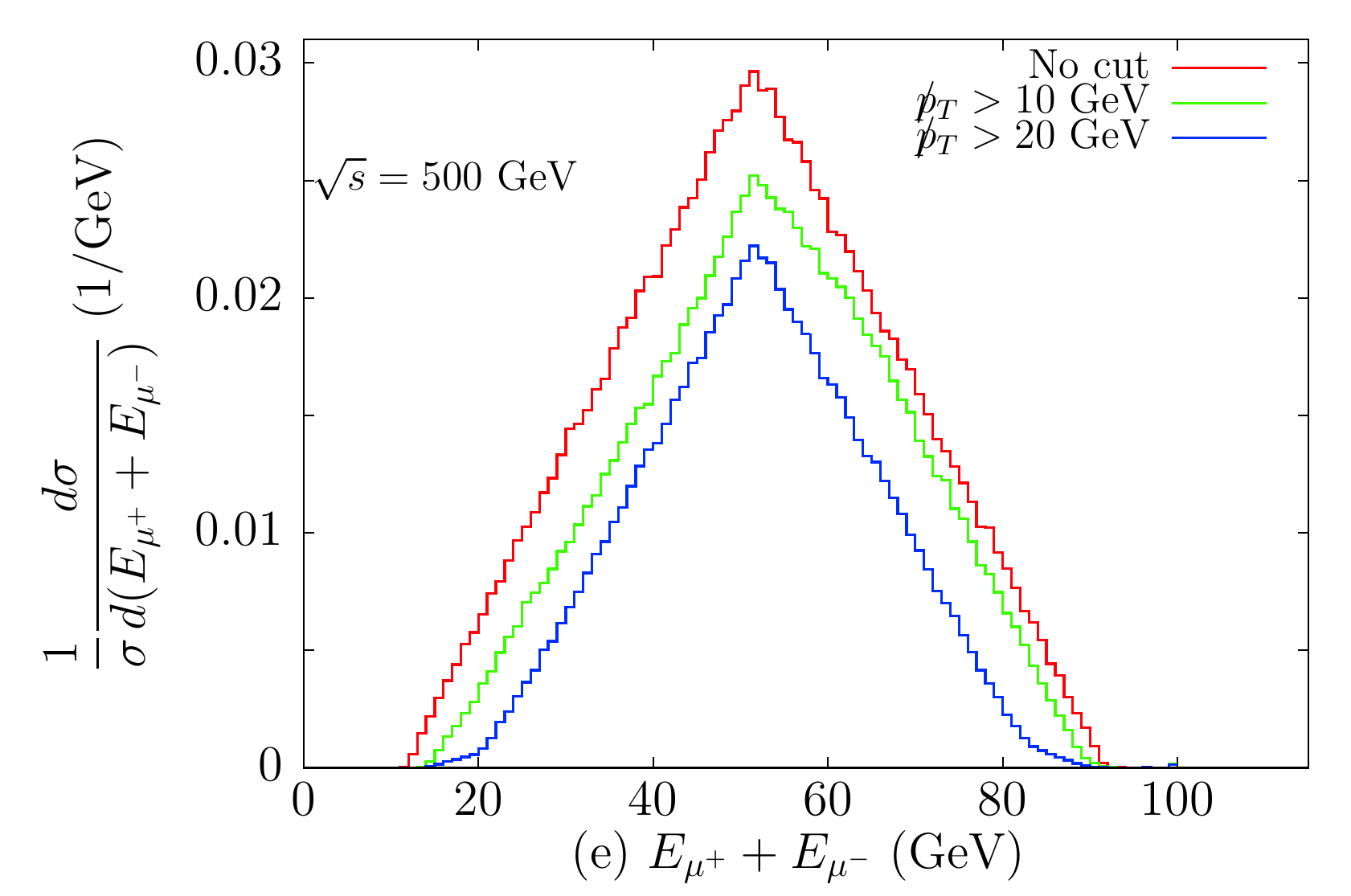}
~~
\includegraphics[width=.47\textwidth]{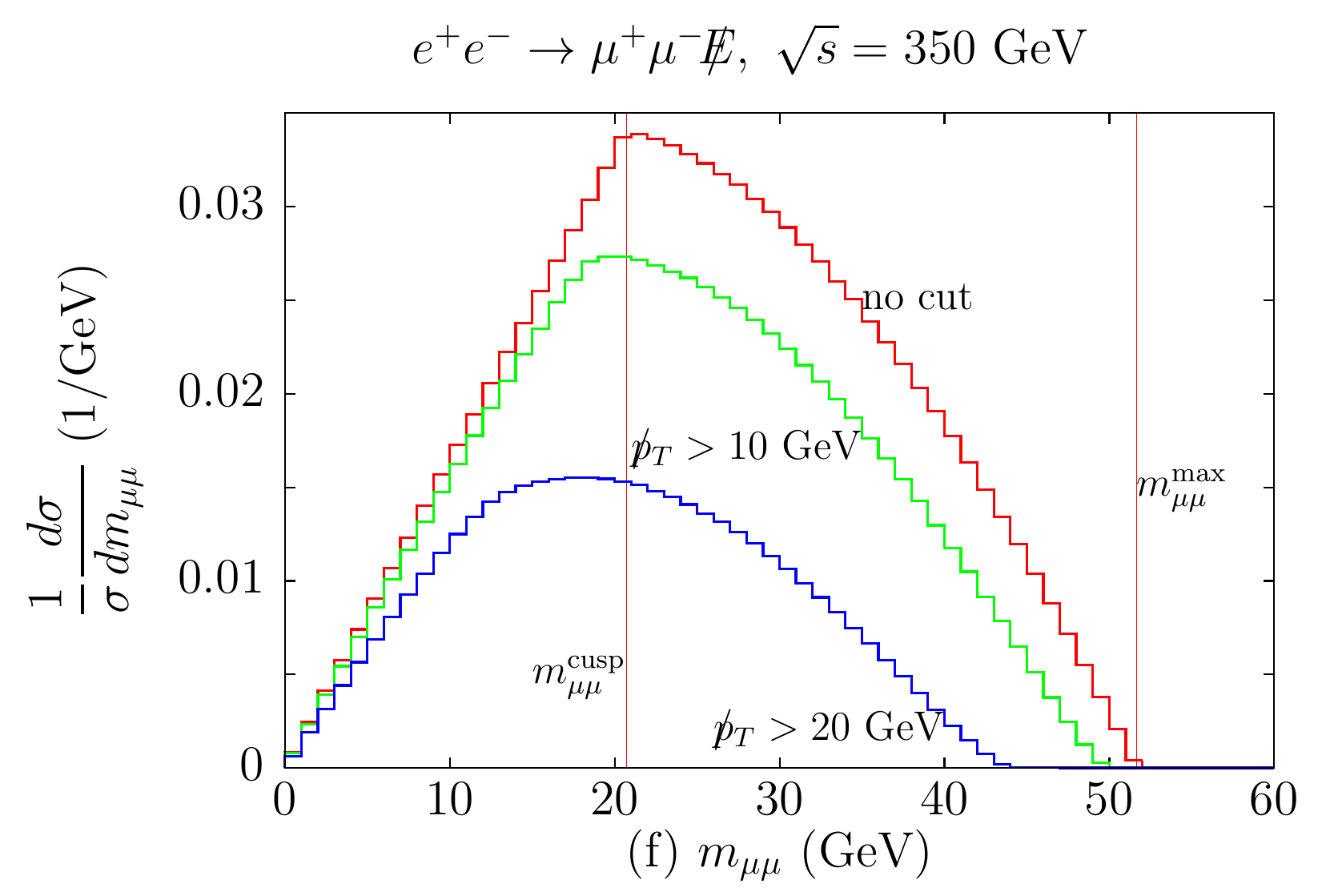}
\caption{\label{fig:ptmisscut}
\casea~for $e^+ e^- \to  \smur\smur\to \mu^+ \mu^- +\me$. Effects due to various $\mpt$ cuts on (a) $m_{\mu\mu}$, (b) $\mrec$, (c) $\cos\Theta$, (d) $E_{\mu}$,
and
(e) $E_{\mu^+}+E_{\mu^-}$
distributions without spin-correlation and other realistic effects at $\sqrt{s}=500\gev$.
Each distribution is normalized by the total cross section.
Panel (f) for the $m_{\mu\mu}$ distribution is set to 350 GeV for comparison.
}
\end{figure}

In a realistic experimental setting, the previously discussed kinematical features may be smeared, rendering the cusps and endpoints less effective for extracting the mass parameters.
We now study the effects of the acceptance cuts.

%

We first explore the effects due to a missing transverse momentum ($\mpt$) cut, which is essential
to suppress the dominant SM background of $e^+ e^- \to e^+ e^- \mu^+ \mu^-$
with the outgoing $e^+ e^-$ going down the beam line and not detected.
Obviously, the $\mpt$ cut removes some events, reducing the event rate.
In addition,  the $\mpt$ cut does not apply evenly over the distribution.
The positions of the cusp and endpoints can be shifted in some cases.

In Fig.~\ref{fig:ptmisscut},
we show the effects of a $\mpt$ cut on the distributions of
$m_{\mu\mu}$,  $\mrec$,
 $\cos\Theta$,  $E_\mu$, and  $E_{\mu\mu}$.
We normalize each distribution by the total cross section without other kinematic cuts.
First, the $m_{\mu\mu}$ distributions with various $\mpt$ cuts
are shown
in Fig.~\ref{fig:ptmisscut}(a) for $\sqrt{s}=500\gev$
and in Fig.~\ref{fig:ptmisscut}(f) for $\sqrt{s}=350\gev$.
The $m_{\mu\mu}$ cusp in the higher c.m.~energy case does not present a notable feature  while 
the lower energy case with $\sqrt{s}=350\gev$
has a more pronounced cusp shape.
With a $\mpt>10\gev$ cut,
the $\maa$ distribution
retains its triangular shape,
but starts to lose the true cusp and maximum positions.
The shift is a few GeV.
If $\mpt>20\gev$,
the sharp cusp is smeared out and the $\mllmax$ position is shifted by about $10\gev$.
In both cases, the $\mllmin$  remains intact.
The $\mrec$ distribution in Fig.~\ref{fig:ptmisscut}(b), on the contrary, keeps its triangular shape
even with a high $\mpt$ cut.
It is interesting to note that
the $\mpt$ cut shifts the $\mrecmin$ and $\mrecmax$
while keeping the $\mrecc$ position fixed.
Figure \ref{fig:ptmisscut}(e) presents the distribution
of the summed energy of the two visible particles,
which are still triangular after the $\mpt$ cut.
The cusp position is retained,
but the minimum and maximum positions are shifted.

We note that $\mpt$ cut does not affect
the positions of the variables $\mllmin$, $\mrecc$, and $\ellcusp$ appreciably,
which all correspond to the kinematical configurations $(iii)$ and $(iv)$ in Eq.~(17).
Here the two visible particles ($a_1a_2$) move in the same direction,
and two invisible particles ($X_1 X_2$) move also in the same direction, opposite to the $a_1a_2$ system.
A $\mpt$ cut would not change the system configuration. 
In contrast, for the configurations $(i)$ and $(ii)$ in Eq.~(17), 
$a_1$ and $a_2$ are moving in the opposite direction, and a cut on the $X_1 X_2$ system alters the individual particle 
as well as the configuration appreciably. 

\begin{figure}[t!]
\centering
\includegraphics[width=.47\textwidth]{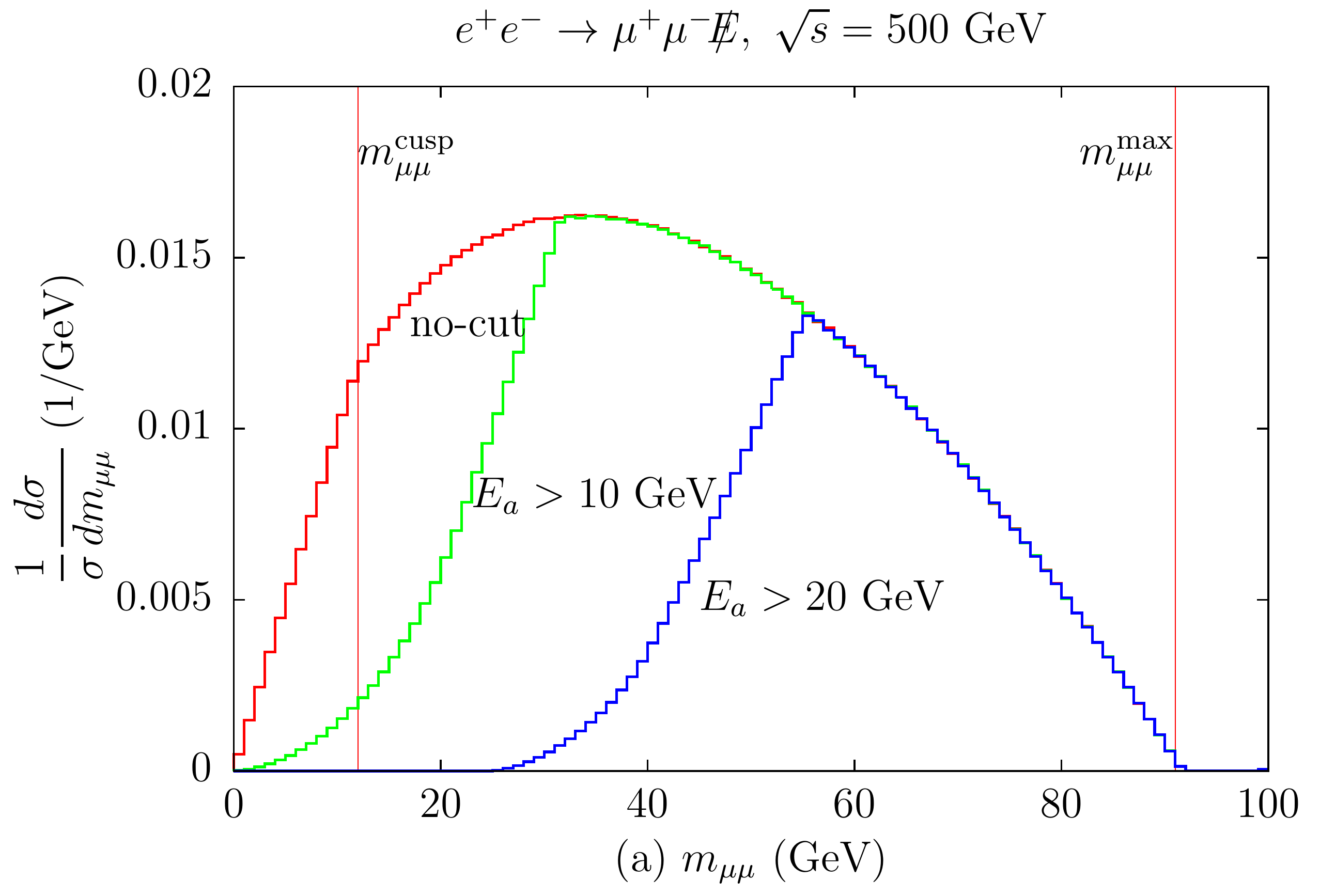}
~~
\includegraphics[width=.47\textwidth]{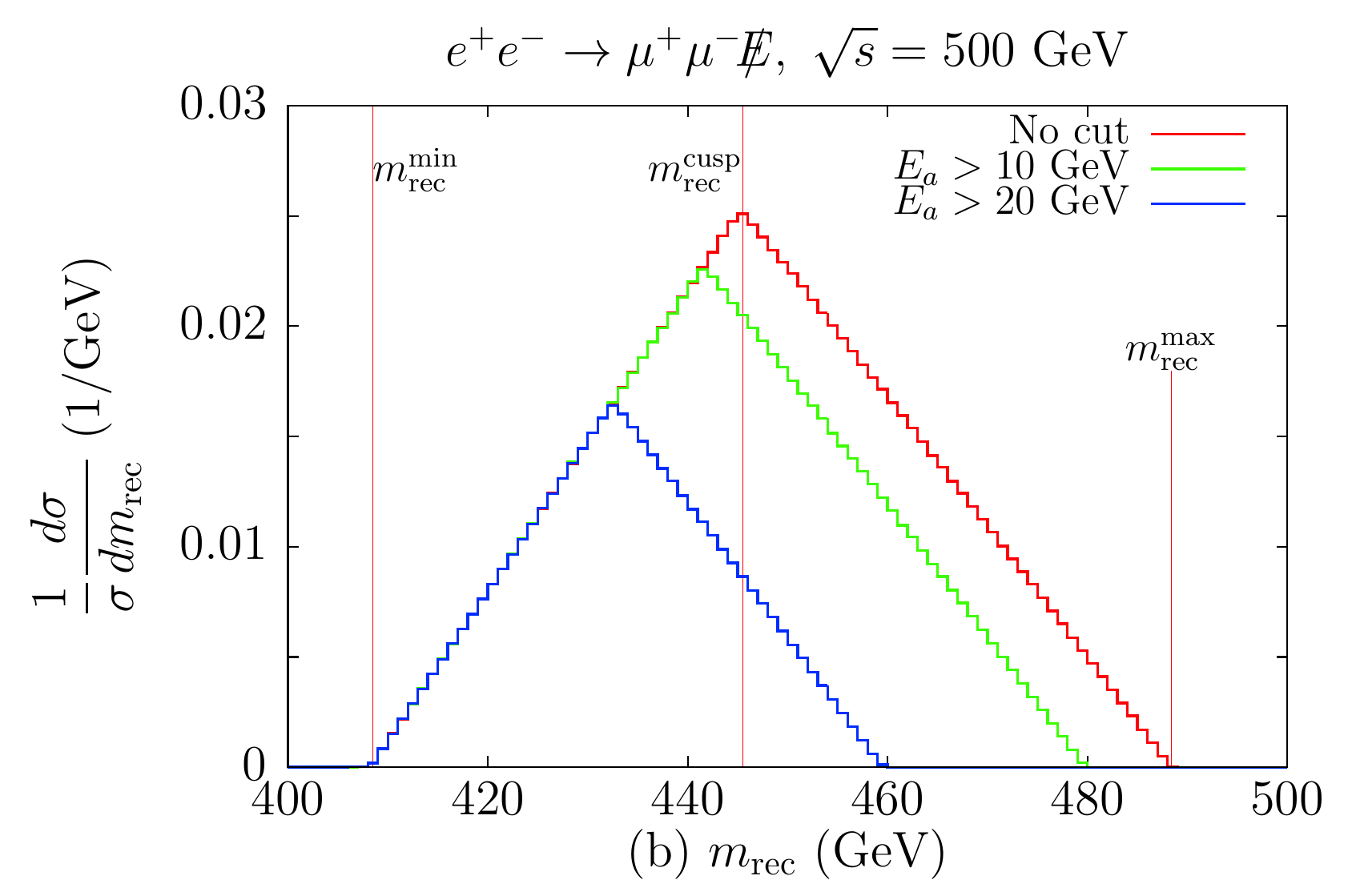}
\\
\includegraphics[width=.47\textwidth]{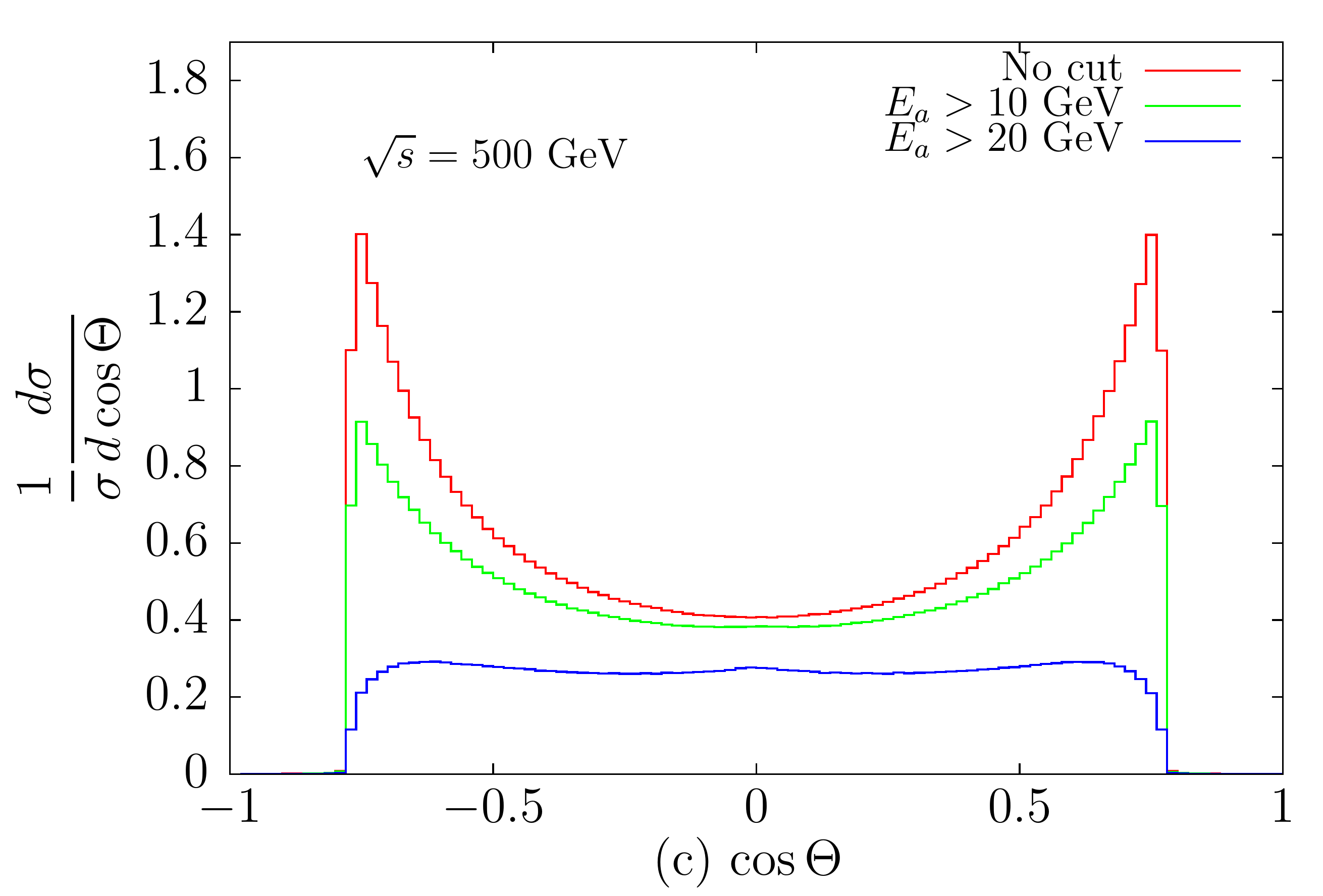}
~~
\includegraphics[width=.47\textwidth]{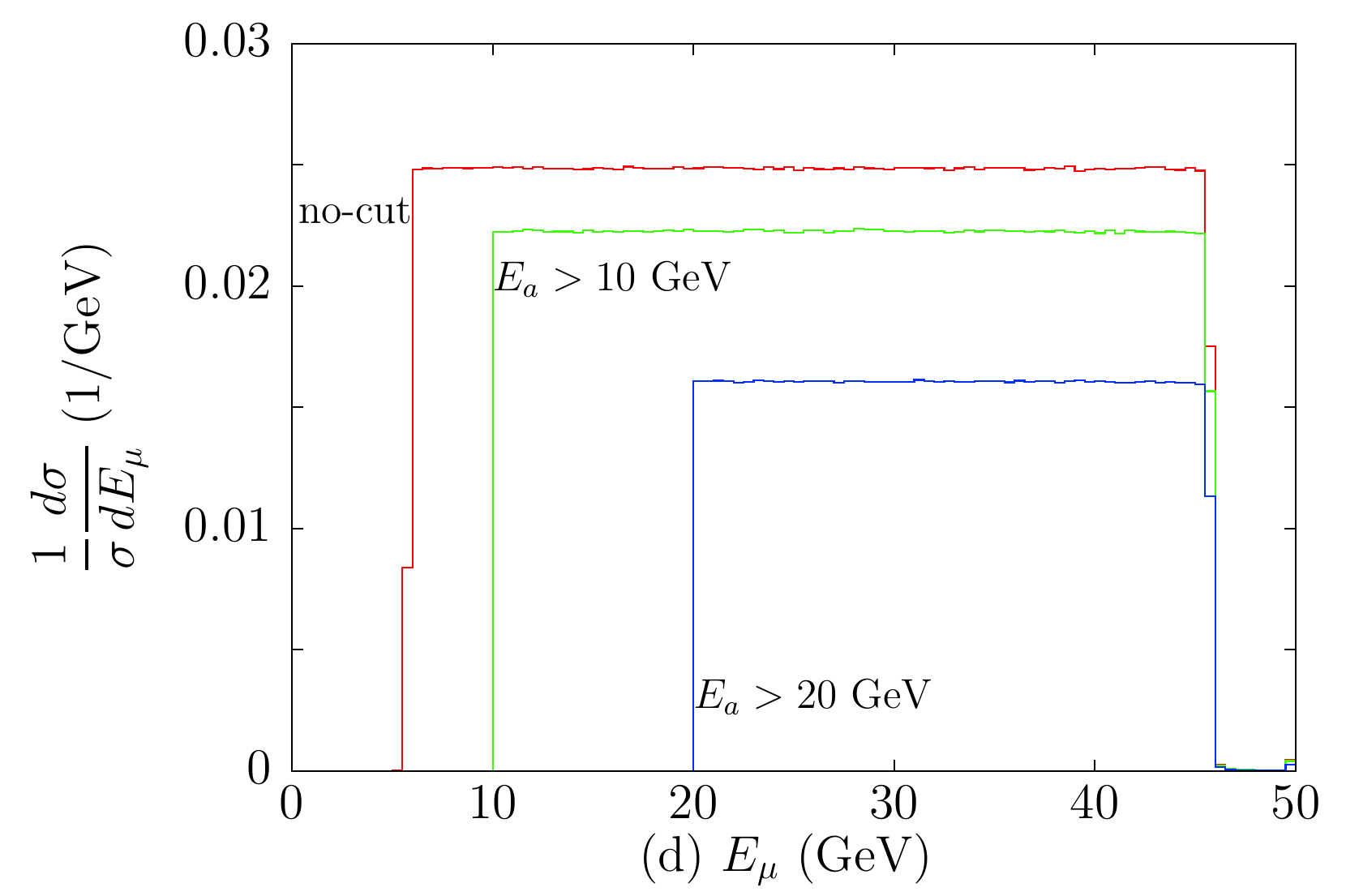}
\\
\includegraphics[width=.47\textwidth]{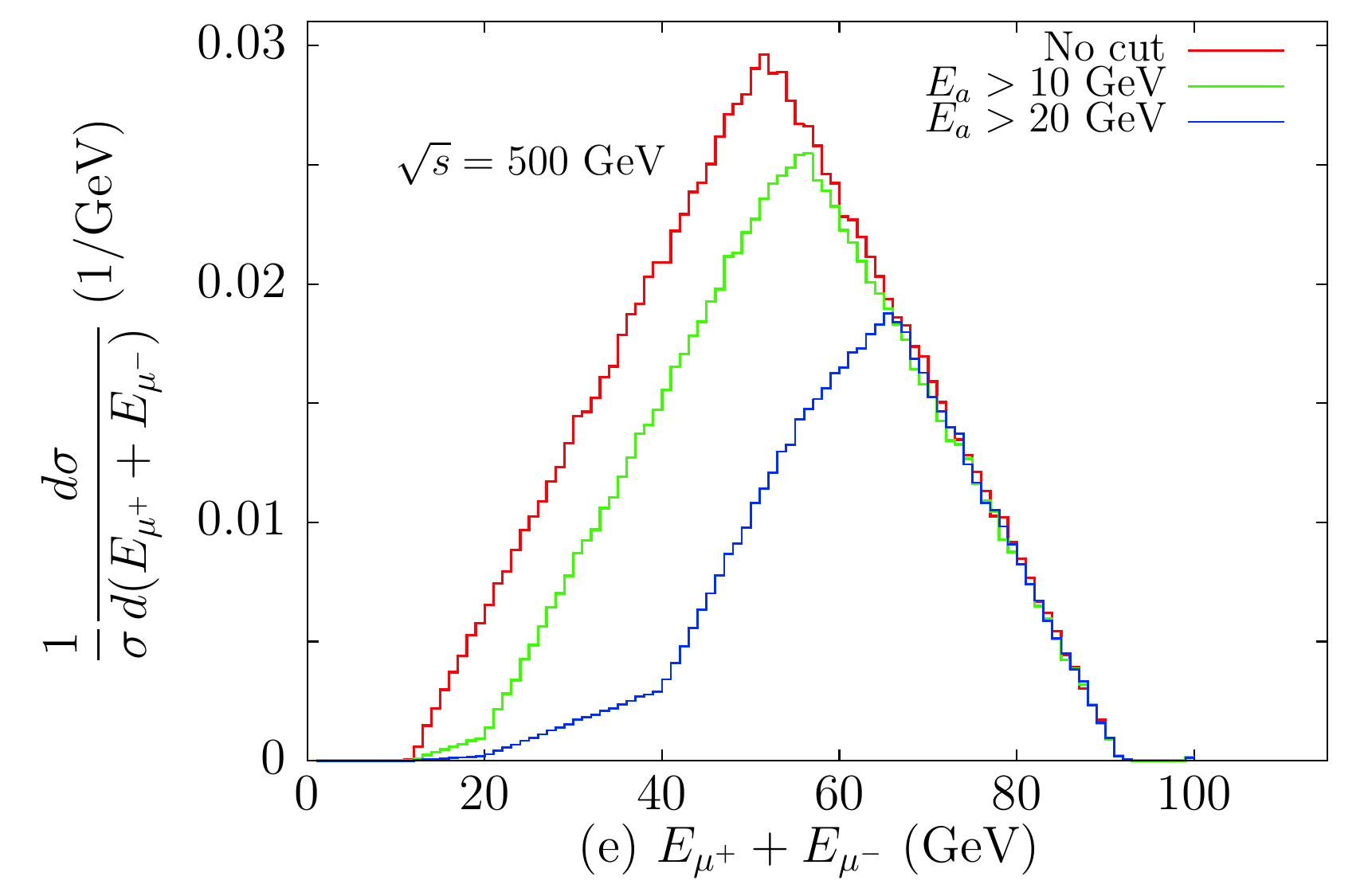}
~~
\includegraphics[width=.47\textwidth]{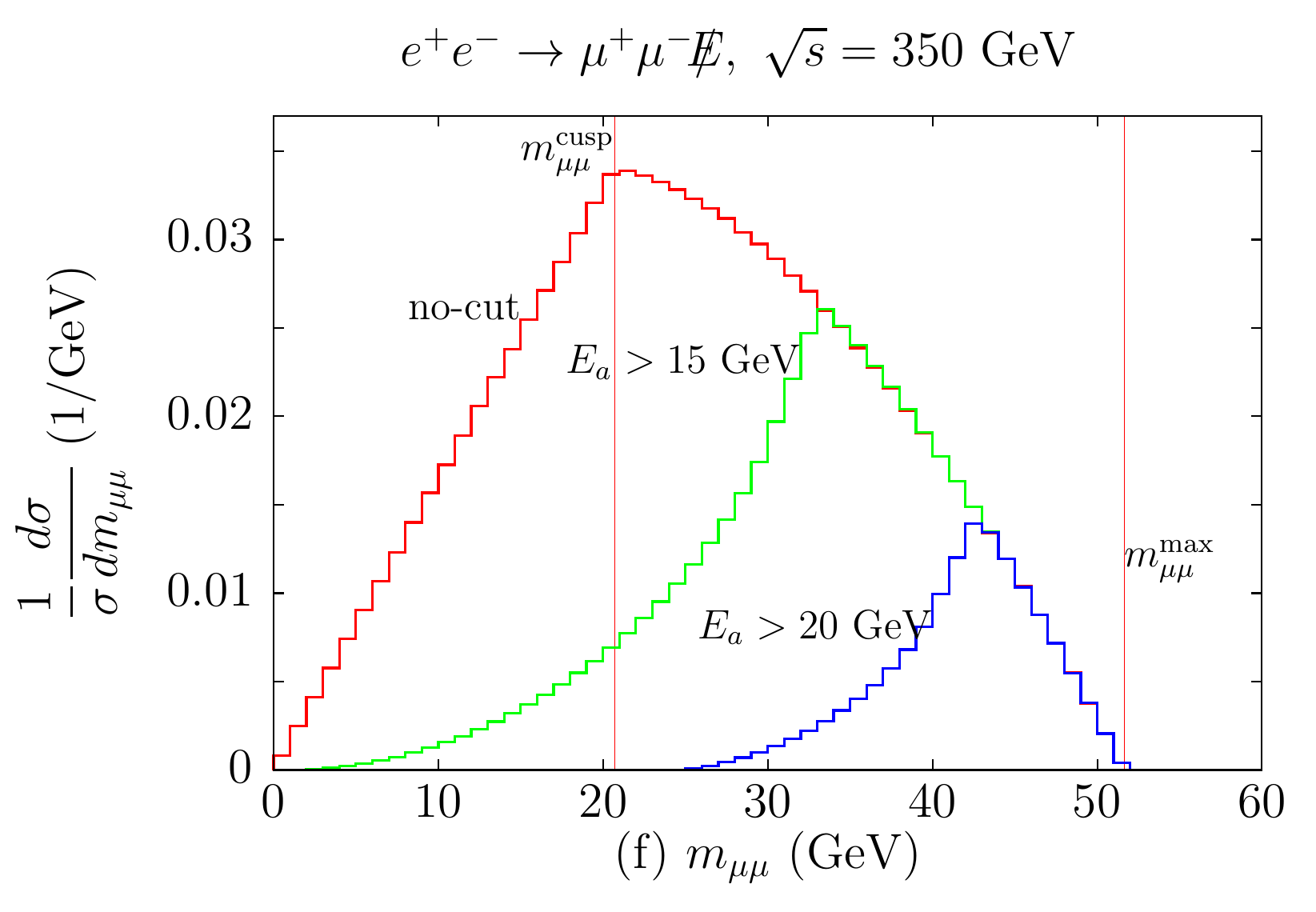}
\caption{\label{fig:Eacut}\casea~for $e^+ e^- \to  \smur\smur\to \mu^+ \mu^- \,\me$. 
Effects due to various $E_a$ cuts on the (a) $m_{\mu\mu}$, (b) $\mrec$, (c) $\cos\Theta$, (d) $E_{\mu}$,
and (e) $E_{\mu^+}+E_{\mu^-}$
distributions without spin-correlation and other realistic effects at $\sqrt{s}=500\gev$.
Each distribution is normalized by the total cross section without any other acceptance cut.
Panel (f) for the $m_{\mu\mu}$ distribution is set to 350 GeV for comparison. 
}
\end{figure}

The least affected variable is the $\cos\Theta$ distribution in Fig.~\ref{fig:ptmisscut}(c).
The $|\cos\Theta|_{\rm max}$ positions remain the same,
and the $\mpt$ cut removes the data nearly evenly all over the distribution.
Figure \ref{fig:ptmisscut}(d) shows the $E_\mu$ distribution  under the $\mpt$ cut effects.
Similar to the case of $\cos\Theta$, the $\mpt$  cut reduces the whole rate roughly uniformly, and the box-shaped distribution is still maintained.

Figure \ref{fig:Eacut} presents the five kinematic distributions with the effects of the $E_a$ cut.
The normalization  is done with the total cross section without any cut.
Two  $m_{\mu\mu}$ distributions are presented, one for $\sqrt{s}=500\gev$  in Fig.~\ref{fig:Eacut}(a) 
and the other for $\sqrt{s}=350\gev$ in Fig.~\ref{fig:Eacut}(f).
Both retain its maximum position after the $E_a$ cut.
However, the $m_{\mu\mu}$ cusp position is shifted by a sizable amount,
approximately 10 GeV for $E_a>15\gev$ cut at $\sqrt{s}=350\gev$.
This behavior is the same for the $E_{\mu\mu}$ distribution in Fig.~\ref{fig:Eacut}(e).
The $\mrec$ distribution in Fig.~\ref{fig:Eacut}(b)
behaves oppositely:
the maximum and cusp positions are shifted while the minimum position is retained.
Therefore, the $E_a$ cut does not change the one-dimensional configuration $(i)$ of
Eq.~(\ref{eq:1d:configuration}).

The $\cos\Theta$ distributions under the $E_a$ cuts are shown in Fig.~\ref{fig:Eacut}(c).
The locations of $|\cos\Theta|_{\rm max}$ remain approximately the same, but the sharp cusps are reduced somewhat.
Finally the $E_a$ distribution in Fig.~\ref{fig:Eacut}(d) shows the expected shift of its minimum into the lower bound on $E_a$.
Note that some data satisfying $E_a>E_a^{\rm cut}$ are also cut off, since the $E_a$ cut has been applied to both of the final leptons.
In summary, the acceptance cut distorts the kinematic distributions,
and shifts the singular positions. When we extract the mass information from the endpoints,
these cut effects must be properly taken into account.


\subsection{\label{sec:realistic considerations}Mass measurements with realistic considerations}

\subsubsection{Backgrounds and simulation procedure}

For our signal of $e^+ e^- \to \mu^+ \mu^- + \me$,
there are substantial SM backgrounds.
The main irreducible SM background is $W$ boson pair production,
$e^+ e^- \to W^+  W^-\to \mu^+ \nu_\mu\mu^-  \bar{\nu}_\mu$.
The next dominant mode is $ZZ$ production,
$e^+ e^- \to Z Z\to \mu^+ \mu^-  \nu_i  \bar{\nu}_i$
where $\nu_i$ denotes a neutrino of all three flavors.
The $\ww$ background is larger than the $ZZ$  background
by a factor of about 20. In the following numerical simulation,
we include the full SM processes for the final state
 $\mu^+ \mu^-   \nu  \bar{\nu}$.

Another substantial SM background is from $e^+ e^- \to e^+ e^- \mu^+ \mu^-$
where the outgoing $e^+$ and $ e^-$ go down the beam pipe
and are missed by the detectors.  It is mainly generated by Bhabha scattering
with the incoming electron and positron through a $t$-channel diagram.
This background could be a few orders of
magnitude larger than the signal.
However, a cut on the missing transverse momentum can effectively remove it.
The maximum missing transverse momentum in this background
comes from the final electron and positron,
each of which retains the full energy ($\sqrt{s}/2$ each)
and moves within an angle of
$1^\circ$ with respect to the beam pipe (at
the edge of the end-cap detector coverage).
As a result, most of these background events lie within
\begin{equation}
\left(\rlap/p_T\right)_{{\rm beam~line~}e^+ e^-} \lesssim 3\times250\gev \times\sin\left(1^\circ\right) \simeq 15\gev.
\end{equation}
We thus design our basic acceptance cuts for the event selection
\bea
\label{eq:basic:cut}
\hbox{ \tt Basic cuts: }
&& E_a \geq 10\gev,
\quad
\mpt \geq 15\gev ,
\\ \no &&
|\cos\theta_\ell^{\rm cm} | \leq 0.9962,
\quad
m_{aa} \geq 1\gev,
\quad
m_{\rm rec} \geq 1\gev.
\eea
The angular cut on $\theta_\ell^{\rm cm}$ requires
that the observed lepton lies within $5^\circ$ from the beam pipe.
This angular acceptance and
the invariant mass cut on the lepton pair regularize the perturbative singularities.
We also find that the $\rlap/p_T$ cut removes the background from $e^+e^-\to e^+e^-\tau^+\tau^-$~\cite{Kalinowski:2008fk}.

In principal, the full SUSY backgrounds should  be included in addition to the $\smur$ and $\smul$ signal pair production.
There are many types of SUSY backgrounds.
The dominant ones are the production of $\neuo\neu_{j\geq 2}$
followed by the heavier neutralino decay of $\neu_{j\geq 2} \to \ell^+ \ell^- \neuo$.
However, their contributions are negligible with our mass point and event selection.

At the ILC environment, it is crucial to consider the other realistic factors in order to reliably estimate the accuracy for the mass determination. These include
the effects of ISR,  beamstrahlung \cite{B:ISR} and detector resolutions.
%
For these purposes, we adopt the ILC-Whizard setup \cite{Whizard}, which accommodates the SGV-3.0 fast detector simulation suitable for the ILC \cite{Berggren:2012ar}. 

\subsubsection{\protect\casea: $\smur\smur$ pair production}
\label{subsec:casea}

\begin{figure}[t!]
\centering
\begin{minipage}{2.95in}
\centering
\includegraphics[height=5.7cm, width=7.2cm]{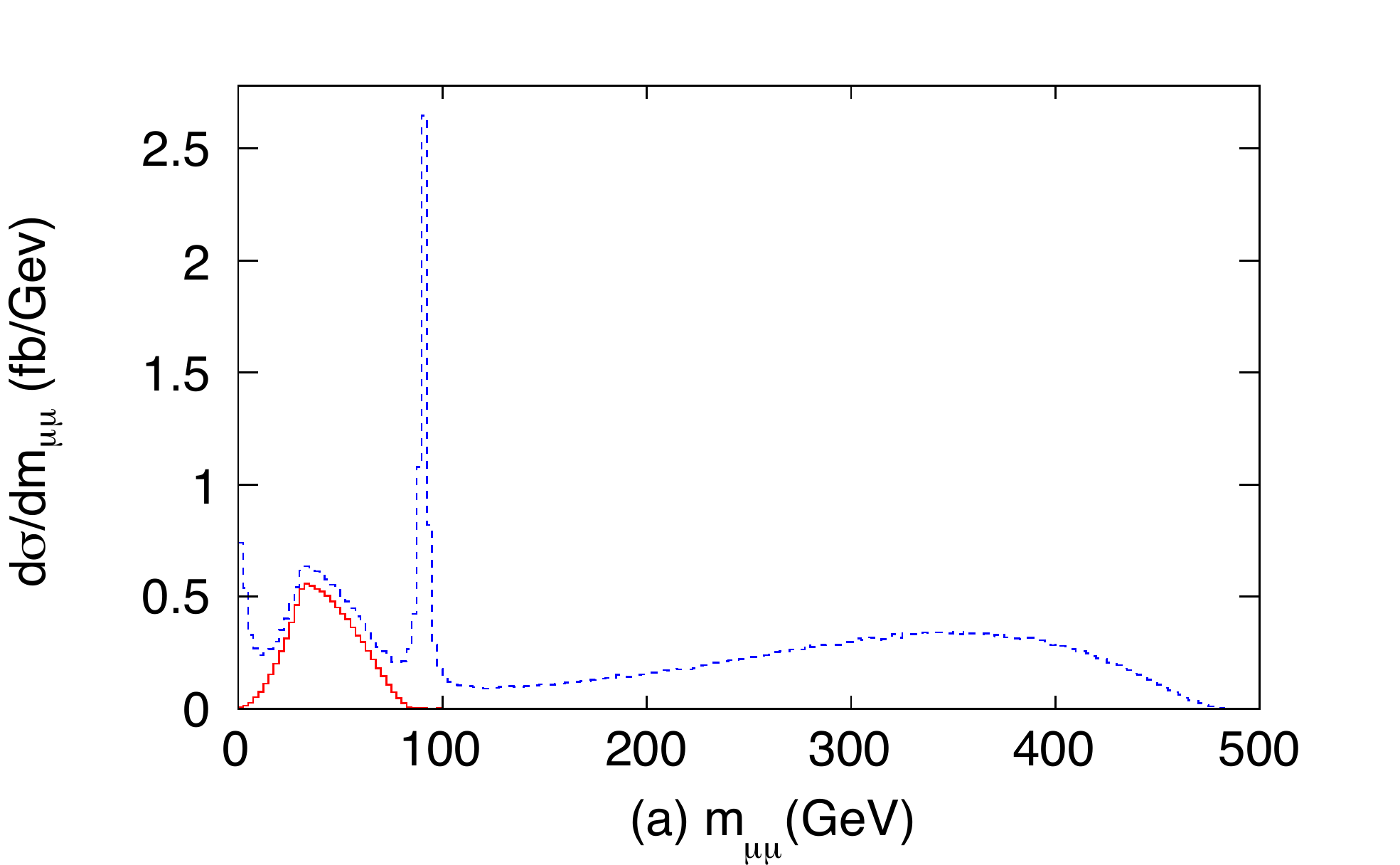}
\end{minipage}
\begin{minipage}{2.95in}
\centering
\includegraphics[height=5.7cm, width=7.2cm]{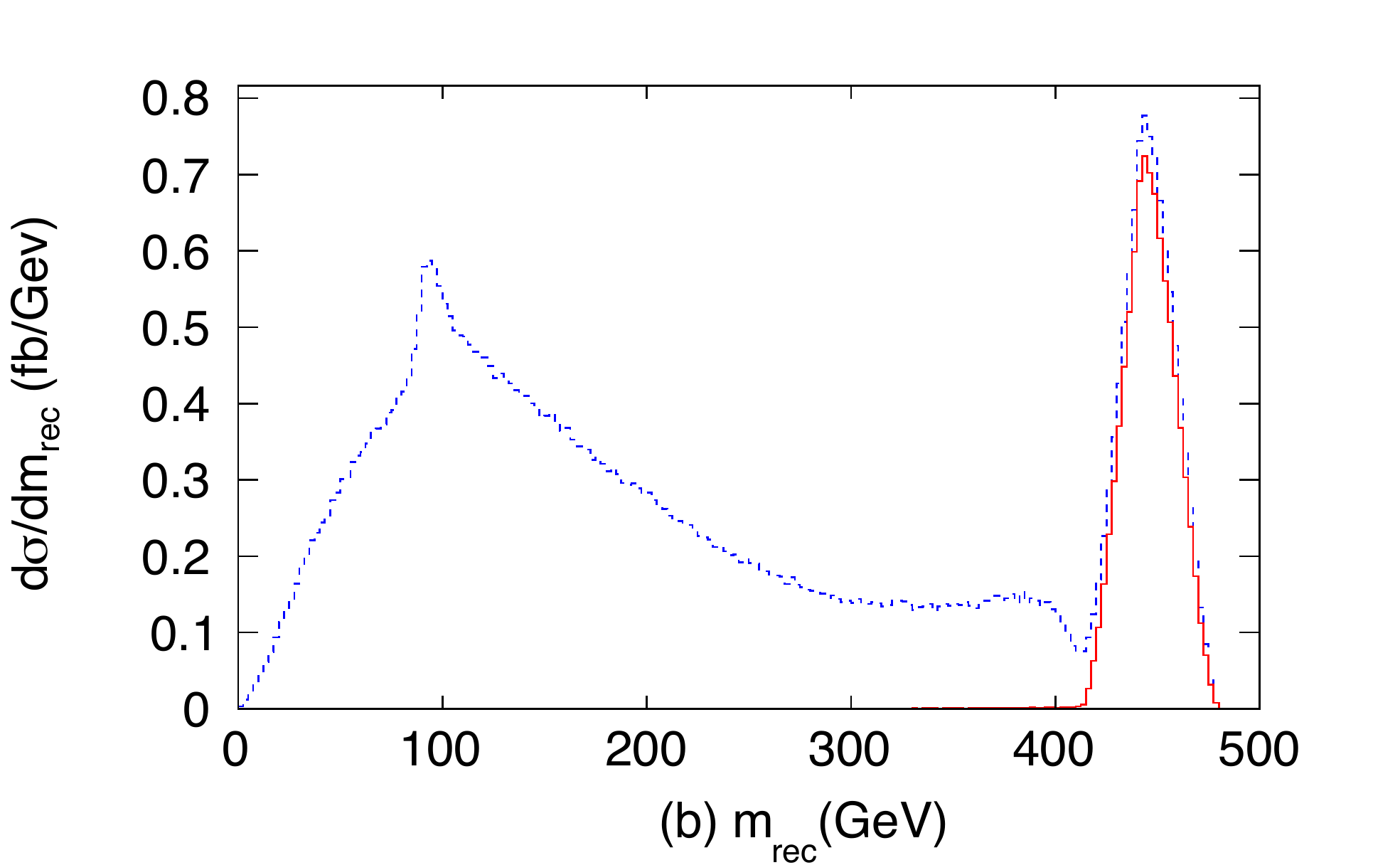}
\end{minipage}
\begin{minipage}{2.95in}
\centering
\includegraphics[height=4.6cm, width=7.2cm]{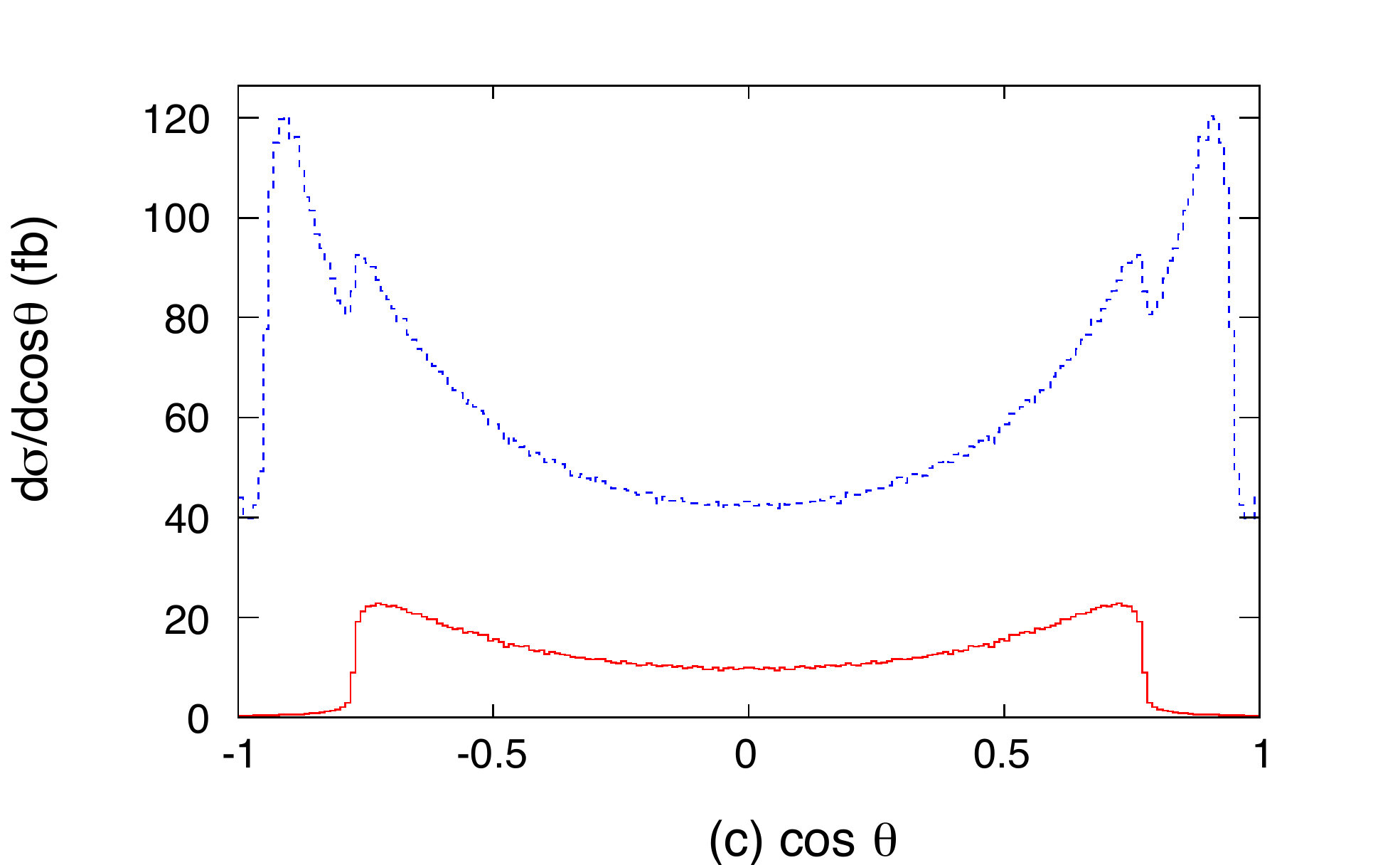}
\end{minipage}
\begin{minipage}{2.95in}
\centering
\includegraphics[height=5cm, width=7.2cm]{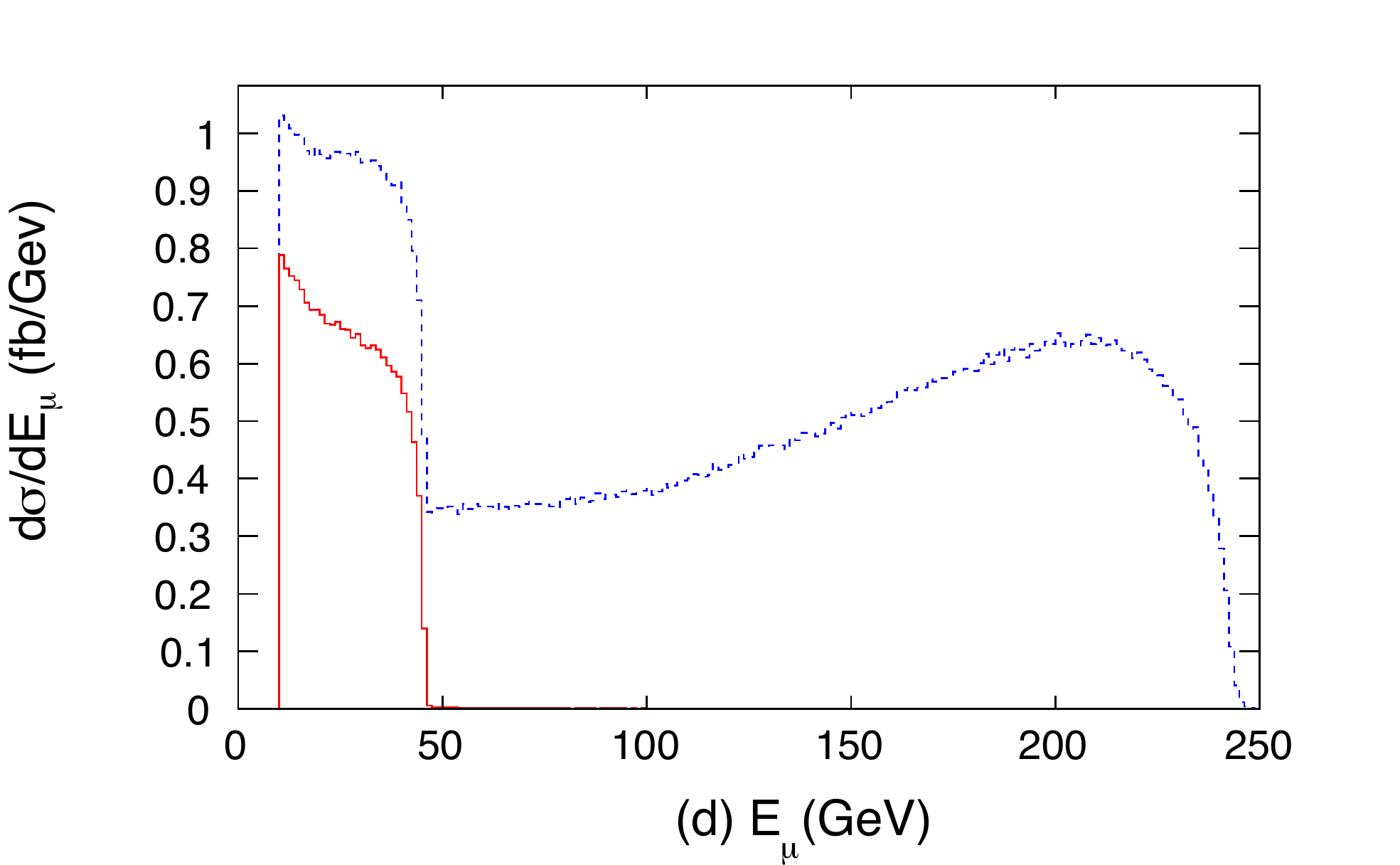}
\end{minipage}
\begin{minipage}{2.95in}
\centering
\includegraphics[height=4.9cm, width=7.2cm]{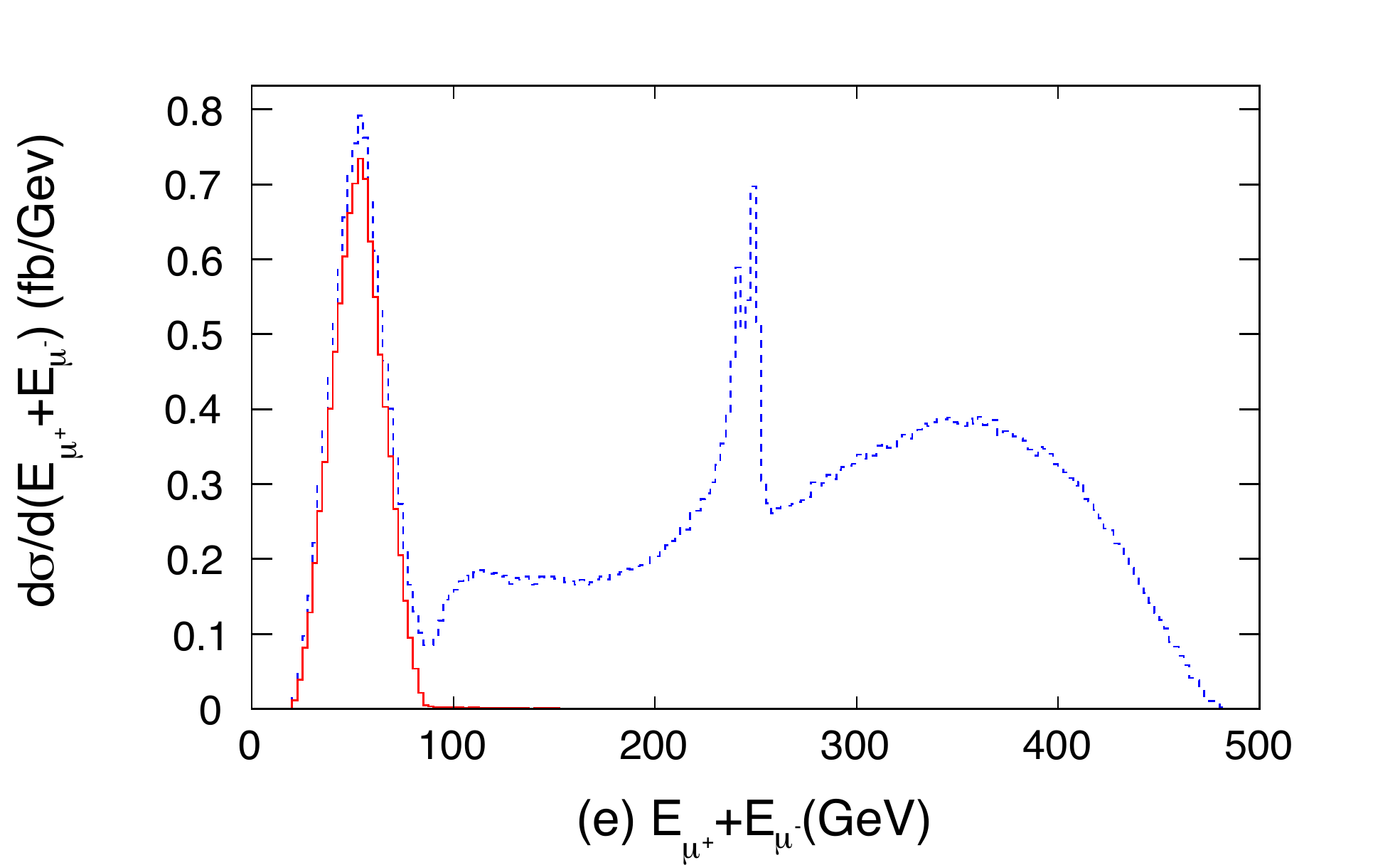}
\end{minipage}
\caption{\label{fig:caseA:basic}
\casea~for $e^+ e^- \to  \smur\smur\to \mu^+ \mu^- \,\me$. Basic acceptance cut on the (a) $m_{\mu\mu}$, (b) $\mrec$, (c) $\cos\Theta$, (d) $E_{\mu}$, and (e) 
$E_{\mu^+}+E_{\mu^-}$
distributions with spin-correlation and other realistic effects.
The c.m.~energy is set to $\sqrt{s}=500\gev$ for all distributions.
The solid (red) line denotes our signal of the resonant production of a $\smur$ pair.
The dashed (blue) line is the total event
including our signal and the SM backgrounds.
%
}
\end{figure}

For the mass spectrum in \casea,
Fig.~\ref{fig:caseA:basic} presents a full simulation of
the five kinematic distributions at $\sqrt{s}=500\gev$
with the basic cuts in Eq.~(\ref{eq:basic:cut}).
The solid (red) line denotes our signal of the resonant production of a $\smur\smur$ pair.
The dashed (blue) line is the total distribution
including our signal and the SM backgrounds.

The $m_{\mu\mu}$ distribution from our signal in Fig.~\ref{fig:caseA:basic}(a)
does not reveal the best feature of the antler process.
Its cusp is not very pronounced and its maximum is submerged
under the dominant $Z$ pole.
As discussed before, this is because the c.m.~energy of 500 GeV is too high compared with the smuon mass. On the contrary, the $\mrec$ distribution  in Fig.~\ref{fig:caseA:basic}(b)
separates our signal from the SM backgrounds well.
A sharp triangular shape is clearly seen above the SM
background tail.
This separation is attributed to the weak scale mass of the missing particle $X$.
If $X$ were much lighter such as $M_X\simeq 10\gev$, the cusp position in the $\mrec$ distribution of the signal would be shifted to a lower value and thus overlap with that of the large $\ww$ background.

Figure \ref{fig:caseA:basic}(c) presents the $\cos\Theta$ distributions
with the $\ww$ background and the $\smur\smur$ signal.
However, the highest point of $\cos\Theta$ (the cusp location) is shifted from the location of the $\cosmax$ in Table \ref{table:cusp:values}, by about $2\sim 3\%$.
This is from the kinematical smearing due to ISR and beamstruhlung effects.

Figure \ref{fig:caseA:basic}(d) shows the muon energy distribution,
which consists of two previously box-shaped distributions.
Our signal distribution,
which is expected to be flat for a scalar boson,
is distorted by ISR.
The SM background, mainly the $\ww$ background,
shows a more tilted distribution,
which has additional effects from spin correlation.
The reason for the tilted distribution
toward higher $E_\mu$
is that the $\ww$ production has the largest contribution
from the production
of $W^-_L W^+_R$
mediated by a $t$-channel neutrino~\cite{Hagiwara}.
Here $W^-_L$ ($W^+_R$) denotes the left-handed (right-handed)
negatively (positively) charged $W$ boson.
$W^-_L$ has the left-handed coupling of $\ell^-_L$-$\bar{\nu}_R$-$W^-_L$
so that the decayed $\ell^-_L$ moves along the parent $W^-$ direction
and the $\bar{\nu}$ in the opposite direction.
The $\ell^-$ tends to have
higher energy.
Even though the $E_\mu$ distribution is not flat both for the signal and the backgrounds,
their maximum positions are the same as predicted in Table \ref{table:cusp:values}.
However, the minimum position for the $W^+W^-$ distribution is below the acceptance cut while the minimum for the $\smur\smur$ signal is approximately the same as the cut.
The measurement of these minima becomes problematic.
 As a result, the other kinematic observables discussed here are essential in the measurement of these masses.

\begin{figure}[t!]
\centering
\includegraphics[width=.47\textwidth]{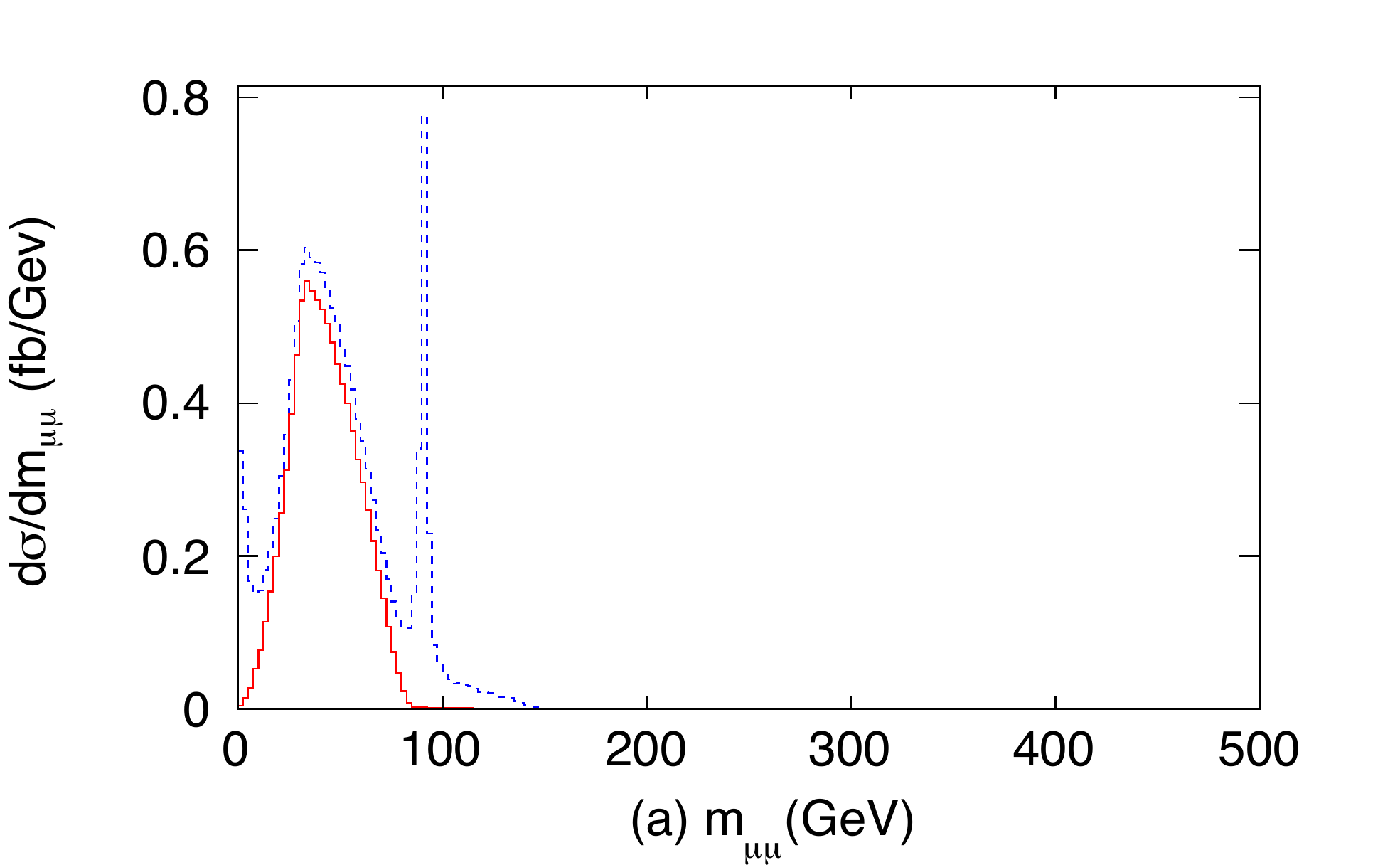}
~~
\includegraphics[width=.47\textwidth]{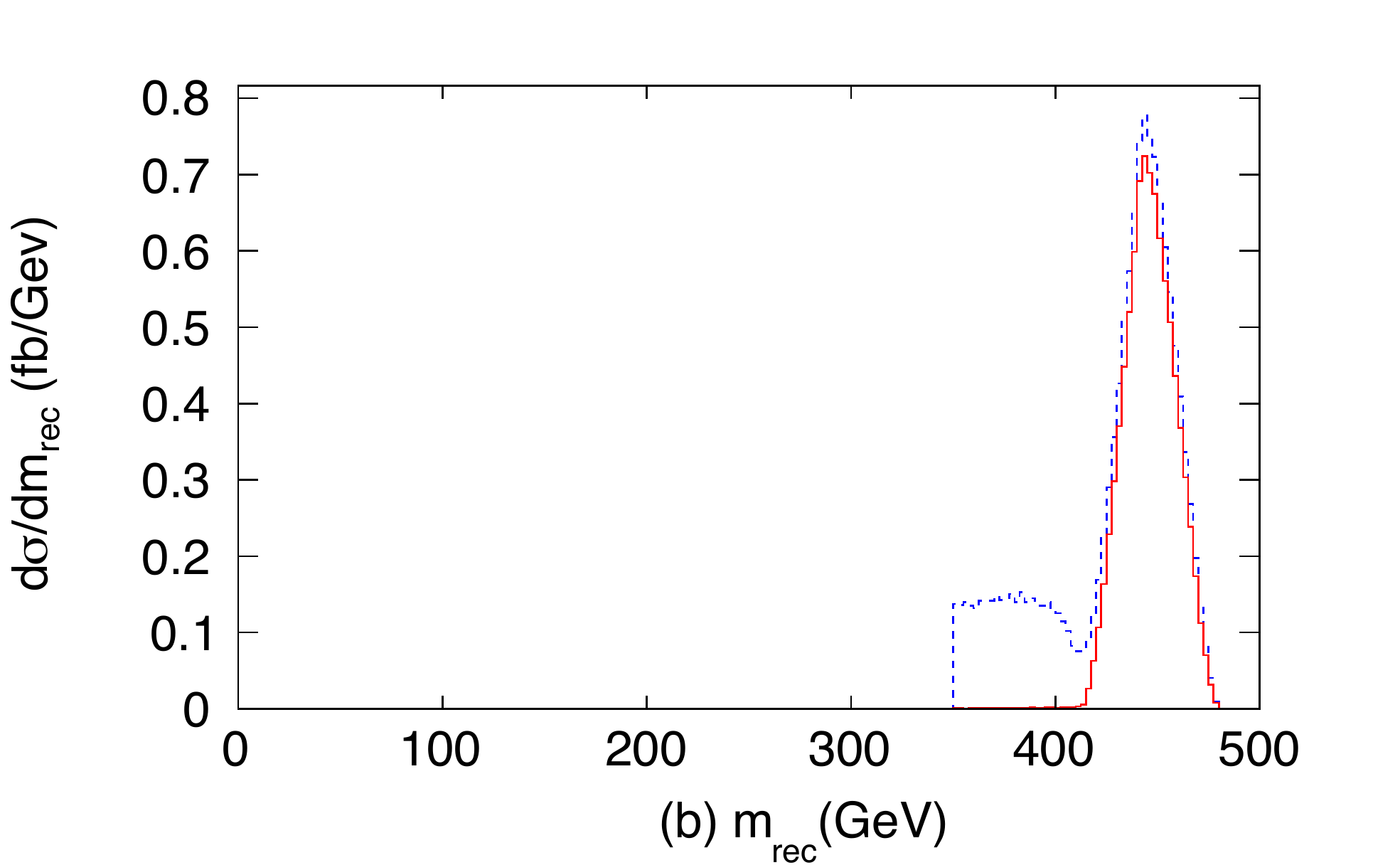}
\\
\includegraphics[width=.47\textwidth]{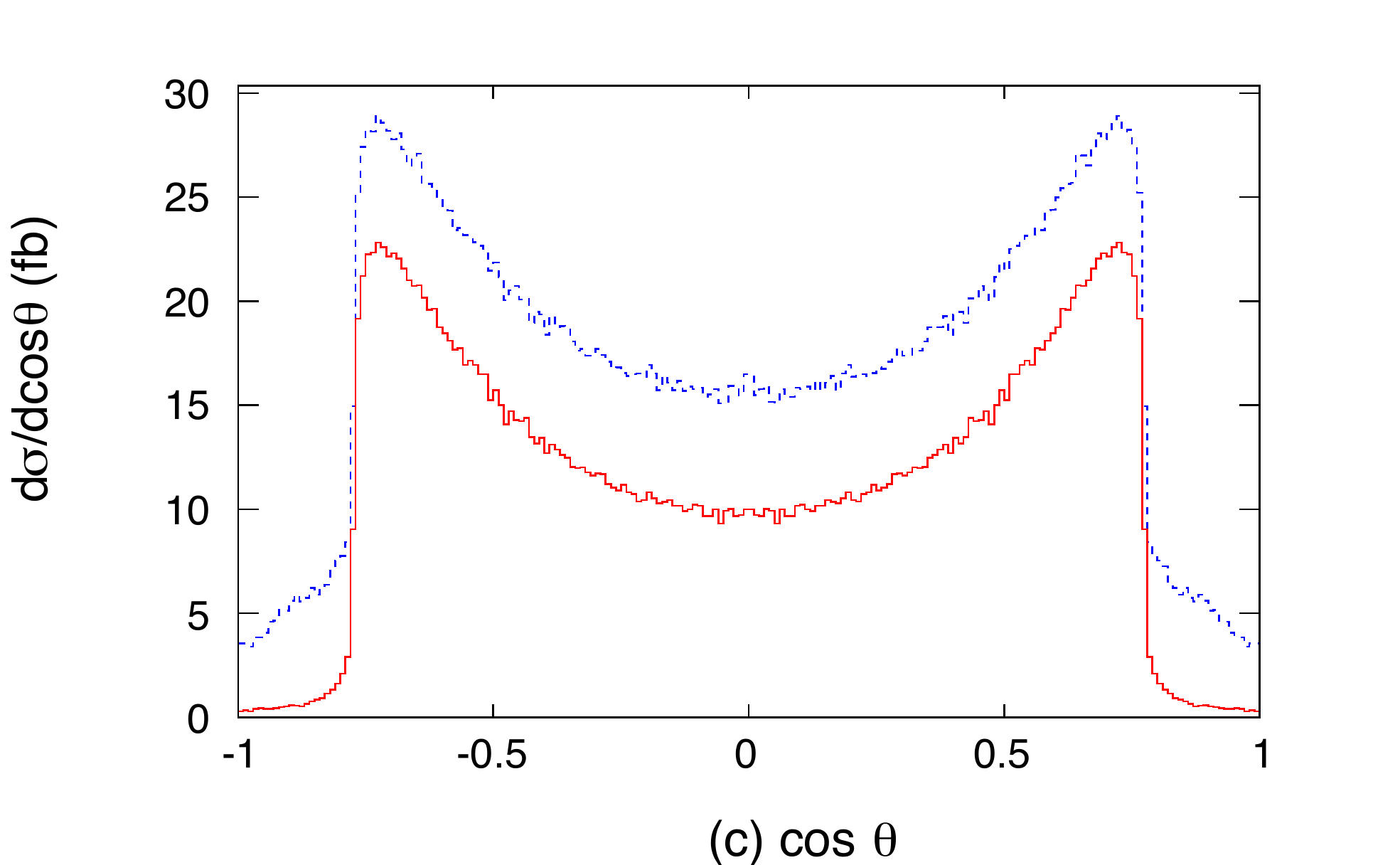}
~~
\includegraphics[width=.47\textwidth]{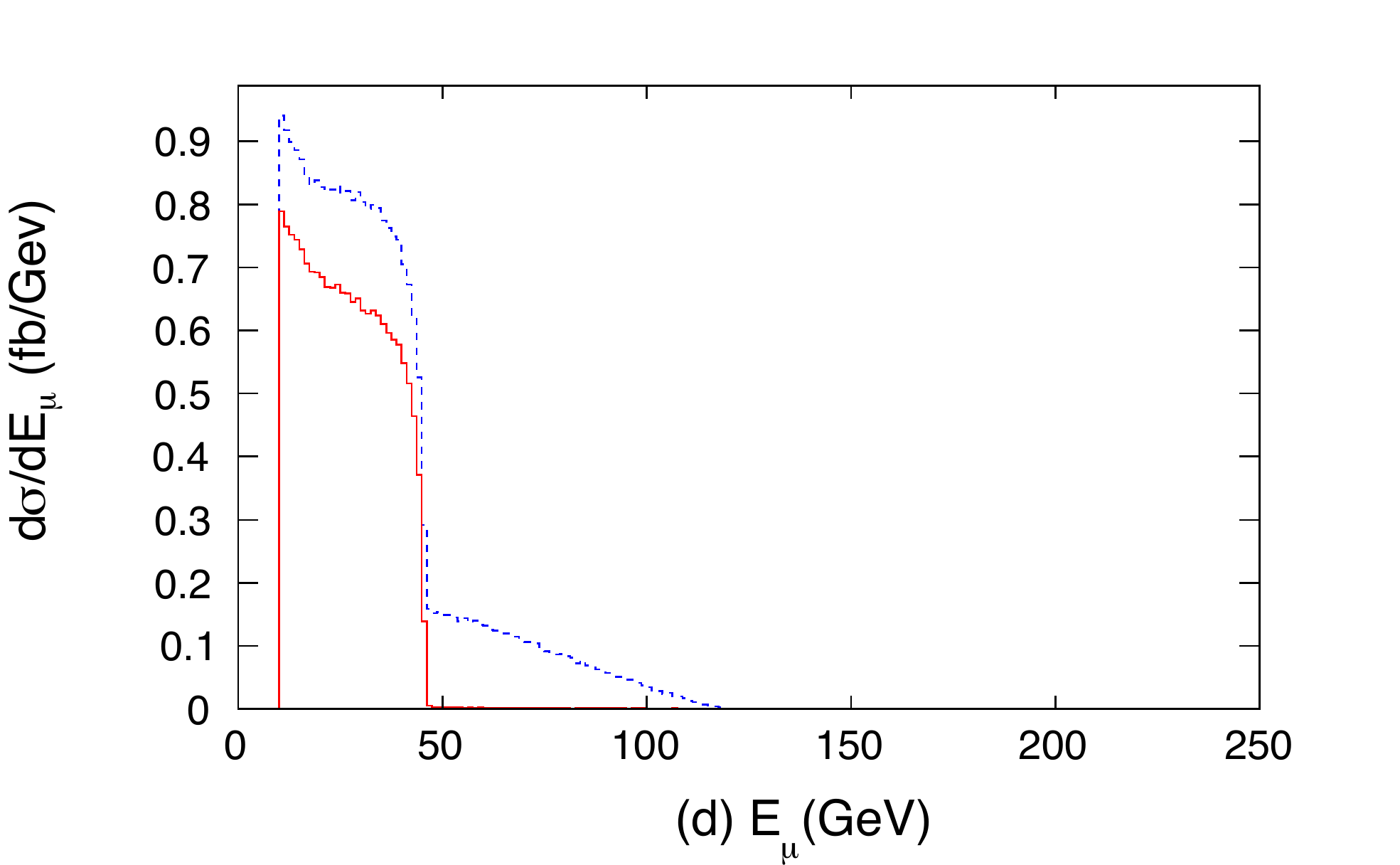}
\\
\includegraphics[width=.47\textwidth]{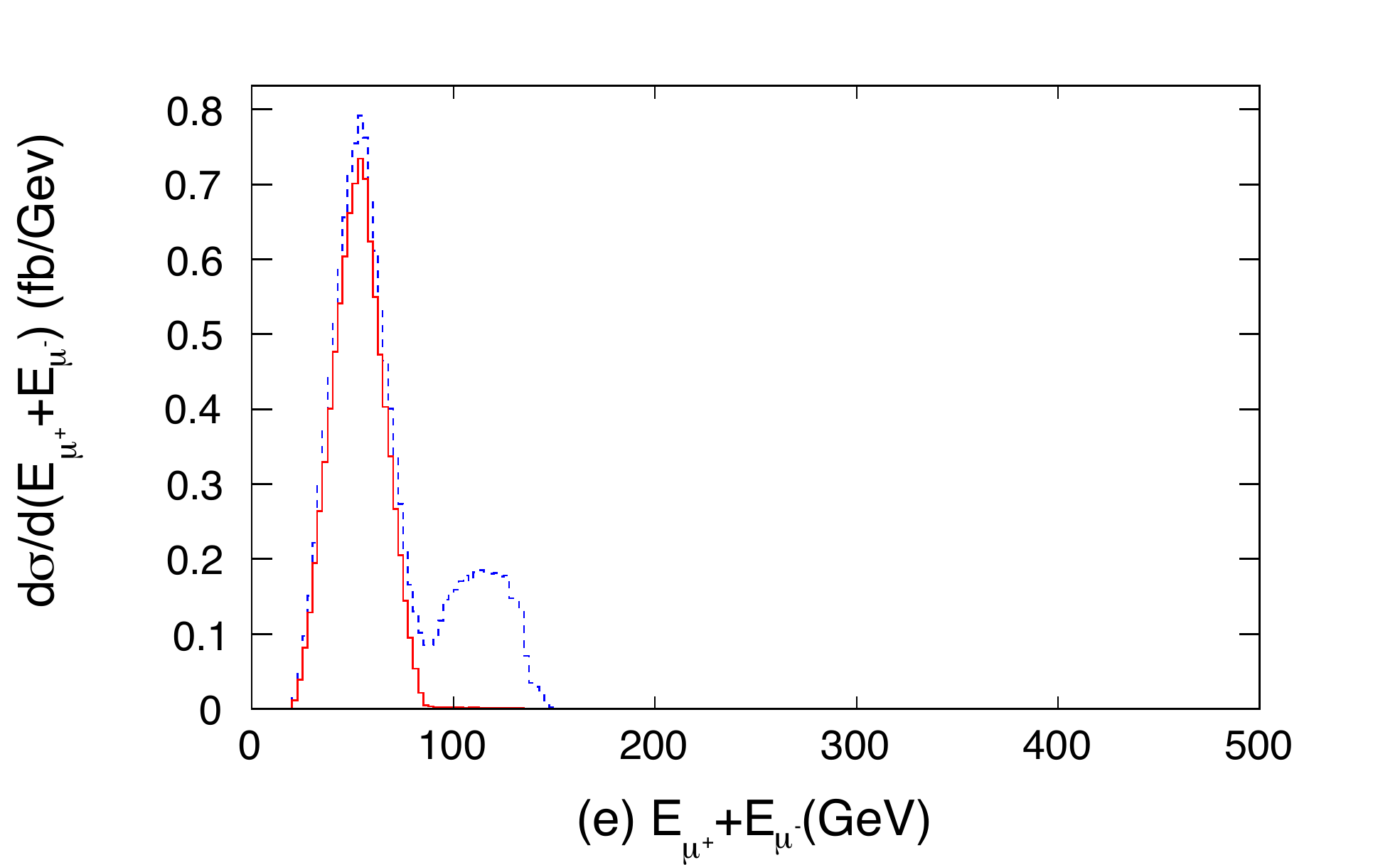}
~~
\caption{\label{fig:caseA:mxxcut}
\casea~for $e^+ e^- \to  \smur\smur\to \mu^+ \mu^- \,\me$. 
The effect of an additional cut of $\mrec>350\gev$ on the (a) $m_{\mu\mu}$, (b) $\mrec$, 
(c) $\cos\Theta$, (d) $E_{\mu}$, and (e) $E_{\mu^+}+E_{\mu^-}$
distributions with spin-correlation and other realistic effects. 
The c.m.~energy is set to $\sqrt{s}=500\gev$ for all distributions.
The solid (red) line denotes our signal of the resonant production of a $\smur$ pair.
The dashed (blue) line is the total differential cross section
including our signal and the SM backgrounds.
%
}
\end{figure}

Finally Figs.~\ref{fig:caseA:basic}(e) presents the energy sum of two visible particles. 
The distribution for our signal is triangular
and separated from the SM backgrounds.
Even in the full and realistic simulation,
the cusps and endpoints of the signal are very visible.
In fact, the signal part of the distribution takes a very similar form to that of $\mrec$.

Understanding those kinematic distributions of our signal
is of great use to suppress the SM background.
For example, we apply an additional cut of
\bea
\label{eq:mrec:cut}
\mrec > 350\gev,
\eea
and present the distributions of the same five kinematic variables
in Fig.~\ref{fig:caseA:mxxcut}.
Our signal, denoted by the solid (red) lines, remains intact
since $\mrecmin=408\gev$ for $\smur\smur$.
On the other hand,
a large portion of the SM background is excluded.
The antler characteristics of our signal emerge in the total distributions.
We can identify all of the cusp structures.

\subsubsection{\protect\caseb: production  of $\smur\smur$ and $\smul\smul$}
\begin{figure}[t!]
\centering
\includegraphics[width=.47\textwidth]{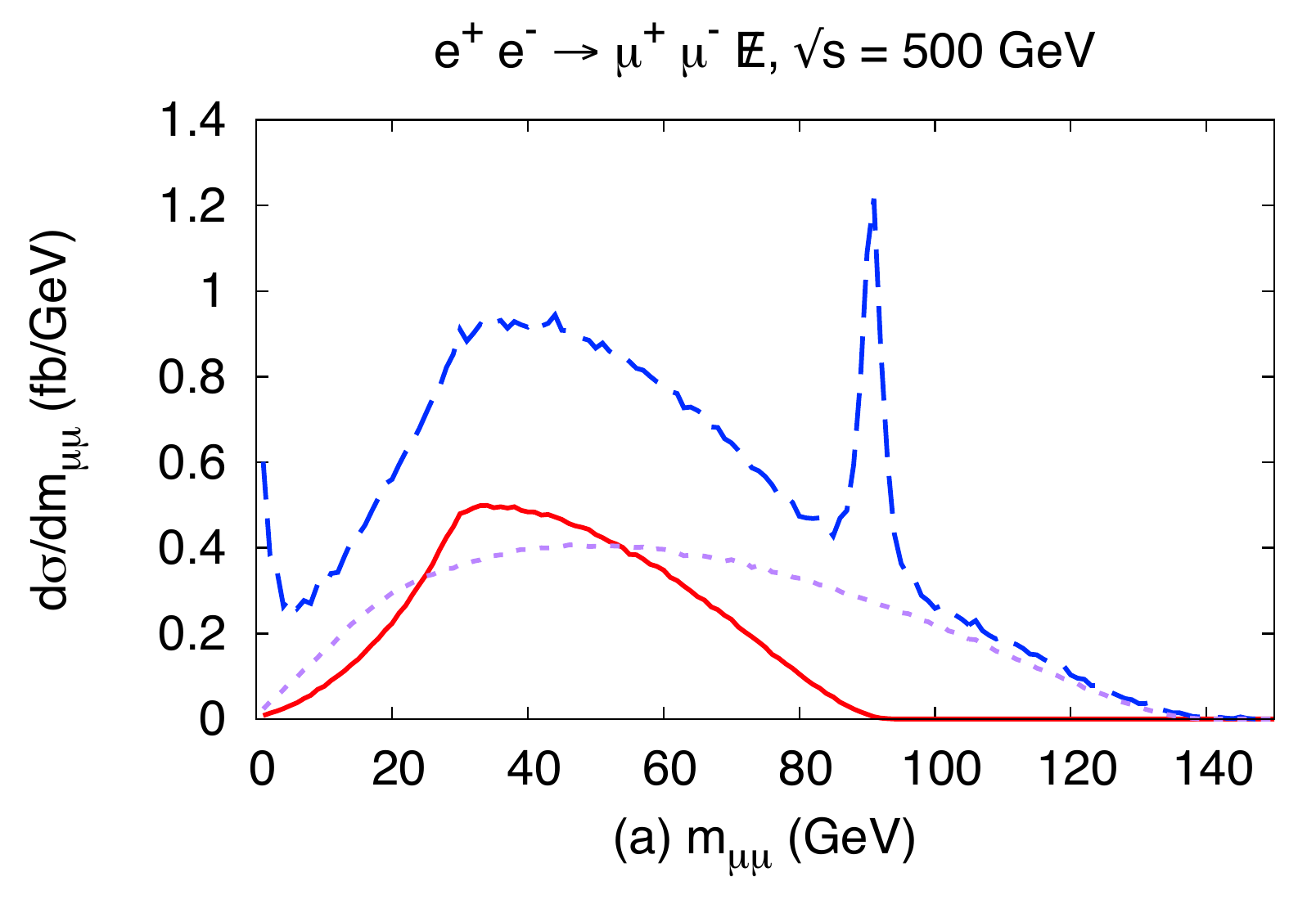}
~~
\includegraphics[width=.47\textwidth]{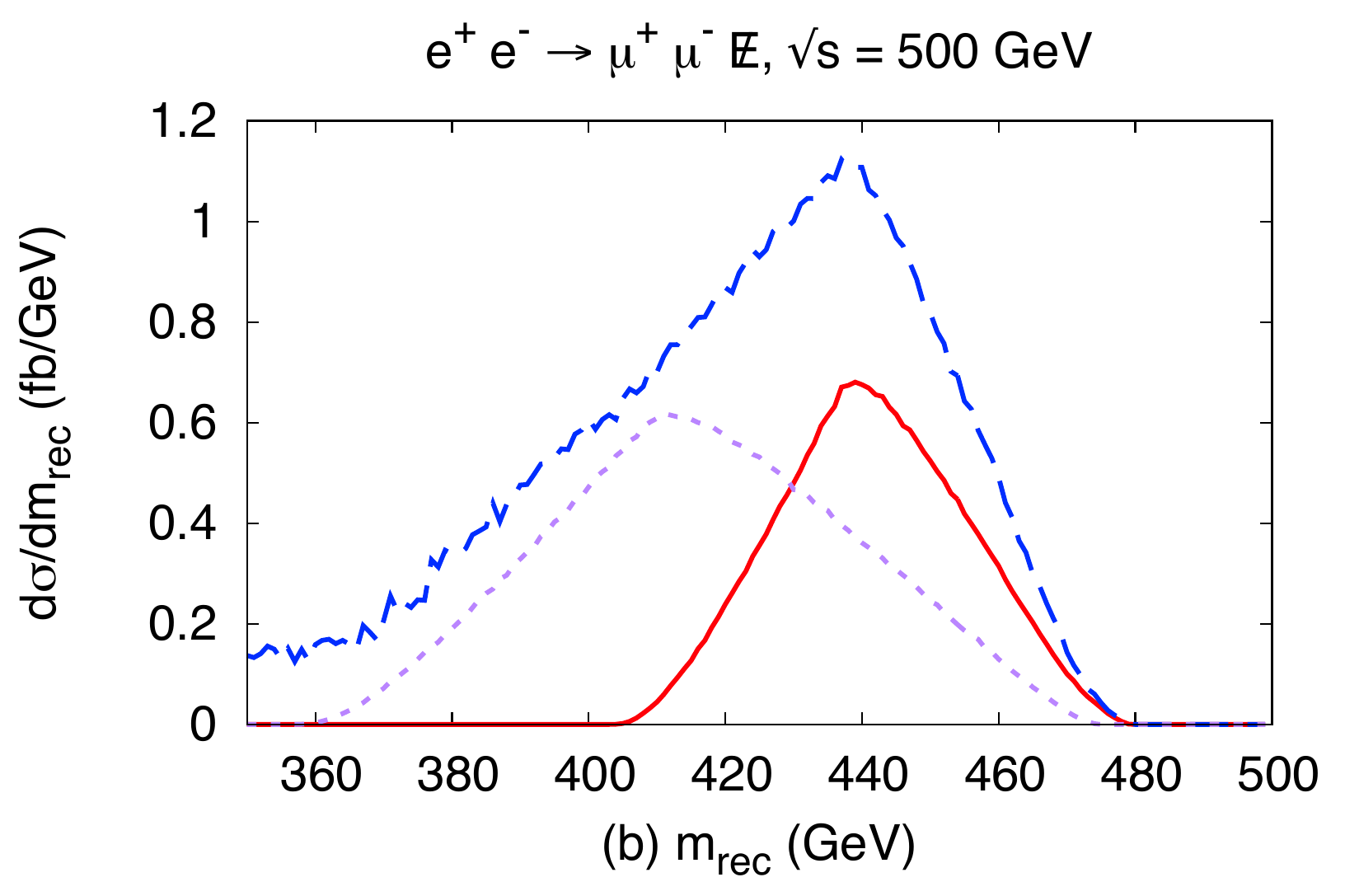}
\\
\includegraphics[width=.47\textwidth]{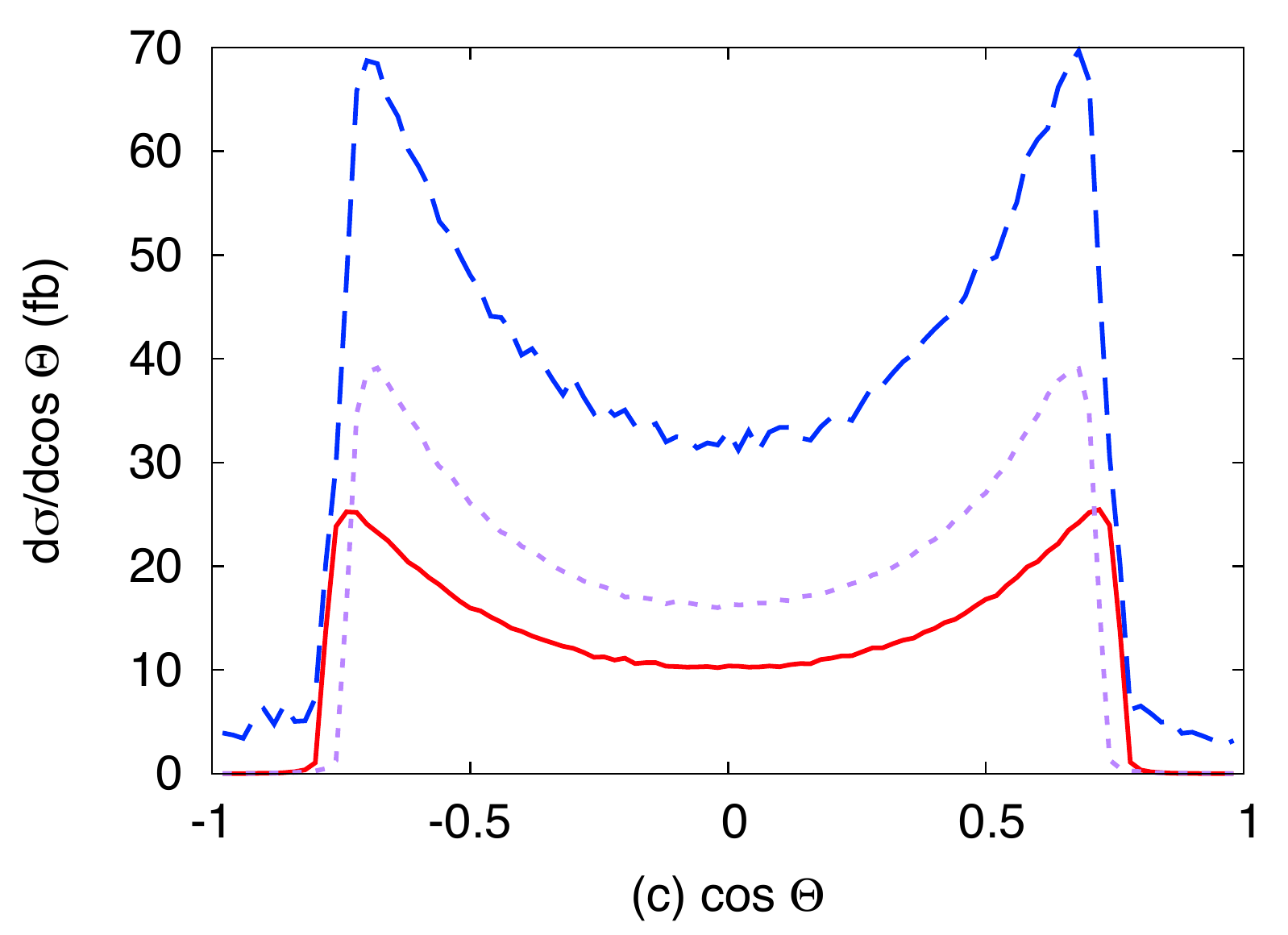}
~~
\includegraphics[width=.47\textwidth]{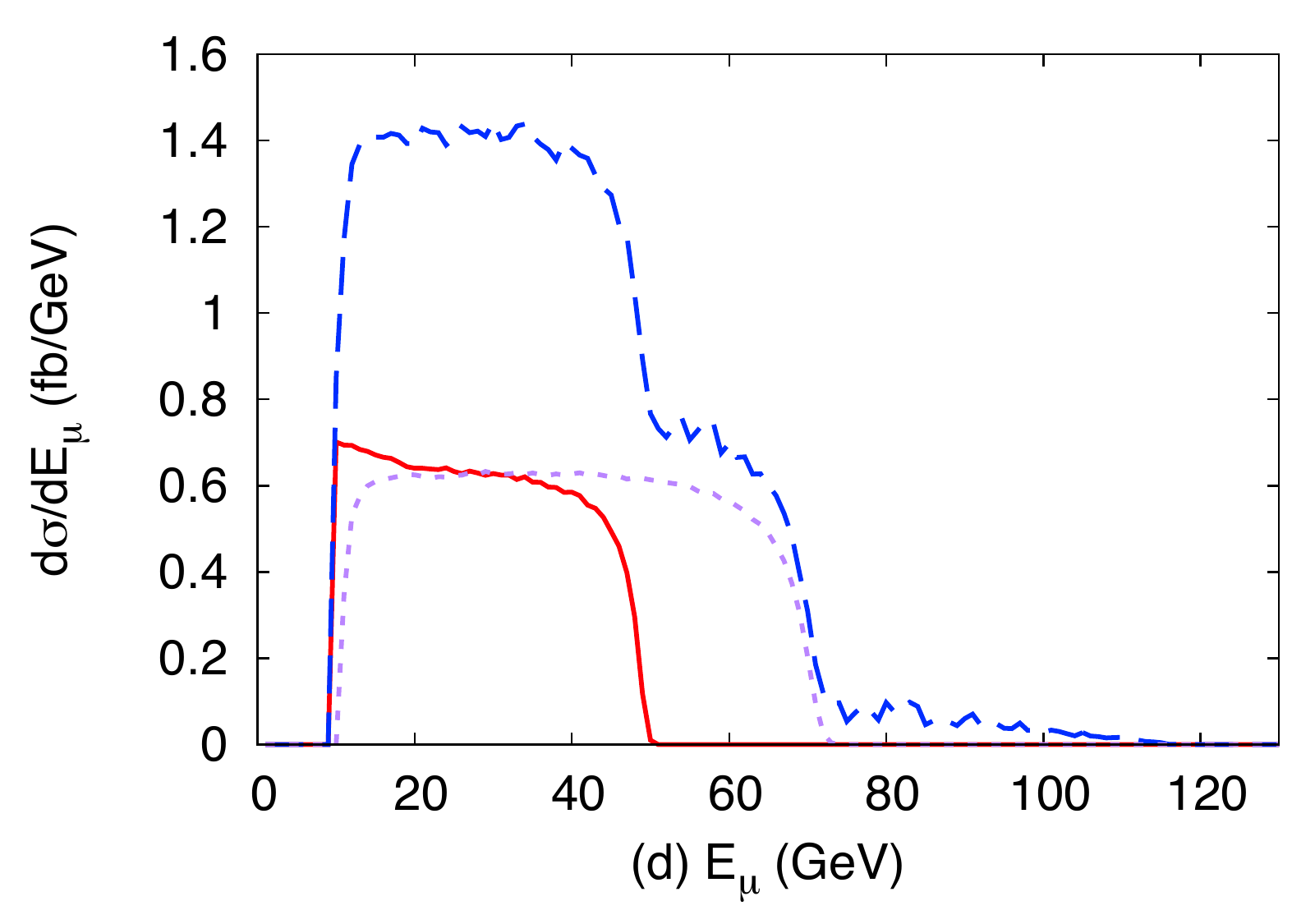}
\\
\includegraphics[width=.47\textwidth]{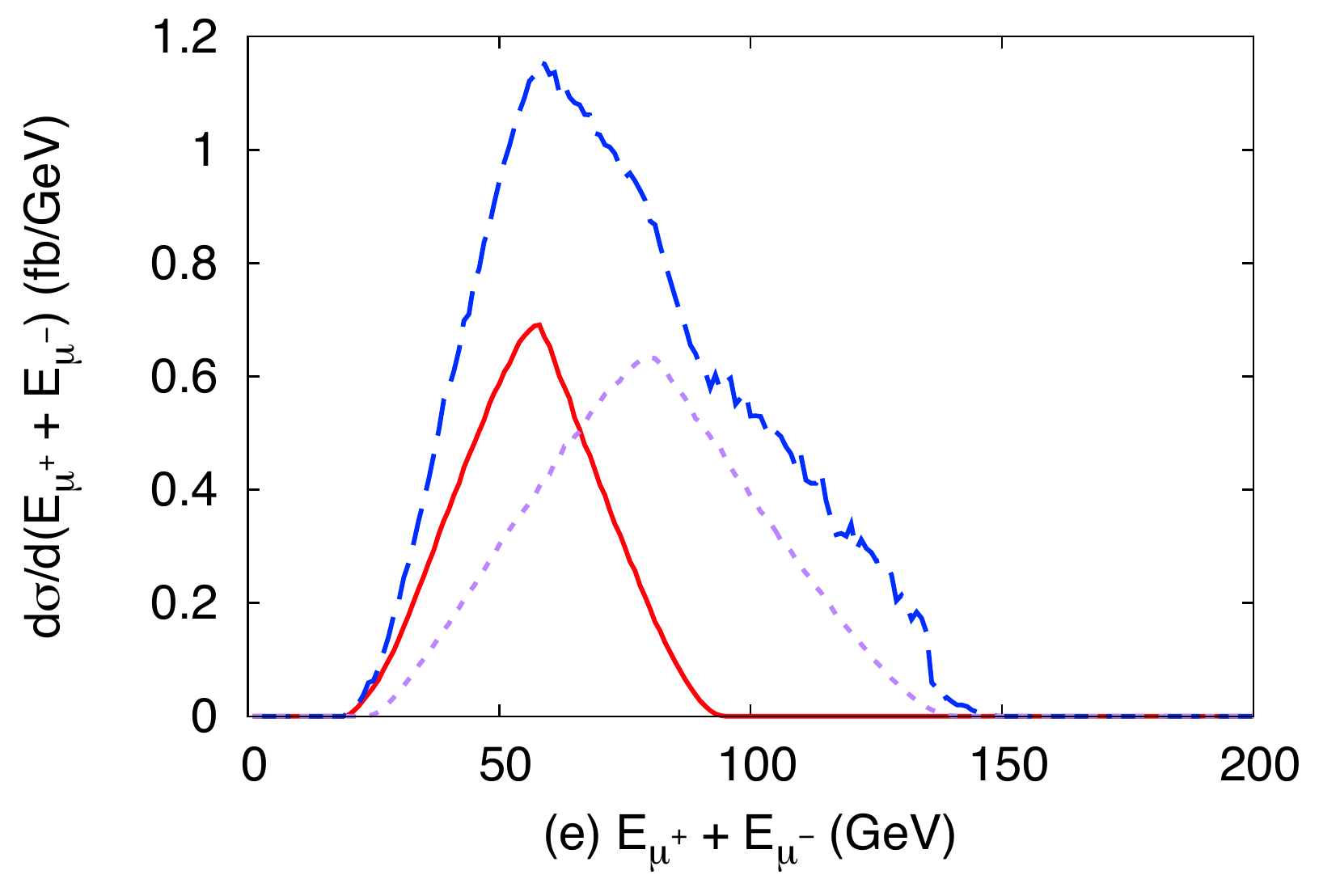}
\caption{\label{fig:caseB:mxxcut}
\caseb~for $e^+ e^- \to  \smul\smul,\ \smur\smur\to \mu^+ \mu^- \,\me$. The additional cut of $\mrec>350\gev$ is included.  We show the 
(a) $m_{\mu\mu}$, (b) $\mrec$, (c) $\cos\Theta$, (d) $E_{\mu}$,
and (e) $E_{\mu^+}+E_{\mu^-}$
distributions with spin-correlation and other realistic effects.
The c.m.~energy is set $\sqrt{s}=500\gev$ for all distributions.
The solid (red) line corresponds to $\smur^+\smur^-$,
the dotted (purple) line to $\smul^+\smul^-$.
The dashed (blue) line is the total differential cross section
including our signal and the SM backgrounds.
%
}
\end{figure}

We now consider the more complex \caseb,
where three different antler processes ($\smur\smur$, $\smul\smul$, and $\ww$) are simultaneously involved.
In Fig.~\ref{fig:caseB:mxxcut},
we present five distributions for \caseb~at $\sqrt{s}=500\gev$.
Here, the $\mrec>350\gev$ cut has been applied to suppress the main SM backgrounds
from $\ww$.
The solid (red) line is the $\smur\smur$ signal,
the dotted (purple) line is from $\smul\smul$.
Finally, the dashed (blue) line is the total differential cross section
including our two signals and the SM backgrounds.
Note that the total rate for $\smur\smur$ is compatible with that for
$\smul\smul$.

In Fig.~\ref{fig:caseB:mxxcut}(a), we show the $m_{\mu\mu}$ distributions.
As expected from the previous analyses, the $\smur\smur$ signal leads to a cusp structure, while $\smul\smul$ and $\ww$ do not due to the specific mass and energy relations.
On the contrary,
the $\mrec$ distribution  for $\smur\smur$ denoted by the solid (red) curve
and that for $\smul\smul$ by the dotted (purple) curve do show a triangle: see
Fig.~\ref{fig:caseB:mxxcut}(b). The SM background is well under-control after the stringent cuts.
The challenge is to extract the hidden mass information from the observed overall (dashed blue) curve as a combination of the twin peaks. It is conceivable to achieve this by a fitting procedure based on two triangles.  Instead, as done below, we demonstrate another approach by taking advantage of the polarization of the beams.

Figure \ref{fig:caseB:mxxcut}(c) presents the $\cos\Theta$ distribution.
The visible $\cos\Theta$ cusp is usually attributed to the lighter intermediate particles ($\smur$ in our case).
A larger $\cosmax$ comes from a smaller $m_B$ with a given c.m.~energy.
We see that, with our parameter choice,
 $\smur\smur$ and  $\smul\smul$ lead to a similar value of $\cosmax$, which differ by about $5\%$.

The $E_\mu$ distribution, with the energy endpoint in Fig.~\ref{fig:caseB:mxxcut}(d),
is known to be one of the most robust variables.
Two box-shaped distributions are added to create a two-step stair. Although ISR and beamstrahlung smear the sharp edges, the observation of the two maxima should be quite feasible. On the other hand, the determination of $\elmin$ could be more challenging if the
 acceptance cut for the lepton lower energy threshold overwhelms $\elmin$ for $\smur\smur$,
and makes it marginally visible for $\smul\smul$.

Finally, we present the energy sum distribution of two visible particles
in Figs.~\ref{fig:caseB:mxxcut}(e).
The individual distribution from $\smur\smur$ and $\smul\smul$ production leads to impressive sharp triangles,
as those in Fig.~\ref{fig:caseB:mxxcut}(b). The challenge is, once again, to extract the two unknown masses from the observed summed distribution.
We next discuss beam polarization as a way to accomplish this.

%

\begin{figure}[t!]
\centering
\includegraphics[width=.47\textwidth]{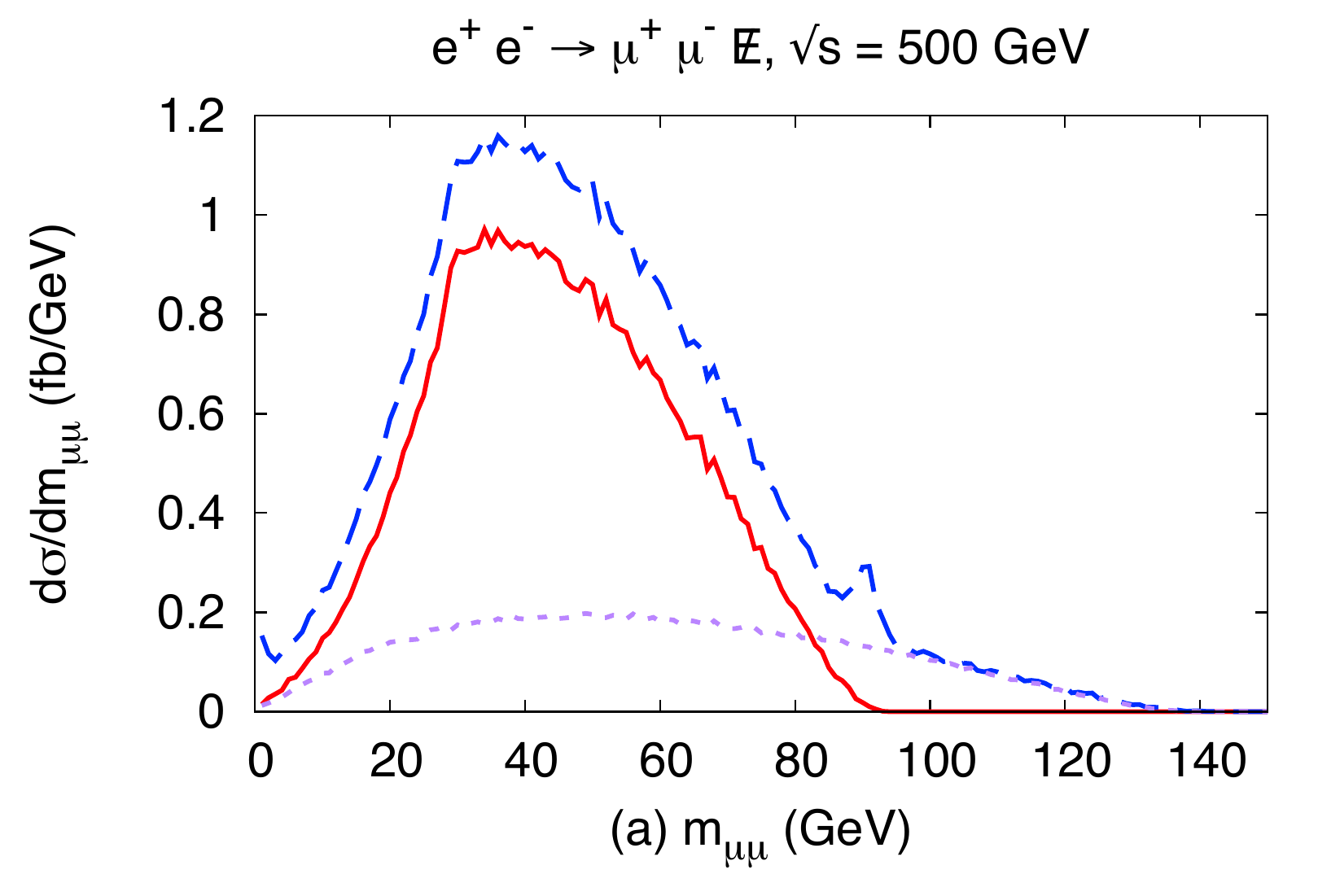}
~~
\includegraphics[width=.47\textwidth]{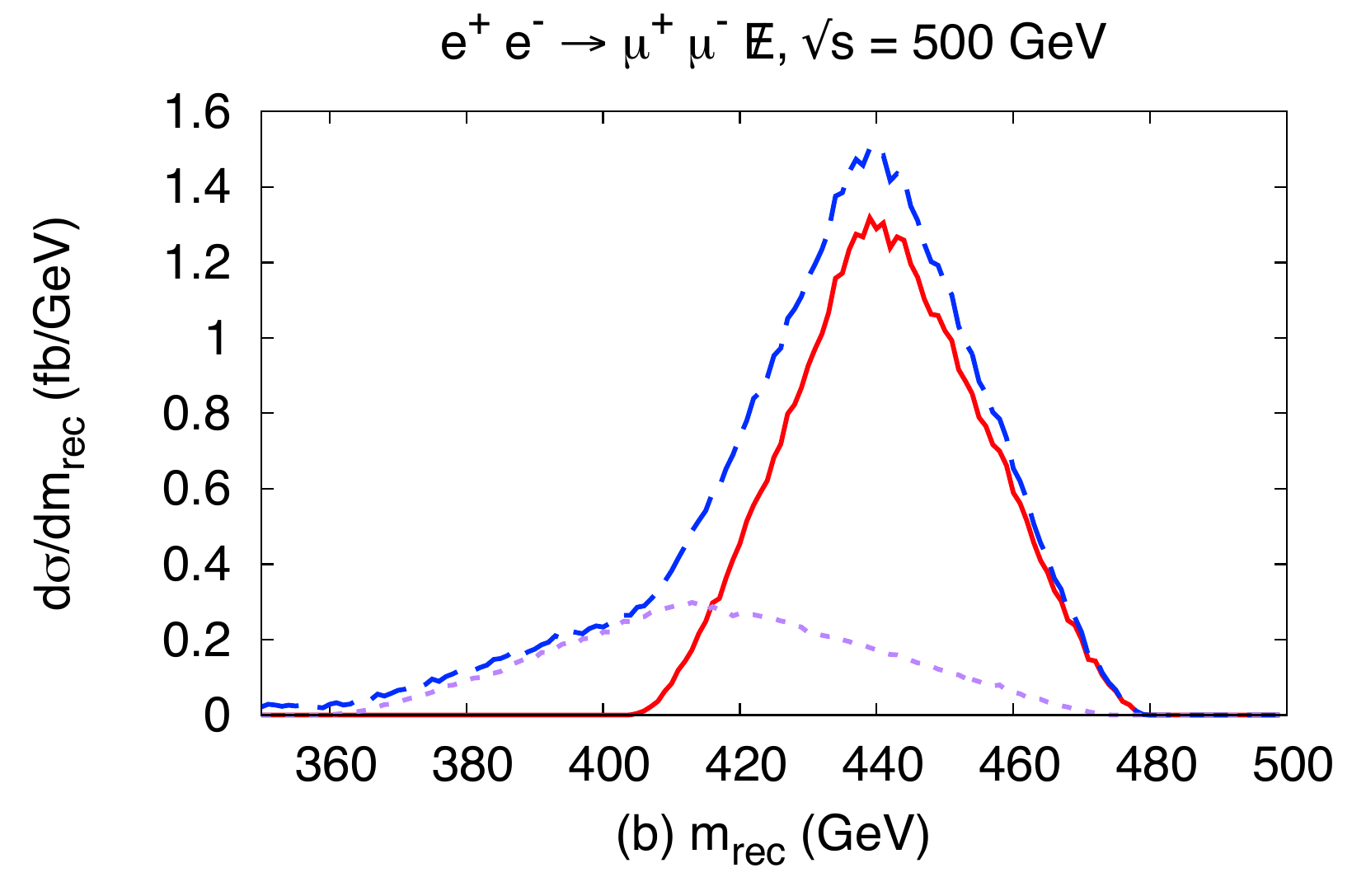}
\\
\includegraphics[width=.47\textwidth]{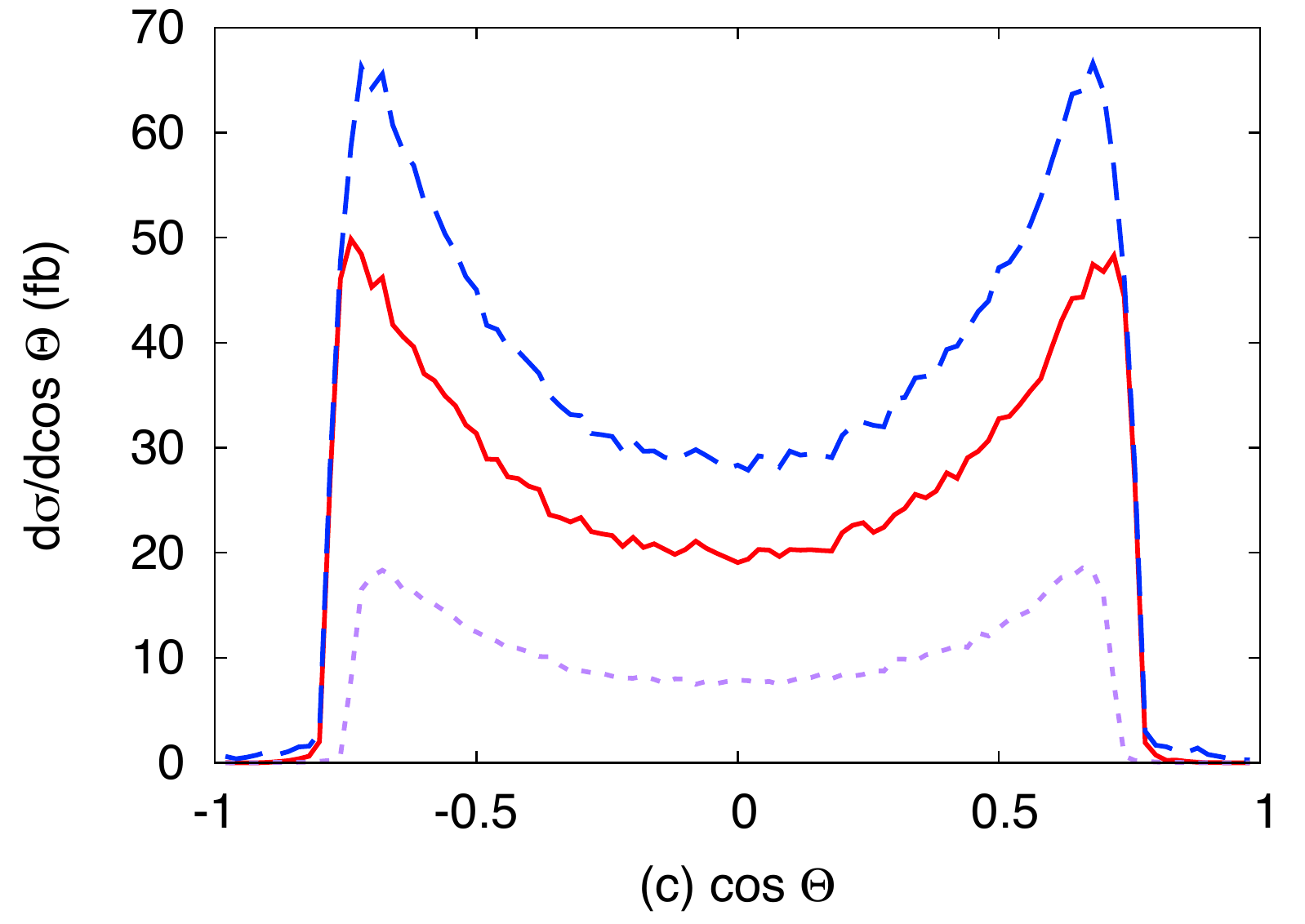}
~~
\includegraphics[width=.47\textwidth]{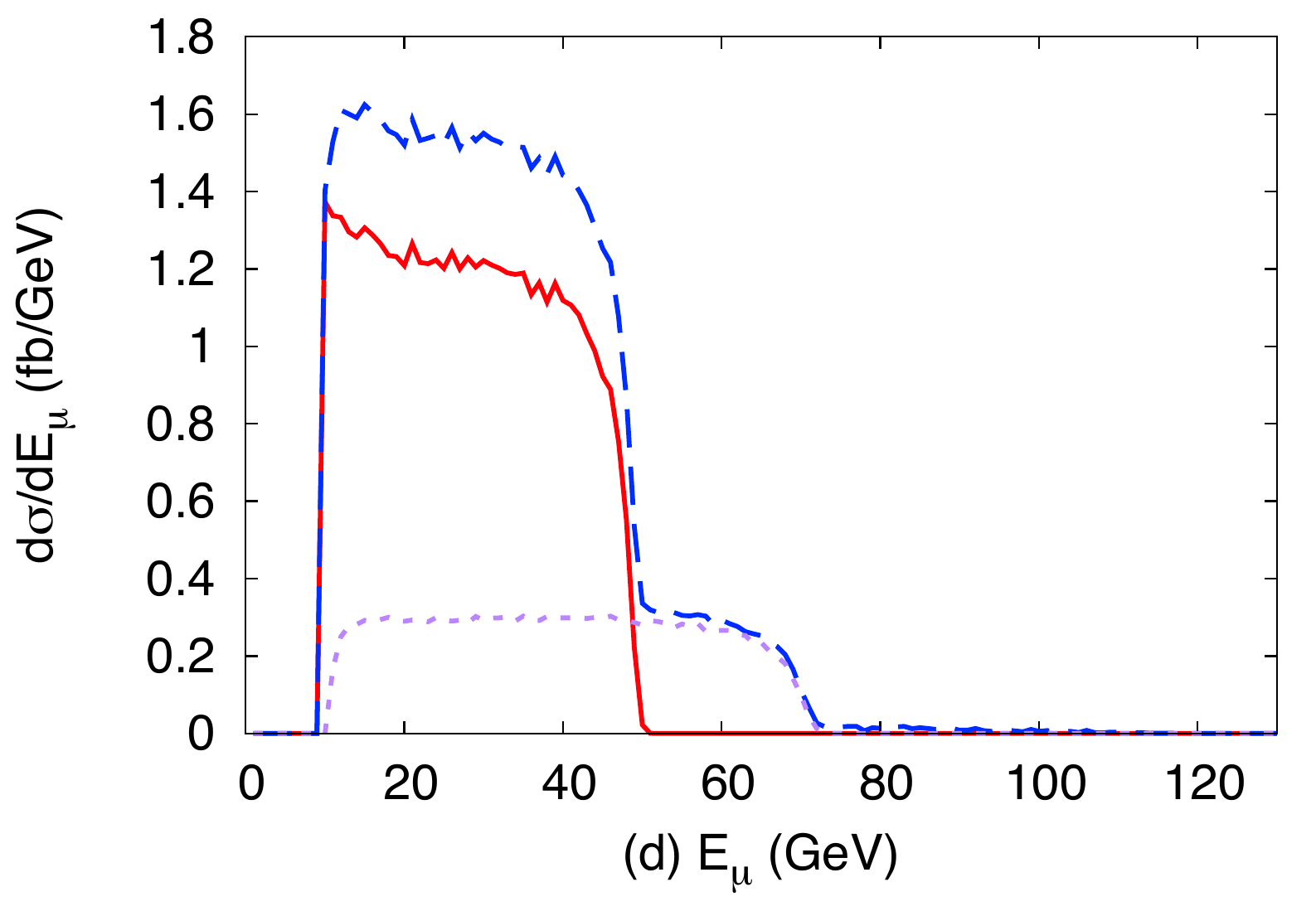}
\\
\includegraphics[width=.47\textwidth]{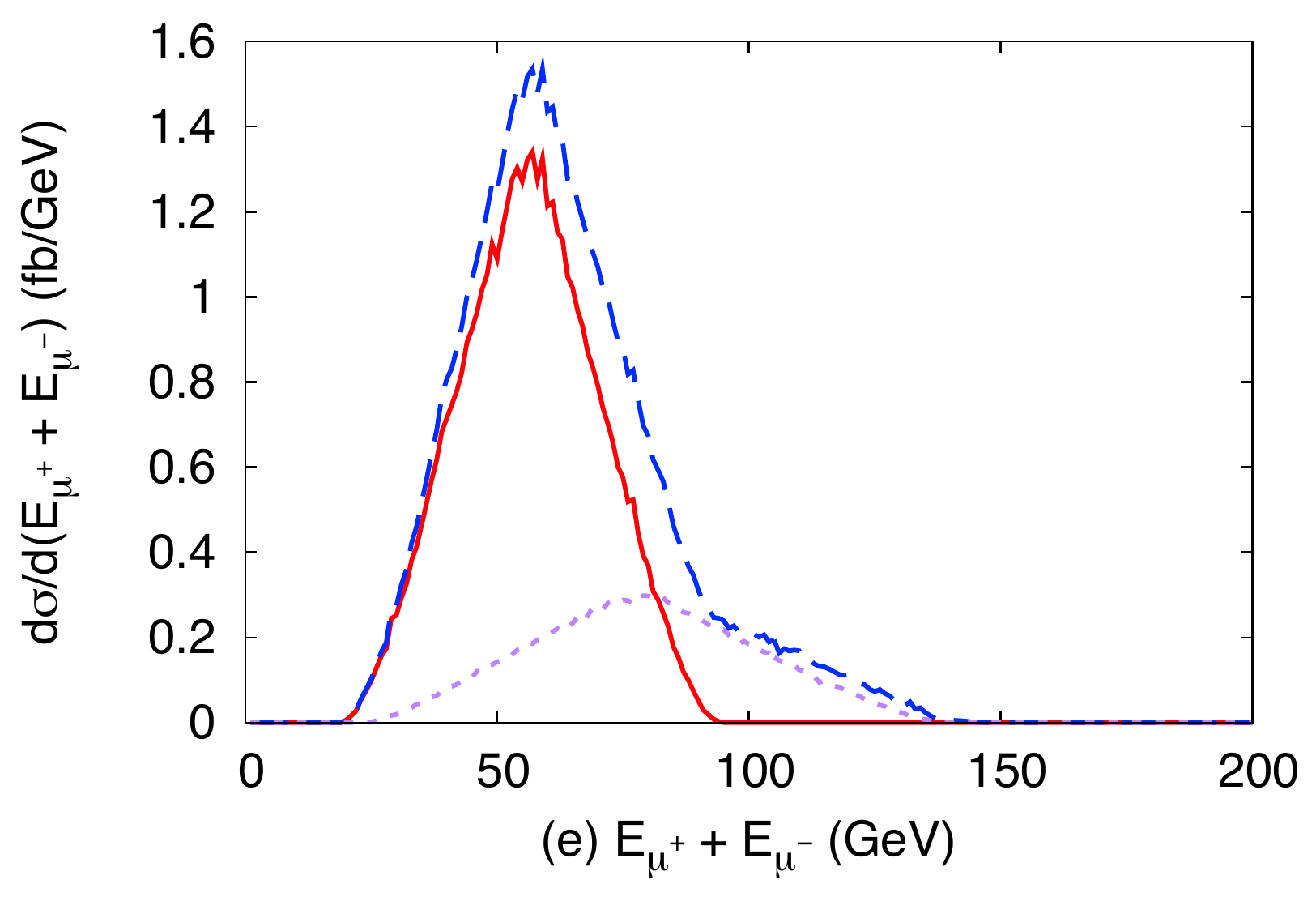}
\caption{\label{fig:caseB:RH}
\caseb~for $e^+ e^- \to  \smul\smul,\ \smur\smur\to \mu^+ \mu^- \,\me$. Effects of an additional cut of $\mrec>350\gev$ and polarizations ${\cal P}_{e^-} = +80\%$ and ${\cal P}_{e^+} = -30\%$ on the 
(a) $m_{\mu\mu}$, (b) $\mrec$, (c) $\cos\Theta$, (d) $E_{\mu}$, and
(e) $E_{\mu^+}+E_{\mu^-}$
distributions with spin-correlation and other realistic effects.
The c.m.~energy is set to $\sqrt{s}=500\gev$ for all distributions.
The solid (red) line corresponds to $\smur^+\smur^-$,
the dotted (purple) line to $\smul^+\smul^-$.
The dashed (blue) line is the total differential cross section
including our signal and the SM backgrounds.
%
}
\end{figure}

All of the distributions
show that the two entangled new physics signals as well as
the SM backgrounds
limit the precise measurements of the cusps and endpoints.
The polarization of the electron and positron beams
can play a critical role in disentangling this information.
The current baseline design of the ILC
anticipates
at least 80\% (30\%) polarization of the electron (positron) beam.
By controlling the beam polarization,
we can suppress the SM backgrounds
and distinguish the two different signals.
For the $\smur\smur$ signal,
our optimal setup is  ${\cal P}_{e^-} = +80\%$ and ${\cal P}_{e^+} = -30\%$,
denoted by $e_R^- e_L^+$,
while for the $\smul\smul$ signal we apply ${\cal P}_{e^-} = -80\%$ and ${\cal P}_{e^+} = +30\%$
denoted by $e_L^- e_R^+$.

Figure \ref{fig:caseB:RH} shows how efficient the right-handed electron beam
is at picking out the $\smur\smur$ signal.
For the suppression of the SM backgrounds, we apply
the cut of $\mrec \geq 350\gev$.
As before,
the solid (red) line corresponds to $\smur^+\smur^-$,
the dotted (purple) line to $\smul^+\smul^-$.
The dashed (blue) line is the total differential cross section
including our signal and the SM backgrounds.
The nearly right-handed electron beam suppresses the
SM background as well as
the $\smul\smul$ signal.
Only the $\smur\smur$ signal stands out.
The main SM background is through the resonant $\ww$ production.
The  left-handed coupling
of $e$-$\nu_e$-$W$ is suppressed by the right-handed electron beam.
Another interesting feature is that the $Z$-pole in the $m_{\mu\mu}$ distribution
is also very suppressed.  A significant contribution to the $Z$-pole is from $e^+ e^- \to \nu_e \bar{\nu}_e Z$
process where $Z$ is via $WW$ fusion.  Again the left-handed coupling of the charged current is suppressed
by the right-handed electron beam.

The advantage of the cusp is clearly shown here.
Its peak structure is not affected.
However, the endpoints $\mrecmin$, $\elmin$,  and $\ellmax$
do overlap with the backgrounds,
although the right-handed polarization removes
a large portion of the SM backgrounds.
We also observe that
$\mrecmax$, $\elmax$,  and $\ellmin$
are not contaminated.
In summary, the mass measurement of $\smur$ and $\neuo$
through the cusps and endpoints
is well benefitted by the right-handed polarization of the electron beam.

\begin{figure}[t!]
\centering
\includegraphics[width=.49\textwidth]{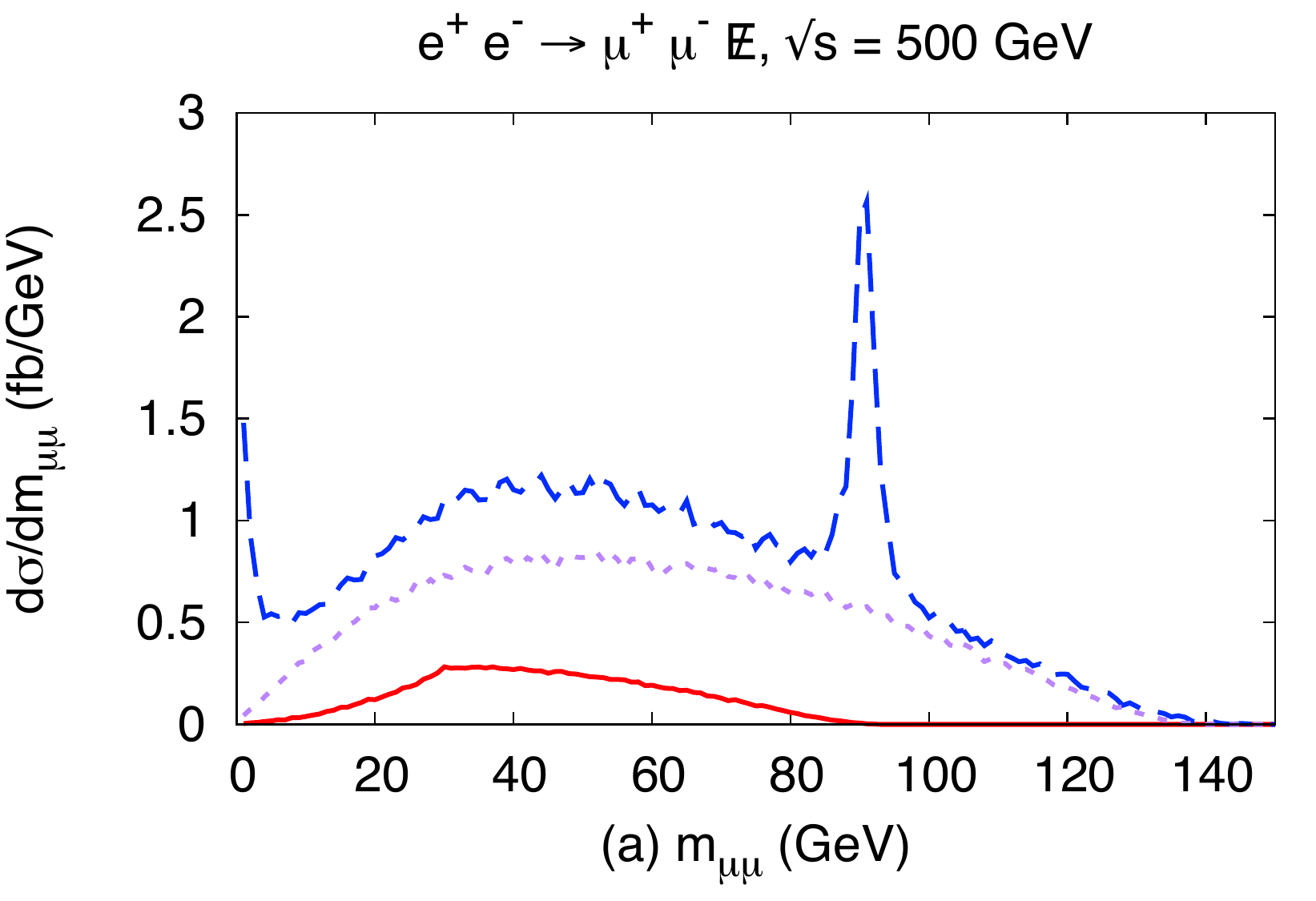}
~~
\includegraphics[width=.47\textwidth]{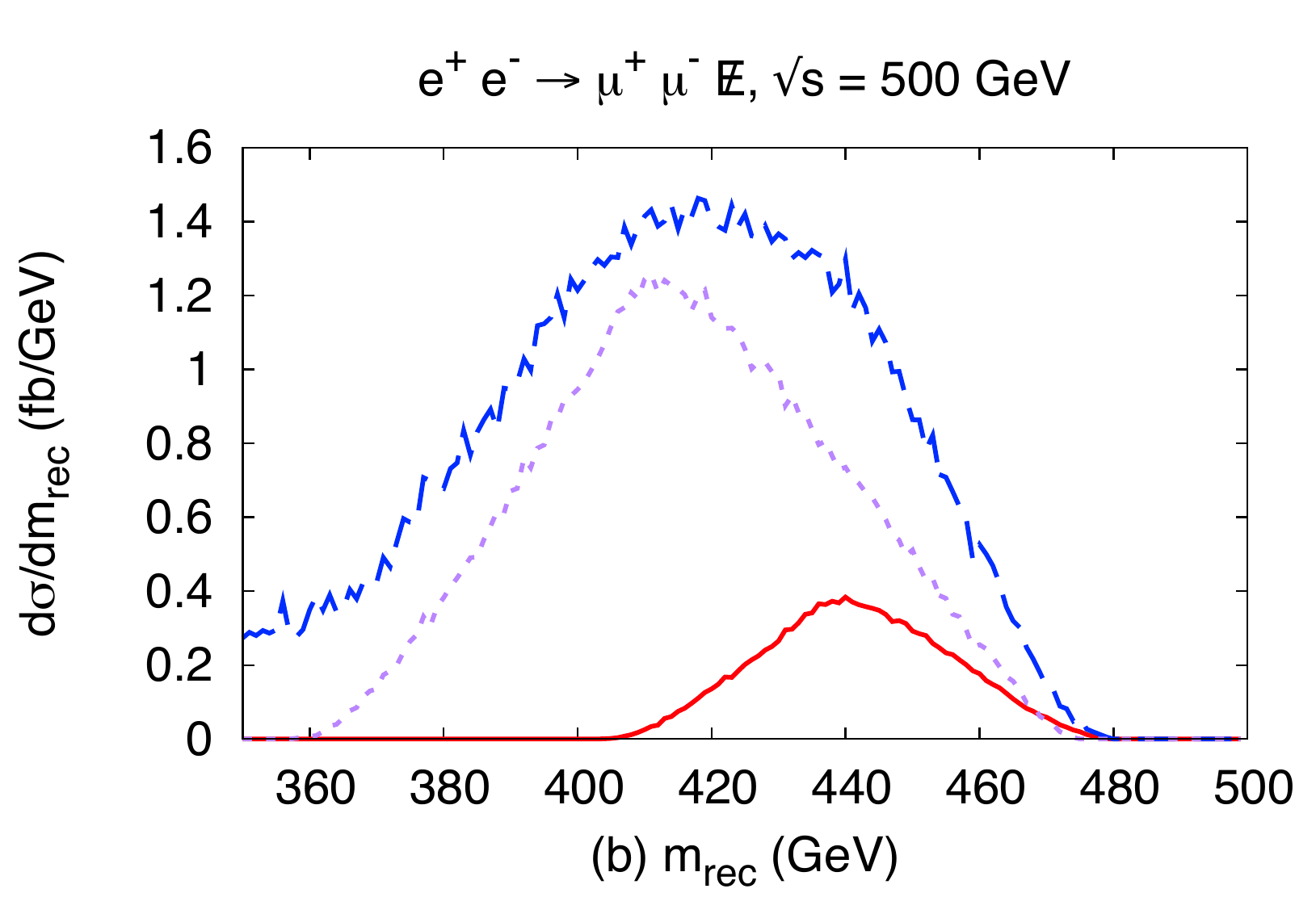}
\\
\includegraphics[width=.49\textwidth]{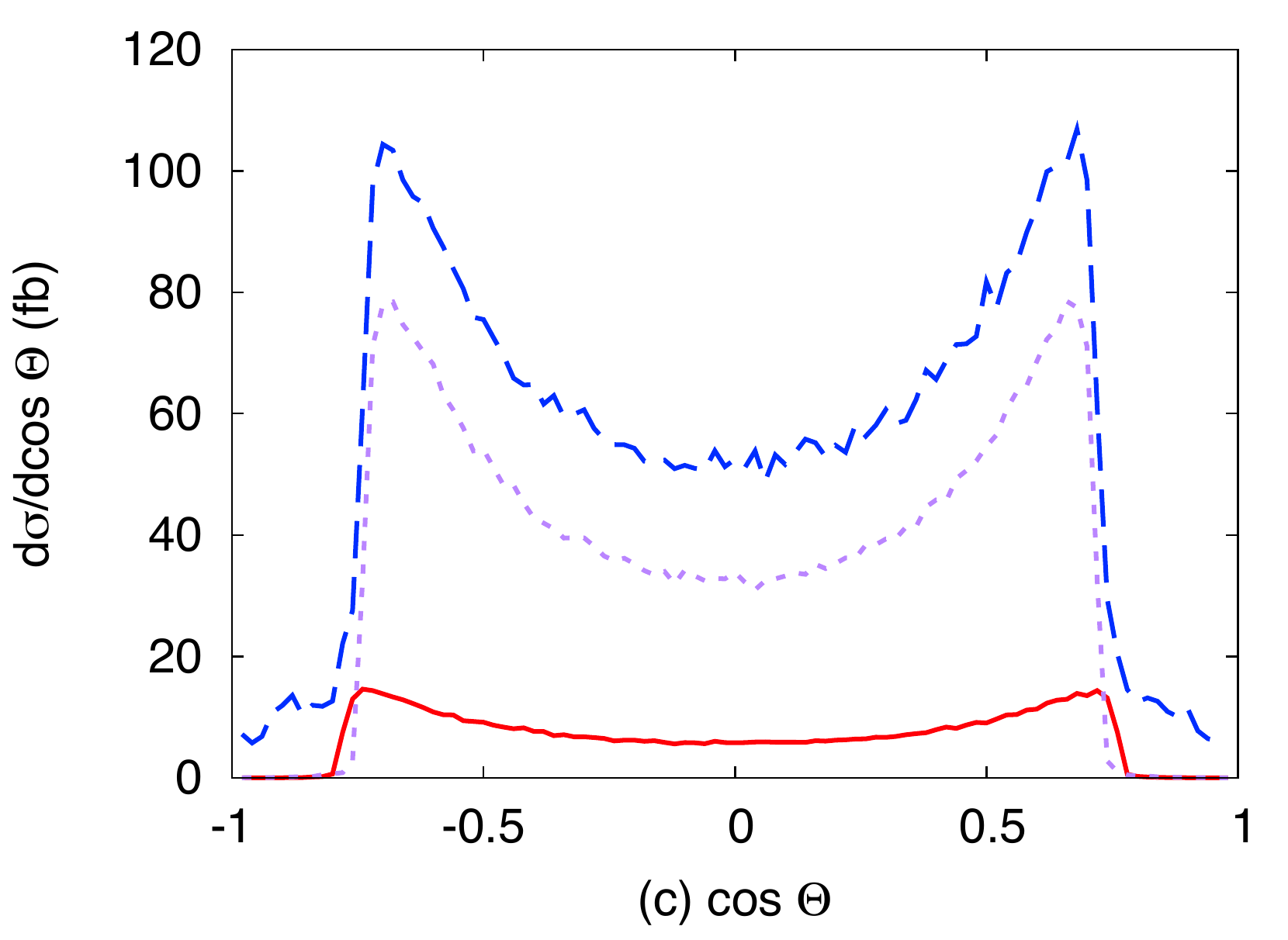}
~~
\includegraphics[width=.47\textwidth]{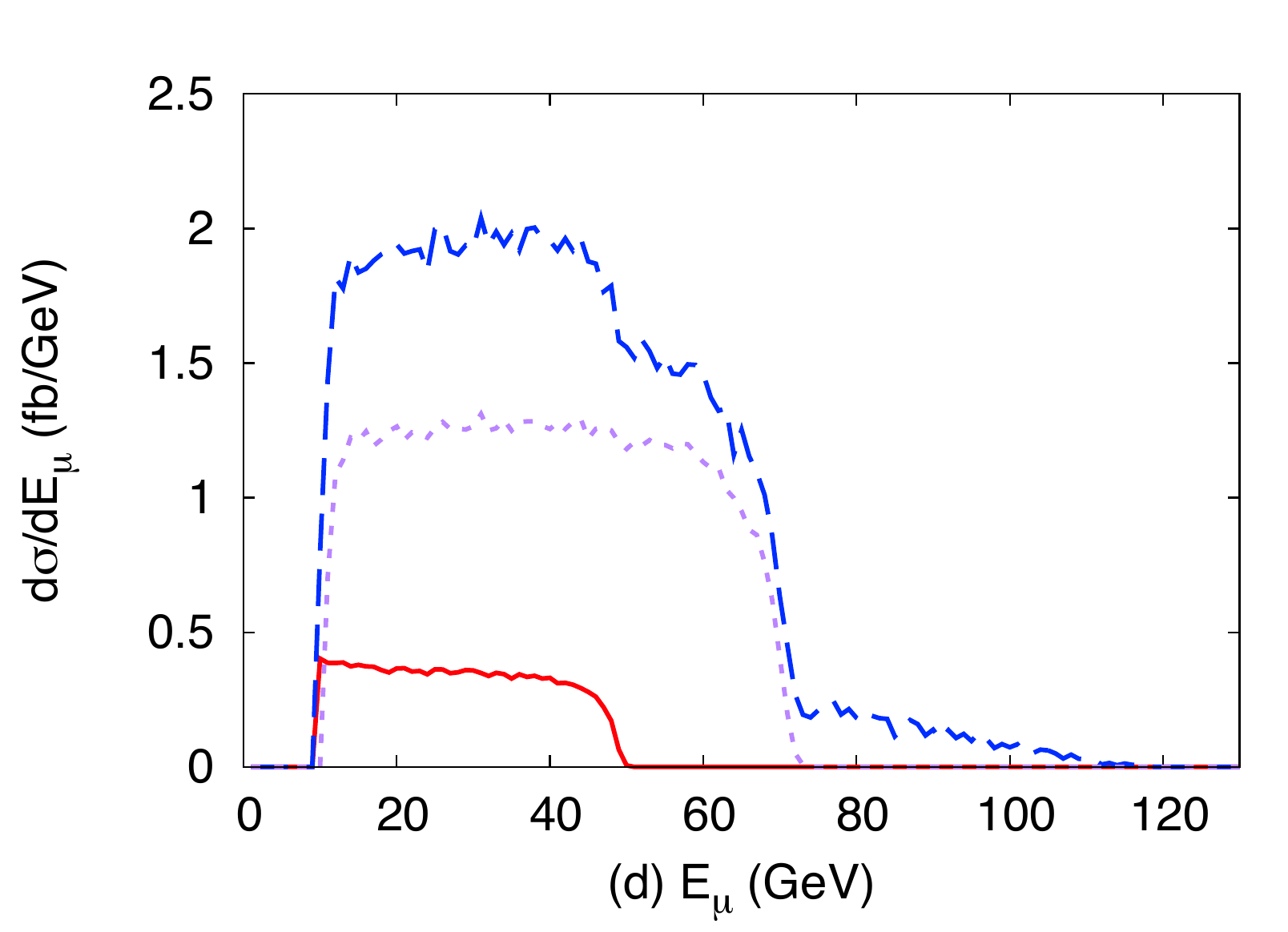}
\\
\includegraphics[width=.49\textwidth]{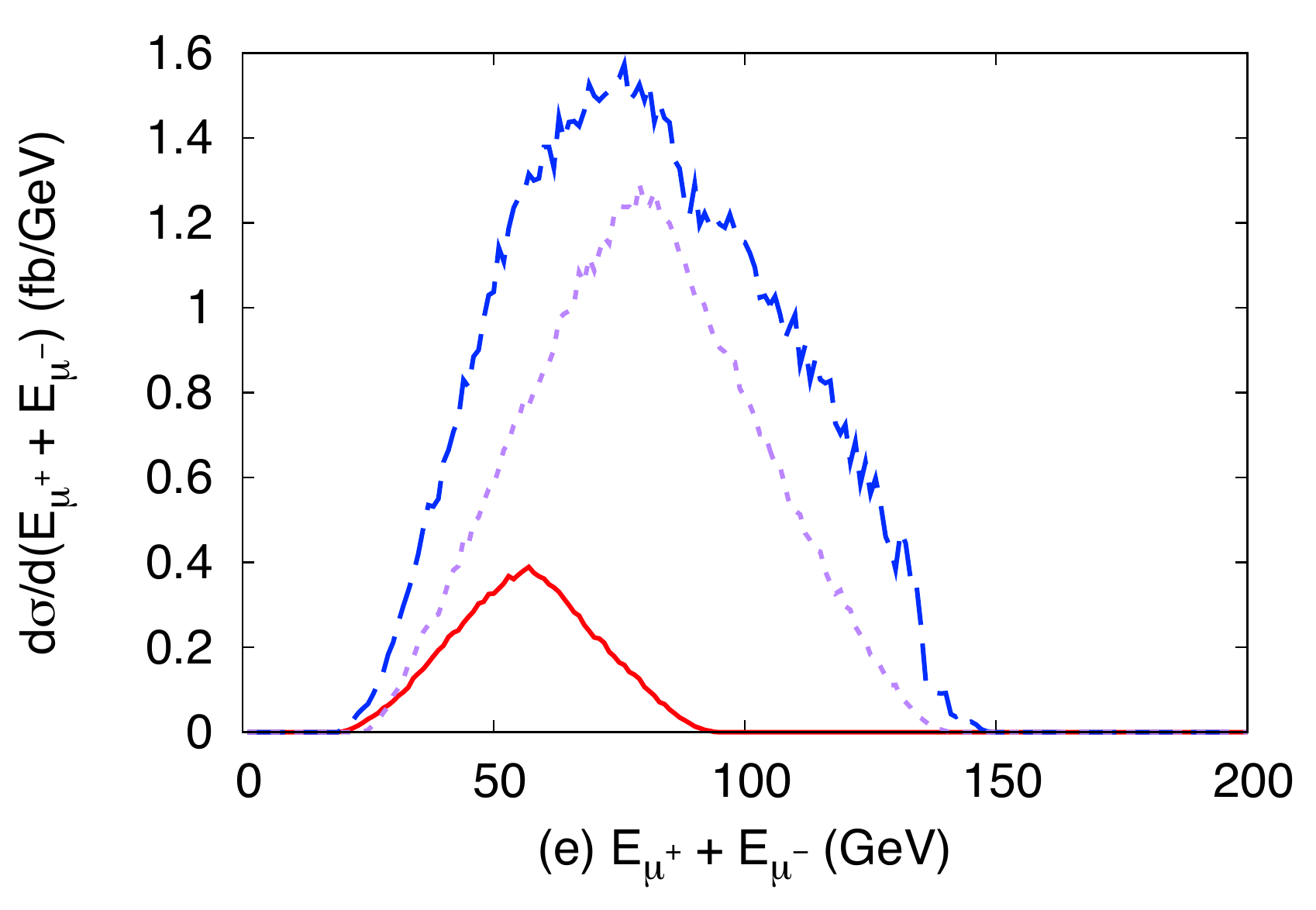}
\caption{\label{fig:caseB:LH}
\caseb~for $e^+ e^- \to  \smul\smul,\ \smur\smur\to \mu^+ \mu^- \,\me$. Effects of an additional cut of $\mrec>350\gev$ and polarizations ${\cal P}_{e^-} = -80\%$ and ${\cal P}_{e^+} = +30\%$ on the  
(a) $m_{\mu\mu}$, (b) $\mrec$, (c) $\cos\Theta$, (d) $E_{\mu}$, and (e)
$E_{\mu^+}+E_{\mu^-}$
distributions with spin-correlation and other realistic effects.
The c.m.~energy is set to $\sqrt{s}=500\gev$ for all distributions.
The solid (red) line corresponds to $\smur^+\smur^-$,
the dotted (purple) line to $\smul^+\smul^-$.
The dashed (blue) line is the total event
including our signal and the SM backgrounds.
%
}
\end{figure}

The left-handed $\smul\smul$ signal is more difficult to probe
since its left-handed coupling is the same as the
SM background.
In Fig.~\ref{fig:caseB:LH},
we set ${\cal P}_{e^-} = -80\%$
and ${\cal P}_{e^+} = +30\%$
with the additional cut of $\mrec>350\gev$.
From the $m_{\mu\mu}$ distribution,
we see that the $Z$-pole is still strongly visible
and the round $\mllc$ for the $\smul\smul$ signal is very difficult to identify.
The total $\mrec$ distribution in Fig.~\ref{fig:caseB:LH}(b)
does not show the sharp triangular shape of the antler decay topology either.
The individual triangular shapes of the $\smur\smur$ and $\smul\smul$ signals
along with the SM background
are combined into a rather featureless bump-shaped distribution.
Although there is a peak point,
it is hard to claim as a cusp.
The $\cos\Theta$ distribution in Fig.~\ref{fig:caseB:LH}(c) shows
one of the most characteristic features of the antler topology.
Two sharp cusps appear,
which correspond to the $\smul\smul$ signal.

The total $E_\mu$ distribution in Fig.~\ref{fig:caseB:LH}(d)
does not provide quite a clean series of rectangular distributions.
The mixture of different contributions from $\smur\smur$, $\smul\smul$ and
$\ww$ along with
the smearing makes reading the maximum points more difficult.
The $\elmin$ position of the $\smul\smul$ signal,
which is near the kinematic cut, is mixed with the SM backgrounds
and the $\smur\smur$ signal.
Finally, 
the total $E_{\mu\mu}$ distribution
loses
the triangular shape of the $\smul\smul$ signal: see Fig.~\ref{fig:caseB:LH}(e).
Nevertheless the peak position coincides with the cusp position for both energy sum distributions.
We can identify them with the cusps.

\subsection{The mass measurement precision}
\label{sec:LL}

In order to estimate the achievable precision of a measurement of the masses in the presence of realistic effects, we analyze the distributions we have discussed here using the log-likelihood method based on Poisson statistics.
A benefit of a log-likelihood analysis is that it compares the full shape of the distribution,
not just the position of the cusps and endpoints which, as we have seen,
can be smeared and even moved due to realistic collider effects.
For our log-likelihood calculation, since we have shown that the background can be almost totally removed by appropriate cuts, we focus on comparing one signal to another with different masses for the smuon
and neutralino.

\begin{figure}[t!]
\centering
\includegraphics[height=9cm]{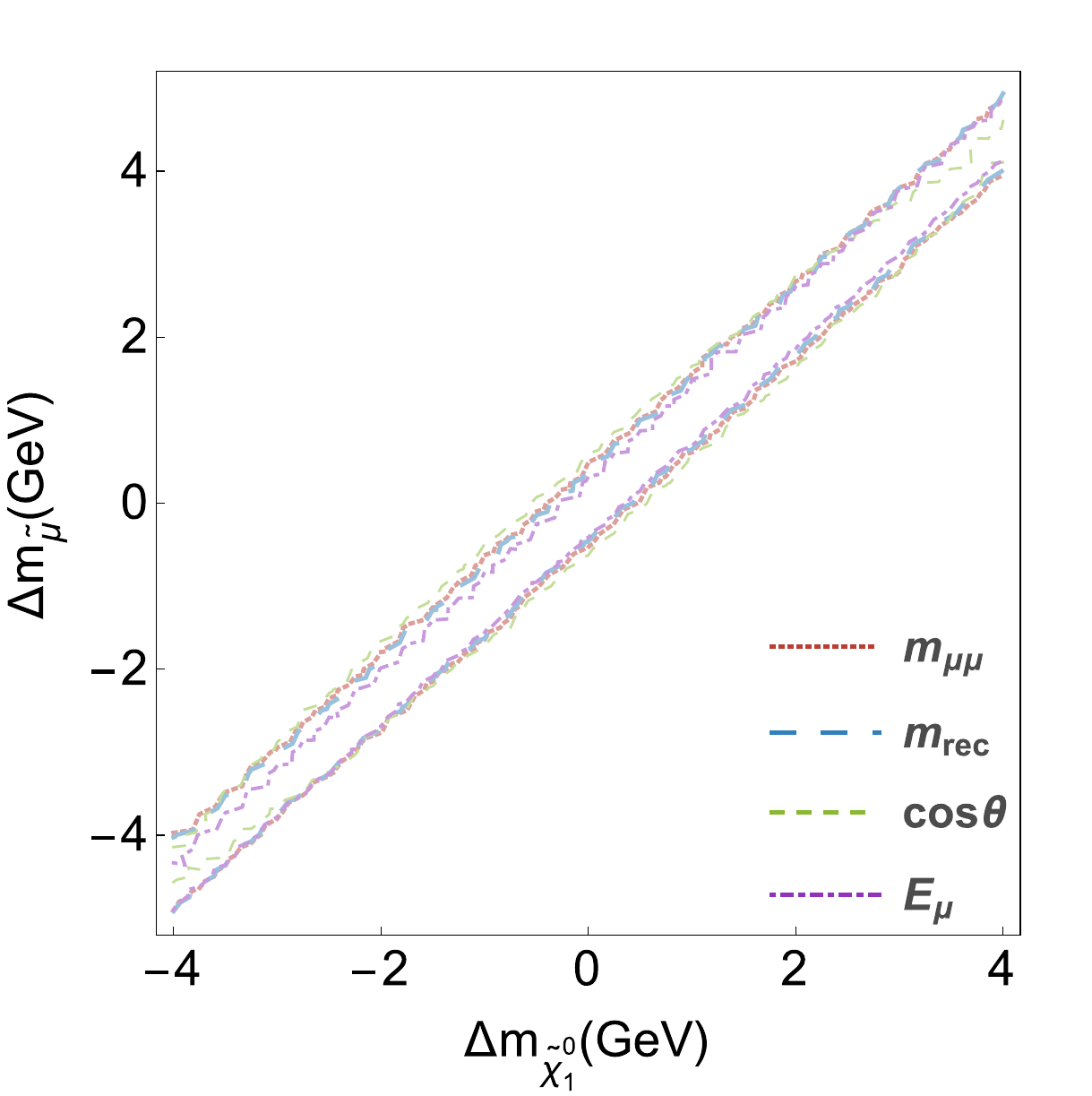}
\caption{\label{fig:contour:caseA}
For \casea~for $e^+ e^- \to  \smur\smur\to \mu^+ \mu^- \,\me$,
the 95\% C.L. contours for the precision
of the mass measurement
in the parameter space of $(\Delta m_\neuo, \Delta m_{\smur})$.
An additional cut of $\mrec>350\gev$ on the distributions with spin-correlation and other realistic effects are included. The c.m.~energy is set to $\sqrt{s}=500\gev$ for all distributions and the integrated luminosity is $100\ifb$.
}
\end{figure}

We calculate the log-likelihood as
\begin{equation}
LL(N;\nu) = 2\sum_i\left[N_i\ \mbox{ln}\left(\frac{N_i}{\nu_i}\right)+\nu_i-N_i\right]
\end{equation}
where $\nu_i$ is the expected number of events in bin $i$ with the masses set according to \casea\ and $N_i$ is the number of events expected in bin $i$ for the alternate mass point.
For each distribution, we use 50 bins.  We take the integrated luminosity to be 100 fb$^{-1}$ and find that the number of signal events is sufficiently large that
the probability distribution of
the log-likelihood approximates well a $\chi^2$ distribution.  We then find that the 95\% confidence level value for each log-likelihood is $LL_{95\%}=67.5$.  We scan over the masses of the smuons and neutralinos in steps of $0.25\gev$,
calculate the log-likelihood for each mass point, and plot the contour where it is equal to $67.5$ in Fig.~\ref{fig:contour:caseA} for four kinematical variables assuming \casea.  These are the 95\% confidence lines for each kinematical variable considered separately.

Considering the kinematics variables of 
$m_{\mu \mu}$ (red), $m_{rec}$ (blue), $\cos\Theta$ (green), and $E_\mu$ (purple), 
we present the 95\% C.L. allied contours in
the parameter space of $(\Delta m_\neuo, \Delta m_{\smur})$ in Fig.~\ref{fig:contour:caseA}.
All the variables are roughly equally good at measuring the two masses, leading to an accuracy of approximately $\pm0.5\gev$ (for clarity of the presentation, we have left out the contours for $E_{\mu \mu}$ and $\erec$). 

%

We also find that our kinematical variables are very sensitive if we vary one mass parameter with the other fixed.
However, the determination for the two masses is correlated,
as seen from Fig.~\ref{fig:contour:caseA} with a linear band rather than a closed ellipse in the plotted region. This is due to the fact that the cusps and endpoints depend on the masses mainly as a ratio rather than independently, as can be seen in Eqs.~\eqref{eq:maa:min:c:max}, \eqref{eq:beta}, and \eqref{eq:Eaa:min:cusp:max}. The ellipse shape of the contour will become manifest when extending to larger regions.

We have also considered the effect of combining these measurements in a joint test-statistic including a calculation of the correlation between these variables. The magnitude of the correlation is quantified by the ratio
of the off-diagonal term to the diagonal  term of the covariance matrix.
We found that the correlation among $\mrec$, $E_\mu$ and $\cos\Theta$
was negligible (the off-diagonal terms of the covariance matrix was a few percent or smaller compared to the diagonal terms), the correlation between $\mrec$ and $E_{\mu\mu}$ was small but non-negligible (the off-diagonal term was approximately 8\% of the diagonal terms), and  $E_{\mu\mu}$ and $\erec$ were fully correlated as expected (the off-diagonal term was the same size as the diagonal term).    However, we did not find appreciable improvement in the precision of the mass measurements
by combining the log-likelihoods.  This is due partly to the correlation between these variables, partly to the differences in how the log-likelihood depends on each of these variables, and partly to the properties of the $\chi^2$ distribution when test statistics with a large number of degrees of freedom are combined as we briefly explain in Appendix \ref{app:non combination improvement}.


\section{Massive visible particle case: chargino pair production}
\label{sec:massive}

It is quite likely that the DM particles will be accompanied by other massive observable final states in the decay process. Although the nature of the cusps is similar to the previous discussions, the characteristic features and their observability may be different.  An important example of this type of kinematics is in chargino pair production followed by the chargino's decay into a $W$ and a $\neuo$.
This process is a typical antler process,
which is different from the smuon pair production
in that the visible particle $W$ is massive.
In order to fully reconstruct the kinematics of the $W$,
we consider the case where the $W$ boson decays hadronically.
Our signal event selection is
\bea
e^+ e^- \to \chaop\chaom \to W^+ W^- \neuo\neuo \to jj, jj +\neuo\neuo.
\eea
For illustrative purposes, we consider the \casec ~in Table \ref{table:sps1b:mass}.

For the LHC searches of gaugino production, there is no sensitivity with the current data yet \cite{LHCInos} 
for the parameter choices under consideration, 
due to the disfavored kinematics of the small mass
difference and the large SM backgrounds. The upcoming Run
II at 13 TeV will likely reach the sensitivity to cover this parameter
region \cite{Inos}. It is thus exciting to look forward to the LHC
outcome. Should a SUSY signal be observed at the LHC, it would strongly
motivate the ILC experiment to further study the SUSY property and to
determine the missing particle mass as proposed in this work.

{\renewcommand{\arraystretch}{1.1}
\begin{table}[t!]
\centering
\begin{tabular}{|c|c|c|c|c|}
\hline
~~~$\sqrt{s}$~~~ & Channel & $(m_B,m_X,m_a)$ & $(\mwwmin,\mwwc,\mwwmax)$ &
$(\mrecmin,\mrecc,\mrecmax)$
\\ \hline \hline
\multirow{3}{*}{500}   & \multirow{3}{*}{$\chaop\chaom$}& $(235,139,m_W)$ &
 $(161,171,221)$ & $(279,296,338)$ \\  \cline{3-5}
 \cline{3-5}
 & & $(\ewmin,\ewmax)$ &
$(\ewwmin,\ewwcusp,\ewwmax)$ & $(\exxmin,\exxcusp,\exxmax)$ \\ \cline{3-5}
& &
$(81,111)$ & $(162, 190, 221)$ &
$(278, 309, 338)$
\\ \hline
\end{tabular}
\caption{\label{table:chargino:cusp:values}
The values of various kinematic cusps and
endpoints for the mass parameters in the \casec.
All the masses and energies are in units of GeV.
}
\end{table}
}

The distributions of the invariant mass
of $\ww$ and $\neuo\neuo$
follow the same characteristic function
where now the visible particle $W$ is massive.
The cusp and endpoint positions of these distributions can be obtained from Table \ref{table:cusp:massive}.
The $\cos\Theta$ distribution for the massive visible particle case
does not present a sharp cusp or endpoint.
The $E_W$ distribution has a minimum and a maximum as in the massless visible particle case.
The distribution of $E_{WW}=E_{W^+}+E_{W^-}$
also accommodates the maximum, cusp and minimum.
In Table \ref{table:chargino:cusp:values},
we present the values of the cusps and endpoints for \casec.

\begin{figure}[t!]
\centering
\includegraphics[width=.47\textwidth]{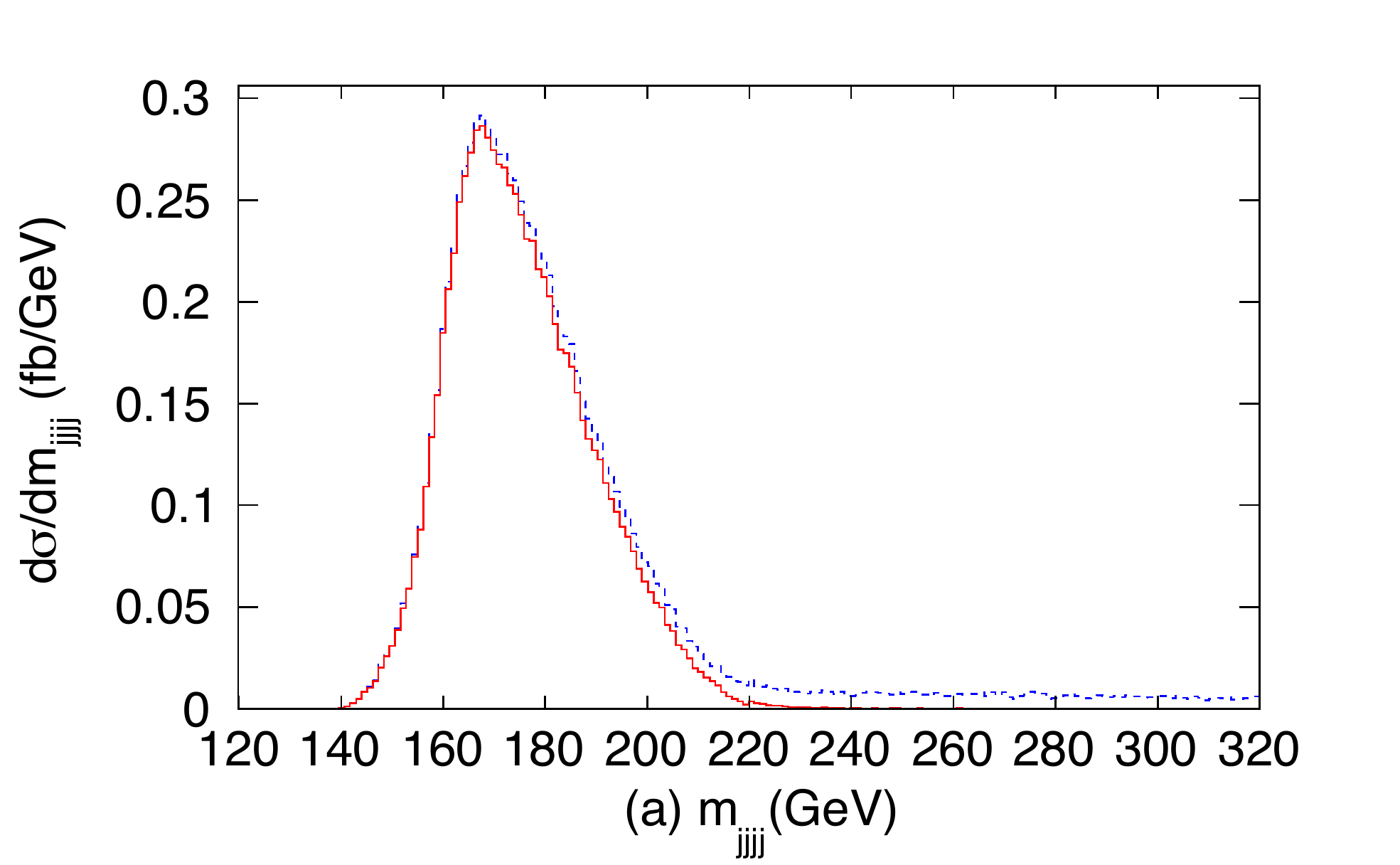}
~~
\includegraphics[width=.47\textwidth]{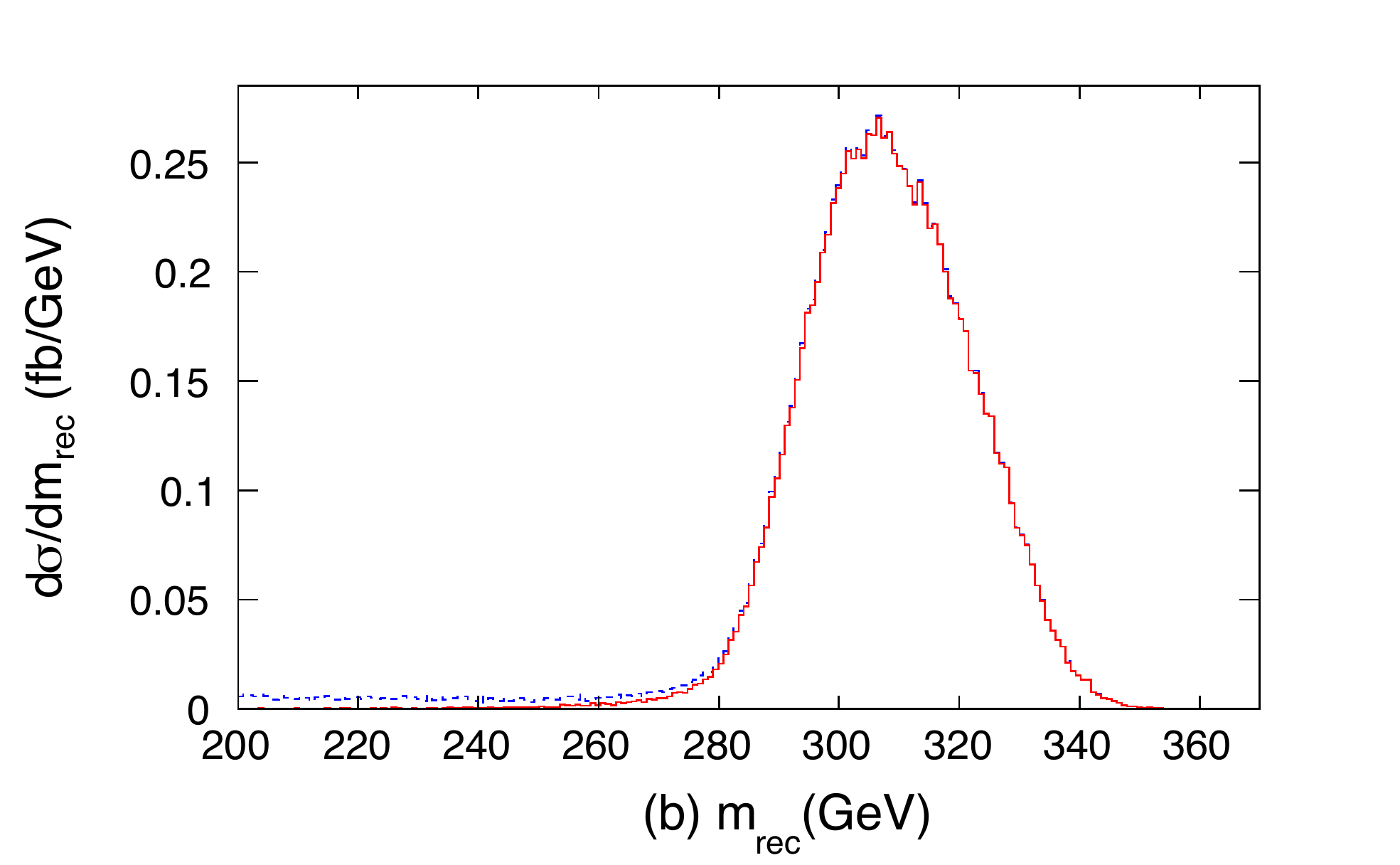}
\\
\includegraphics[width=.47\textwidth]{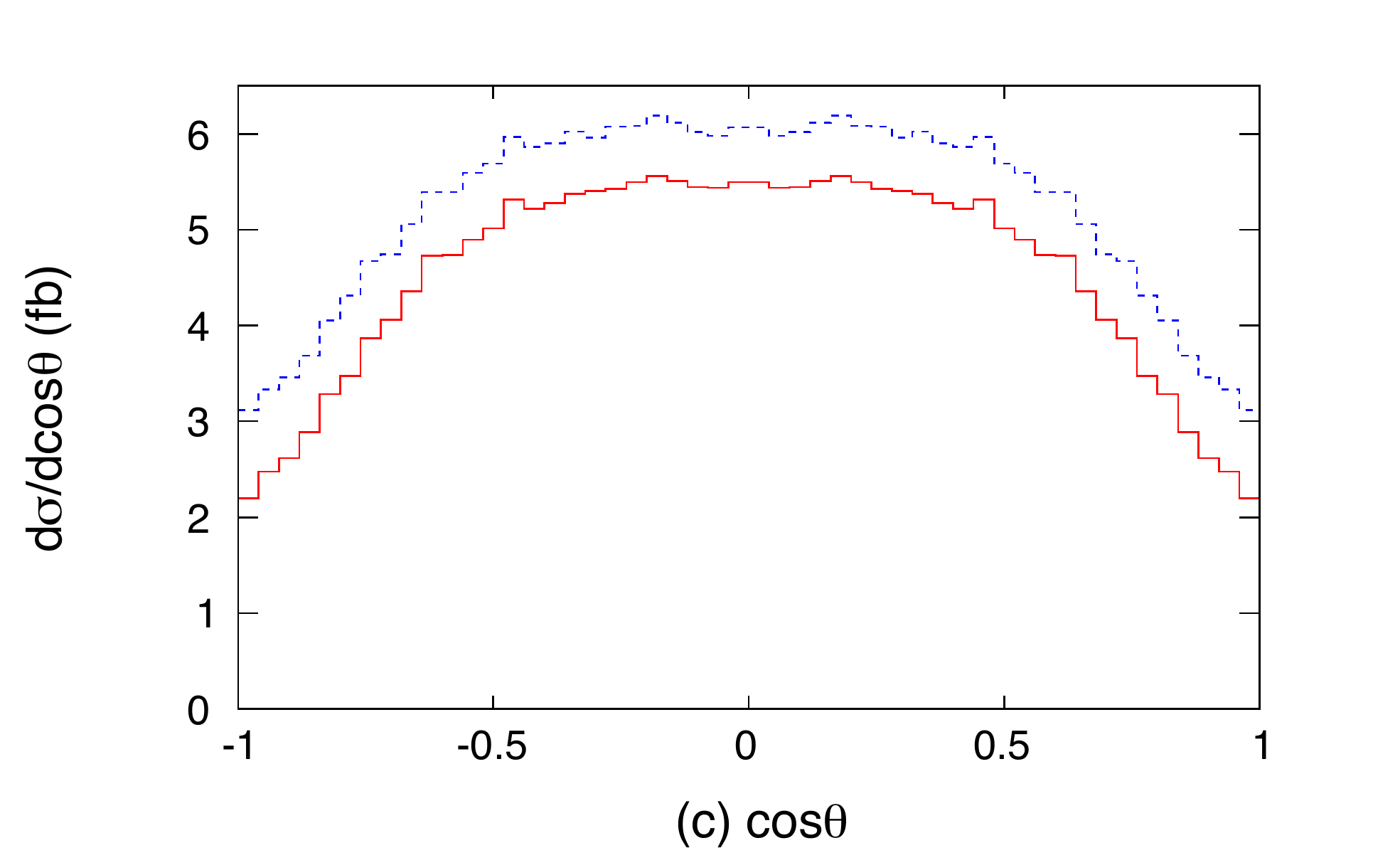}
~~
\includegraphics[width=.47\textwidth]{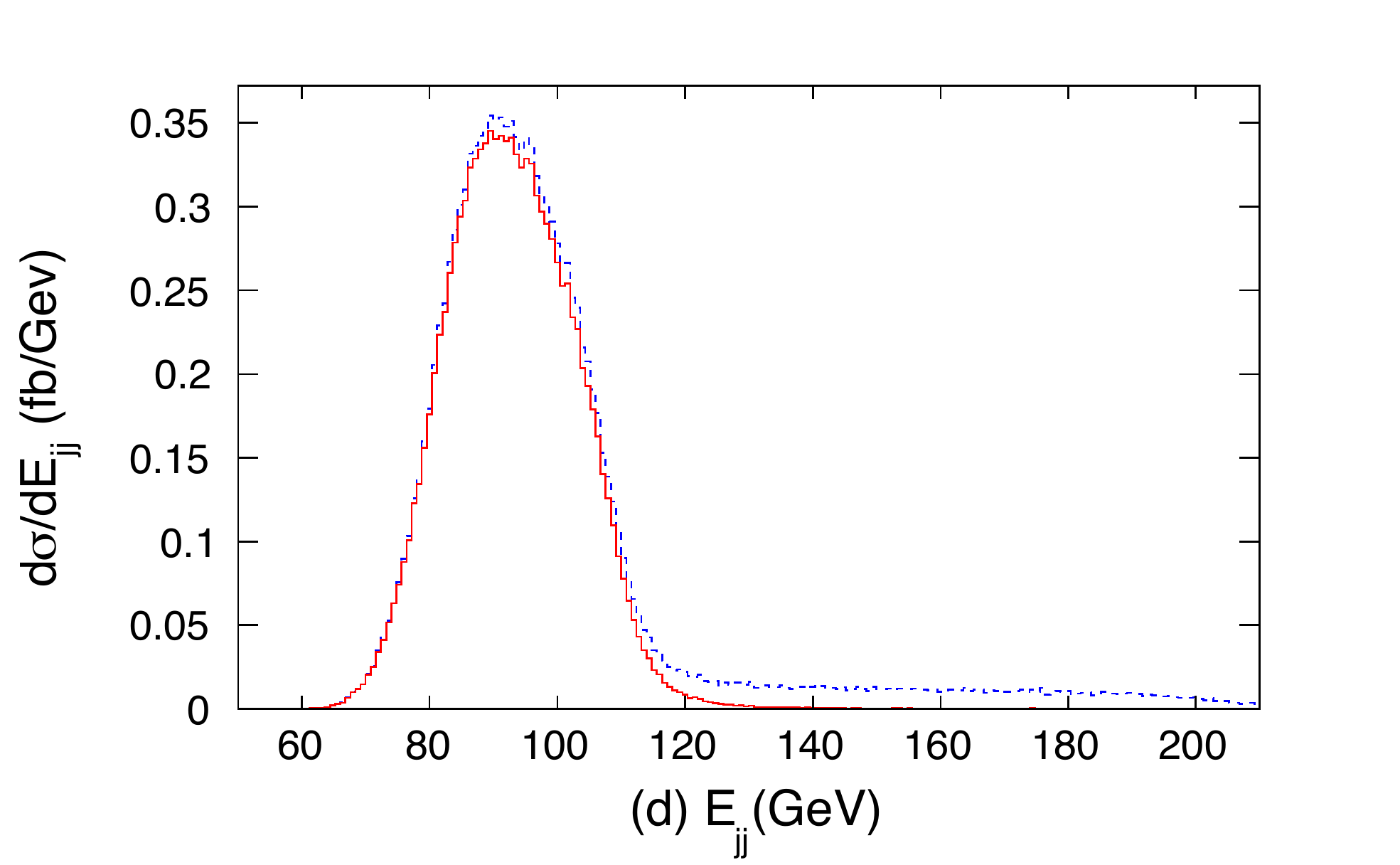}
\\
\includegraphics[width=.47\textwidth]{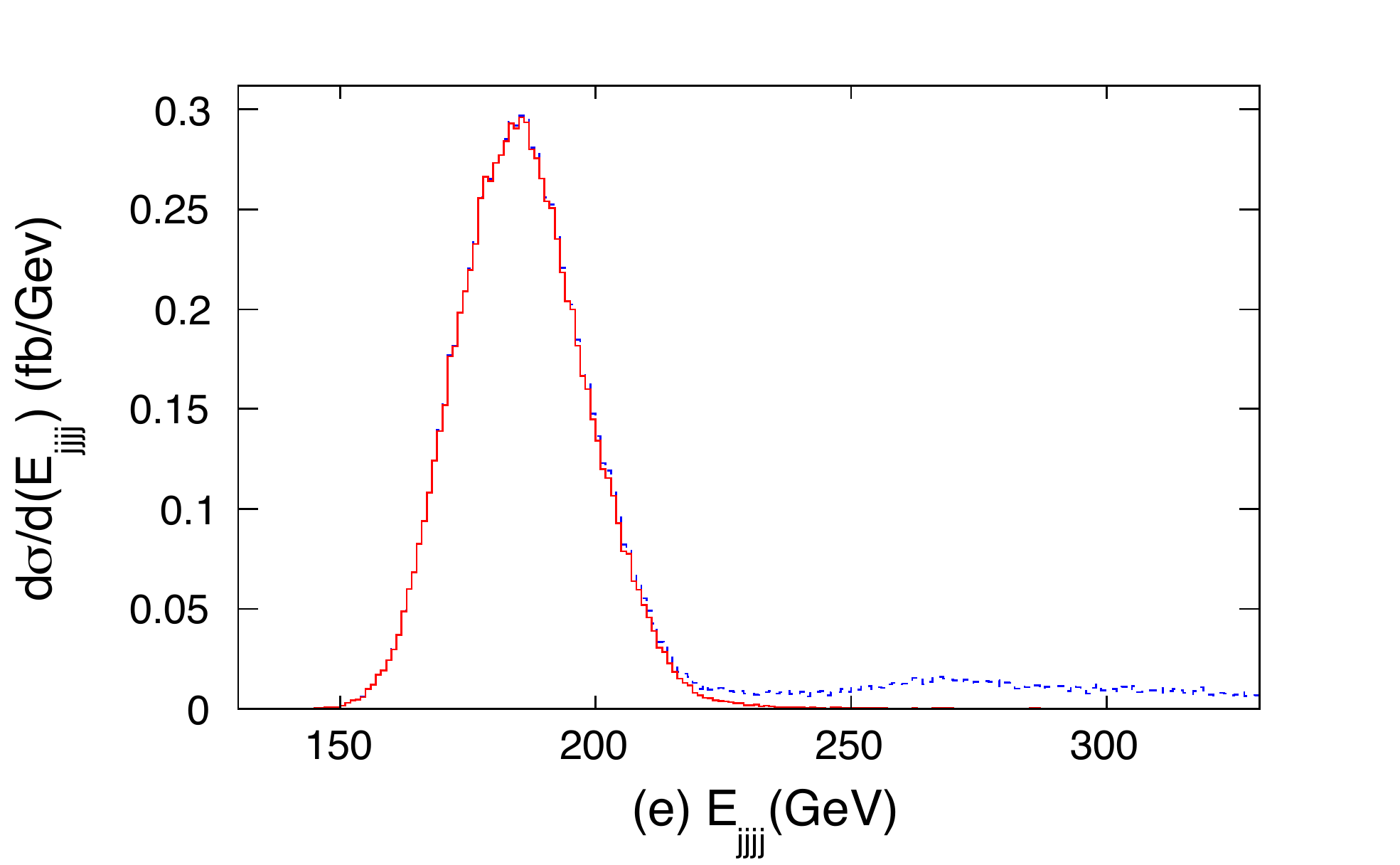}
~~
\caption{\label{fig:cha}
\casec~for $e^+ e^- \to jj,jj+ \me$ with an additional cut of $\mrec \geq 120\gev$ and $|m_{jj} - m_W|<5\Gamma_W$. 
 We show the 
 (a) $m_{jjjj}$, (b) $\mrec$, (c) $\cos\Theta$, (d) $E_{jj}$, and (e)
$E_{jjjj}$
distributions with spin-correlation and other realistic effects.
The c.m.~energy is set to $\sqrt{s}=500\gev$ for all distributions.
The solid (red) line denotes our signal of the resonant production of a chargino pair.
The dashed (blue) line is the total differential cross section
including our signal and the SM backgrounds.
}
\end{figure}

The reconstruction of the variables $m_{WW}$, $\mrec$, 
and $E_{WW}$ is straightforward in terms of the jets and the known collision frame.
In order to reconstruct $E_W$ and $\cos\Theta$, we
split the jets into two pairs and require
each pair to reconstruct an invariant mass near $m_W$. We then note that due to the symmetry of the antler decay topology, the $E_{W^+}$ and $E_{W^-}$ distributions are equal to each other and the $\cos\Theta$ distribution is symmetric with respect to an interchange of $W^+$ and $W^-$.  As a result, the $E_W$ and $\cos\Theta$ distributions can be obtained by averaging the distributions for each $W$.

In addition to our basic cuts outlined in Eq.~\eqref{eq:basic:cut}, we have applied the following cuts
\bea
\label{eq:cut:case}
&\Delta R_{jj}\equiv
\sqrt{\left(\Delta\eta_{jj}\right)^2+\left(\Delta\phi_{jj}\right)^2} \geq 0.4\ , &
\\ \no
&|m_{jj} - m_W |<5\Gamma_W\ , \quad \mrec > 120\gev\ , &
\eea
where the jet separation $\Delta R_{jj}$ is between all pairs of jets, $m_{jj}$ is only between pairs of jets identified with the $W$, and the $\mrec>120\gev$ cut removes most of the remaining SM background.
%
%
Again, we adopt the standard simulation packages ILC-Whizard setup \cite{Whizard}, including the SGV-3.0 fast detector simulation suitable for the ILC \cite{Berggren:2012ar}.

In Fig.~\ref{fig:cha}, the solid (red) lines denote our chargino signal.
The dotted (blue) lines give the total differential cross section including our signal and the SM backgrounds.
The SM backgrounds are computed through the full two-to-six processes
$e^+ e^- \to jjjj\nu \bar\nu$ which includes the full spin correlation.

Figures \ref{fig:cha}(a) and (b) show the invariant mass distributions
of four jets and two invisible particles, respectively.
Realistic effects smear the sharp $m_{jjjj}$ and $\mrec$ distributions significantly.
In particular, the locations of $m_{jjjj}^{\rm min}$
and $\mrecmin$ are shifted to lower values by about 20 GeV from the expected values
with kinematics alone in Table~\ref{table:chargino:cusp:values}.
This is mainly due to detector smearing.
The $m_{jjjj}^{\rm cusp}$ and $m_{jjjj}^{\rm max}$ are
respectively in agreement with the $\mwwc$ and $\mwwmax$ values in Table \ref{table:chargino:cusp:values}
but are significantly smeared.
The $\mrecc$ and $\mrecmax$ are larger by about 10 GeV than the expected values.
As commented earlier, the $\cos\Theta$ distribution in Fig.~\ref{fig:cha}(c) does not have a sharp cusp even before including realistic effects.

Figure \ref{fig:cha}(d) presents the $E_{jj}$ distribution
which
is significantly smeared and the sharp edges are no longer visible
due to jet energy resolution effects.
The expected values of $E_W^{\min}$ and $E_W^{\max}$ cannot be read from this distribution.
 In Fig.~\ref{fig:cha}(e),
we show the distribution of $E_{jjjj}$.
The expected triangular shapes can be seen but the sharp features are smeared due to the
realistic considerations.
Their minimum and maximum positions are moved to approximately 10 GeV lower and higher values, respectively, while
the cusp position identified with the peaks remains near the expected values.

\begin{figure}
\centering
\includegraphics[height=8cm]{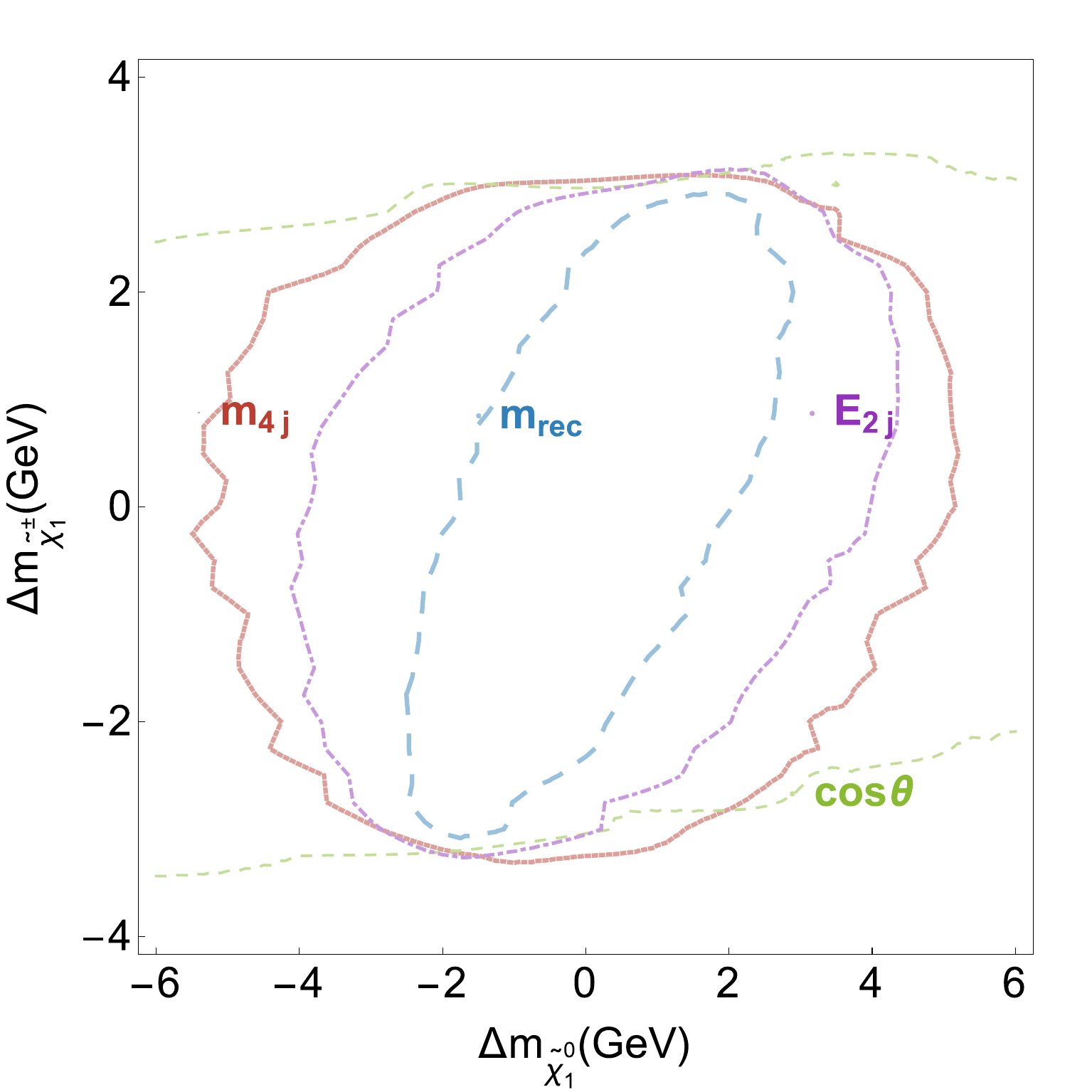}
  \caption{\label{fig:contour:caseC}
\casec~for $e^+ e^- \to jj,jj +\me$,
the 95\% C.L. contours for the precision
of the mass measurement
in the parameter space of $(\Delta m_\neuo, \Delta m_{\chao})$.
The additional cuts of $\mrec \geq 120\gev$ and $|m_{jj} - m_W|<5\Gamma_W$ are included in the distributions as well as spin-correlation and other realistic effects. The c.m.~energy is set to $\sqrt{s}=500\gev$ for all distributions and the integrated luminosity is $100\ifb$.
%
}
\end{figure}


We perform a log-likelihood analysis for the massive visible particle case and present the $95\%$ C.L.~contours for the mass measurement of $\neuo$ and $\chao$
in Fig.~\ref{fig:contour:caseC}.
Remarkable is that $\mrec$ 
leads to the most precise mass measurement,
not the commonly considered variable $E_W$,
especially on the missing particle mass.
The $E_W$ measurement leads to about $\Delta m_\neuo \simeq \pm 4\ \gev$ precision 
while the $\mrec$ improves into $\pm 2\ \gev$.
This is due to the fact that the cusp peak position is more stable with respect to detector smearing effects,
compared with the sharp energy endpoint.
The intermediate chargino mass precision is about 2 GeV both by $E_W$ and $\mrec$.
The mass measurement precision is not as good as that of the smuon pair production, because of inferior hadronic four jet measurement here.


To appreciate the improvement for the missing mass measurement with our antler approach,
we have compared it with the standard ``mono-photon'' signal,
$e^+ e^- \to \gm \,\me$~\cite{eeXXgamma,matchev}. Although this is the most model-independent method, the measurement of the endpoint in a slowly-varying $E_{\gamma}$ spectrum results in rather poor sensitivity. Besides the potential model-dependence of the signal cross section, we find that the background $e^+ e^- \to \gm\nu\bar{\nu}$ is about 100 times larger than the signal for
the benchmark point of Ref.~\cite{matchev}.
We have performed the log-likelihood analysis and find that
the best accuracy for the lightest neutralino mass determination would be no better than about $50\gev$.

\section{Summary and conclusions}
\label{sec:conclusions}

WIMP dark matter below or near the TeV scale remains a highly motivated option. To convincingly establish a WIMP DM candidate, it is ultimately important to reach consistency between direct searches and collider signals for the common parameters of mass, spin and coupling strength~\cite{Baltz:2006fm}.

Through the processes of antler decay topology at a lepton collider,
$e^+ e^- \to B_1 B_2 \to X_1 a_1 + X_2 a_2$,
we studied a new method for measuring the missing particle  mass ($m_{X}^{}$) and the intermediate particle mass ($m_{B^{}}$): the cusp method.
With this special and yet common topology,
we explored six kinematic experimentally accessible observables, $m_{aa}$, $\mrec\equiv m_{XX}$, $\cos\Theta$,
$E_a$, $E_{aa}$ and $\erec\equiv E_{XX}$.
Each of these distributions accommodates singular structures: a minimum, a cusp and  a maximum.
Their positions are determined by the kinematics only, \ie the masses of $B$, $a$, $X$
and $\sqrt{s}$, providing a powerful method to measure the particle masses $m_B^{}$ and $m_X^{}$.
We presented the analytic expressions for their positions
in terms of their masses in section \ref{sec:review}.
We chose to study the accuracy for the mass determination at a lepton collider with three benchmark scenarios in the framework of the MSSM, as listed in Table \ref{table:sps1b:mass}, and named \casea, \caseb, and \casec.

\casea~is the simplest illustration where only a right-handed smuon ($\smur$) pair is kinematically accessible.
\caseb~is slightly more complicated since both right-handed and left-handed ($\smul$) smuon pairs can be produced.
We consider the clean leptonic final state of $\mu^+\mu^- \me$ from the smuon decays.
By presenting the signal kinematics, we first confirmed the analytic expressions numerically in Fig.~\ref{fig:cusp:generic}.
We showed that, except for $m_{aa}$, due to an anticipated kinematical reason, all the other variables yield the pronounced features of a cusp distribution.
Although the SM background $e^+ e^- \to W^+W^-\to \mu^+\nu_\mu\mu^-\bar{\nu}_\mu$ also results in the antler topology, the positions of the cusps are significantly different due to the massless missing particles, the neutrinos.
This difference is used to separate the SM background very efficiently.
Furthermore, we pointed out that the experimental acceptance cuts on the observable leptons may change the positions and the shapes of the cusps in a systematic and predictable way, as seen in Figs.~\ref{fig:ptmisscut} and \ref{fig:Eacut}.

Through a full simulation including spin correlation, the SM backgrounds,
and other realistic effects, we studied how much of the idealistic features of the cusps and endpoints
survive, and how well the cusp method determines the missing particle mass for a 500 GeV ILC.
We found that the inevitable experimental effects of ISR, beamstrahlung and detector resolutions
not only distort the characteristic distributions but also shift the cusp and endpoint positions, as seen in Figs.~\ref{fig:caseA:basic}, \ref{fig:caseA:mxxcut}  and \ref{fig:caseB:mxxcut}.
The beam polarization may be used to effectively separate the final state $\smur\smur$ and $\smul\smul$, as shown in Figs.~\ref{fig:caseB:RH} and \ref{fig:caseB:LH}.
To optimize our statistical treatment, we exploited the log-likelihood method based on the Poisson probability function. The precisions for the mass measurement with various variables in \casea~were shown in Fig.~\ref{fig:contour:caseA}.
The accuracy could reach approximately $\pm 0.5\ \gev$ for smuon pair production, and was comparable for the
muon energy endpoint $E_{\mu}$ and the cusp in $\mrec$, $E_{\mu\mu}$ or $E_{XX}$.

In \casec, we studied the chargino pair production with $\chao \to W^{\pm} \neuo.$ We focused on the hadronic decay $W \to jj$ in order to effectively reconstruct the kinematics, and to explore the detector effects on the hadronic final state.
The poor energy resolution for the hadronic final state of the $W$ decay smears
the cusp and endpoint quite significantly, as shown in Fig.~\ref{fig:cha}.
We found that the $\mrec$, $E_{jjjj}$ and $\erec$ cusps are more stable than the energy endpoint $E_{jj}$ against realistic experimental effects, and thus provided a more robust mass determination reaching approximately $\pm2\ \gev$. 
In the previous section, we also made a comparison with the other proposed methods for determining the missing mass at a lepton collider. We see the merits of our approach.

Under the clean experimental environment and well-defined kinematics, a future high energy lepton collider may take advantage of the antler decay topology and provide an accurate determination for the missing particle mass consistent with the WIMP DM candidate.

\acknowledgments
This work is supported in part by the U.S.~Department of Energy under grant No.~DE-FG02-95ER40896, and
in part by PITT PACC. The work of JS is supported by NRF-2013R1A1A2061331.

\appendix

\section{\label{app:non combination improvement}Log-likelihood combination}
We have found that combining the log-likelihoods for our kinematic variables did not significantly improve the achievable accuracy of the mass measurement.  The reason for this was a combination of the correlation between the variables, the slight differences in how the log-likelihood depended on each kinematic variable, and how the combination is affected by having a large number of bins in each log-likelihood, as we will now explain.

We have found that the log-likelihood for the variables $m_{\mu\mu}$, $\mrec$, $E_\mu$, $E_{\mu\mu}$ and $\erec$
depends approximately quadratically on the mass difference $\Delta m$, where $\Delta m$ is defined to be along the diagonal line with negative slope in Fig.~\ref{fig:contour:caseA},
\begin{equation}
LL = \alpha_{kv} \left(\Delta m\right)^2\ ,
\end{equation}
where $\alpha_{kv}$ is a constant to be determined for each kinematic variable.  We will consider the optimal situation where the kinematic variables are completely uncorrelated and $\alpha_{kv}$ is the same for each kinematic variable and set $\alpha_{kv}=\alpha$.  In this case, the joint test statistic is the sum of the $N$ individual test statistics
\begin{equation}
t_{N} = N\alpha\left(\Delta m\right)^2\ .
\end{equation}
If the number of bins $n$ is large (which is a good approximation in our case with 50 bins for each log-likelihood), then the individual log-likelihoods and the joint test-statistic are well-approximated by Gaussian distributions with mean $\mu_{N}=Nn$ and standard deviation $\sigma_{N}=\sqrt{2Nn}$, where the individual log-likelihoods have $\mu_1=n$ and $\sigma_1=\sqrt{2n}$.  This means that the joint test-statistic gives a $2\sigma_{N}$ measurement in the mass difference as
\begin{equation}
N\alpha\left(\Delta m\right)_{2\sigma_{N}}^2 = Nn+2\sqrt{2Nn}
\end{equation}
while that for an individual log-likelihood has $N=1$.  Solving this for $\Delta m$ gives
\begin{equation}
\left(\Delta m\right)_{2\sigma_N} = \sqrt{\frac{n}{\alpha}+\frac{2}{\alpha}\sqrt{\frac{2n}{N}}}\ .
\end{equation}
If we take the ratio of this with an individual log-likelihood measurement, we have
\begin{equation}
\frac{\left(\Delta m\right)_{2\sigma_N}}{\left(\Delta m\right)_{2\sigma_1}} =
\sqrt{\frac{n+2\sqrt{2n/N}}{n+2\sqrt{2n}}}\ ,
\end{equation}
where $\alpha$ has dropped out.
We can use this formula to note a few things.  First of all, we see that the maximum improvement in the sensitivity achievable asymptotically approaches 0 for the large number of bin $n$ limit, independent of the number of log-likelihoods $N$ combined in this way.  Second, for $n=50$ bins, the maximum improvement in the combined measurement sensitivity is 14.5\% in the limit 
that the number of combined log-likelihoods, $N$, approaches infinity. 
Third, if we only combine $N=2$ or $3$ log-likelihoods, the maximum sensitivity improvement is only 4.3\% and 6.2\%, respectively.  This is in the best case scenario where all the variables are uncorrelated and each $\alpha_{kv}$ is identical.  In the realistic cases in this paper, the sensitivity improvement from combination is no more than a few percent.


\end{document}